\documentclass[12pt]{iopart}
\usepackage{iopams}  
\usepackage{bm}
\usepackage{graphicx}
\usepackage{latexsym}
\def\beq{\begin{equation}}
\def\eeq{\end{equation}}

\def\bea{\begin{eqnarray}}
\def\eea{\end{eqnarray}}
\usepackage[colorlinks=true]{hyperref} 
\hypersetup{
    bookmarks=true,         
    unicode=false,          
    pdftoolbar=true,        
    pdfmenubar=true,        
    pdffitwindow=false,     
    pdfstartview={FitH},    
    pdftitle={My title},    
    pdfauthor={Author},     
    pdfsubject={Subject},   
    pdfcreator={Creator},   
    pdfproducer={Producer}, 
    pdfkeywords={keyword1} {key2} {key3}, 
    pdfnewwindow=true,      
    colorlinks=true,       
    linkcolor=magenta, 
    citecolor=blue,        
    filecolor=magenta,      
    urlcolor=blue       
} 
\usepackage{mathrsfs}

\begin{document}
\begin{flushright}
\href{http://arxiv.org/abs/1801.01125}{arXiv:1801.01125}
\end{flushright}
\title[Topological order, emergent gauge fields, and Fermi surface reconstruction]{Topological order, emergent gauge fields,\\ and Fermi surface reconstruction}

\author{Subir Sachdev}

\address{Department of Physics, Harvard University, Cambridge MA 02138, USA}
\address{Perimeter Institute for Theoretical Physics, Waterloo, Ontario, Canada N2L 2Y5}
\address{Department of Physics, Stanford University, Stanford CA 94305, USA}
\ead{sachdev@g.harvard.edu}
\vspace{10pt}
\begin{indented}
\item[]December 2017
\end{indented}

\begin{abstract}
This review describes how topological order associated with the presence of emergent gauge fields can reconstruct Fermi surfaces of metals,
even in the absence of translational symmetry breaking. We begin with an introduction to topological order
using Wegner's quantum $\mathbb{Z}_2$ gauge theory on the square lattice: the topological state is characterized
by the expulsion of defects, carrying $\mathbb{Z}_2$ magnetic flux.
The interplay between topological order
and the breaking of global symmetry is described by the non-zero temperature 
statistical mechanics of classical XY models in dimension $D=3$;
such models also describe the zero temperature quantum phases of 
bosons with short-range interactions on the square lattice at integer filling. The topological state is again
characterized by the expulsion of certain defects, in a state with fluctuating symmetry-breaking order, along with the presence of emergent gauge fields. The phase diagrams of the
$\mathbb{Z}_2$ gauge theory and the XY models are obtained by embedding them in U(1) gauge theories, and by
studying their Higgs and confining phases. These ideas are then applied to the single-band Hubbard model on the square lattice. A SU(2) gauge theory describes
the fluctuations of spin-density-wave order, and its phase diagram is presented by analogy to the XY models.
We obtain a class of zero temperature metallic states with fluctuating spin-density wave order, topological order associated with defect expulsion, deconfined emergent gauge fields, reconstructed Fermi surfaces (with `chargon' or electron-like quasiparticles), but no broken symmetry.
We conclude with the application of such metallic states to the pseudogap phase of the cuprates,
and note the recent comparison with numerical studies of the Hubbard model and photoemission observations of the electron-doped cuprates. 
In a detour, we also discuss the influence of Berry
phases, and how they can lead to deconfined quantum critical points: this applies to bosons on the square lattice
at half-integer filling, and to quantum dimer models.
\end{abstract}

%
\vspace{2pc}

\noindent
{\tt Partly based on lectures at the 34th Jerusalem Winter School in Theoretical Physics: New Horizons in Quantum Matter,
December 27, 2016 - January 5, 2017}\\

\submitto{\RPP}
%
%
%
\newpage
\tableofcontents
\title[Topological order, emergent gauge fields, and Fermi surface reconstruction]{}

\maketitle

\section{Introduction}
\label{sec:intro}

The traditional theory of phase transitions relies crucially on symmetry: phases with and without a spontaneously broken
symmetry must be separated by a phase transition. However, recent developments have shown that the `topological order'
associated with emergent gauge fields
can also require a phase transition between states which cannot be distinguished by symmetry.

Another powerful principle of traditional condensed matter physics is the Luttinger theorem \cite{Luttinger}: 
in a system with a globally conserved U(1) charge, 
the volume enclosed by all the Fermi
surfaces with quasiparticles carrying the U(1) charge, must equal the total conserved density multiplied by a known phase-space factor, and modulo filled bands.
Spontaneous breaking of translational symmetry ({\it e.g.\/} by a spin or charge density wave) can reconstruct the Fermi surface
into small pockets, because the increased size of the unit cell allows filled bands to account for a larger fraction of the fermion density.
But it was long assumed that in the absence of translational symmetry breaking, the Fermi surface cannot reconstruct in a Fermi volume changing transition.

More recently, it was realized that the Luttinger theorem has a topological character \cite{MO00}, and that it is possible for 
topological order associated with emergent gauge fields 
to change the volume enclosed by the Fermi surface \cite{TSMVSS04,APAV04}. So we can have a phase transition
associated with the onset of topological order, across which the Fermi surface reconstructs, even though there is no symmetry breaking
on either side of the transition.
This review will present a sequence of simple models which introduce central concepts in the theory of emergent gauge fields,
and give an explicit demonstration of the reconstruction of the Fermi surface by such topological order \cite{SBCS16}.

Evidence for Fermi surface reconstruction 
has recently appeared in photoemission experiments \cite{ZXtopo} on 
the electron-doped cuprate superconductor Nd$_{2-x}$Ce$_x$CuO$_4$, in a region of electron density without antiferromagnetic order. Given the theoretical arguments \cite{TSMVSS04,APAV04}, this
constitutes direct experimental evidence for the presence of topological order. In the hole-doped cuprates, Hall effect 
measurements \cite{LTCP15} on YBa$_2$Cu$_3$O$_y$ indicate a small Fermi surface at near optimal hole densities without any density wave order, and the doping dependence of the Hall co-efficient fits well a theory of Fermi surface reconstruction by topological order \cite{EMSY16,CSE17}.
Also in YBa$_2$Cu$_3$O$_y$, but at lower hole-doping, quantum oscillations have been observed, and are likely in a region where there is
translational symmetry breaking due to density wave order \cite{SP15}; however, the quantum oscillation \cite{Greven16} and 
specific heat \cite{Boebinger11} observations indicate the presence of only a single electron pocket, and these are difficult to understand in a model without prior Fermi surface reconstruction \cite{ACS14} (and pseudogap formation) due to topological order. 

This is a good point to pause and clarify what we mean here 
by `topological order', a term which has acquired different meanings in the recent literature. Much interest has focused recently
on topological insulators and superconductors \cite{HasanRMP,ZhangRMP,HasanARCMP} such as Bi$_{1-x}$Sb$_x$. The topological order in these materials is associated with protected electronic states on their boundaries, while the bulk contains only `trivial' excitations which can 
composed by ordinary electrons and holes. They are, therefore, analogs of the {\it integer\/} quantum Hall effect, 
but in zero magnetic field and with time-reversal symmetry preserved; we are not interested in this type of topological order here. Instead, our interest lies in analogs of the {\it fractional\/} quantum Hall effect, but in zero magnetic field and with time-reversal symmetry preserved. States with this type of 
topological order have fractionalized excitations in the bulk {\em i.e.\/} excitations which cannot be created individually by 
the action any local operator; protected boundary excitations may or may not exist, depending upon the flavor of the bulk topological order. 
The bulk fractionalized excitations carry the charges of deconfined {\em emergent gauge fields\/} in any effective theory of the bulk.
Most studies of states with this type of topological order focus on the cases where there is a bulk
energy gap to all excitations, and examine degeneracy of the ground state on manifolds with non-trivial topology. However, we are interested here (eventually) in states with gapless excitations in the bulk, including metallic states with Fermi surfaces: the bulk 
topological order and emergent gauge fields remain robustly defined even in such cases.

We will begin in Section~\ref{sec:ising} 
by describing the earliest theory of a phase transition without a local symmetry breaking order parameter:
Wegner's $\mathbb{Z}_2$ gauge theory 
in $D=3$ \cite{wegner71,kogut79}. Throughout, the symbol $D$ will refer to the spatial dimensionality of classical models at non-zero 
temperature,
or the spacetime dimensionality of quantum models at zero temperature; 
we will use $d=D-1$ to specify the spatial dimensionality of quantum models.
Wegner distinguished the two phases of the gauge theory, confining and deconfining, by computing the behavior of Wegner-Wilson loops,
and finding area-law and perimeter-law behaviors respectively. However, this distinction does not survive the introduction of various
dynamical matter fields, whereas the phase transition does \cite{FradkinShenker}. The modern perspective on Wegner's phase transition
is that it is a transition associated with the presence of topological order in the deconfined phase: this will be presented in 
Section~\ref{sec:z2topo}, along with an introduction to the basic characteristics of topological order. In particular, a powerful
and very general idea is that the 
expulsion of topological defects leads to topological order: for the quantum $\mathbb{Z}_2$ gauge theory in $D=2+1$,
$\mathbb{Z}_2$ fluxes in a plaquette are expelled in the topologically ordered ground state, in a manner reminiscent of the
Meissner flux expulsion in a superconductor \cite{FradkinShenker} (as will become clear in the presentation of Section~\ref{sec:z2even}).

Section~\ref{sec:xy} will turn to the other well-known example of a phase transition without a symmetry-breaking
order parameter, the Kosterlitz-Thouless (KT) transition \cite{Berezinskii1,Berezinskii2,KT73,KT74} 
of the classical XY model at non-zero temperature in $D=2$. The low temperature ($T$) phase was explicitly
recognized by KT as possessing topological order due to the expulsion of free vortices in the XY order, and an associated 
power-law decay of correlations
of the XY order parameter. KT stated in their abstract \cite{KT73} ``A new definition of order called topological order is proposed for two-dimensional systems in which no long-range order of the conventional type exists''.
Despite the absence of conventional long-range-order (LRO), KT showed that there was a phase transition,
at a temperature $T_{KT}$, driven by the proliferation of vortices, which led to the exponential decay of XY correlations
for $T> T_{KT}$.

\begin{figure}[htb]
\begin{center}
\includegraphics[height=3.5in]{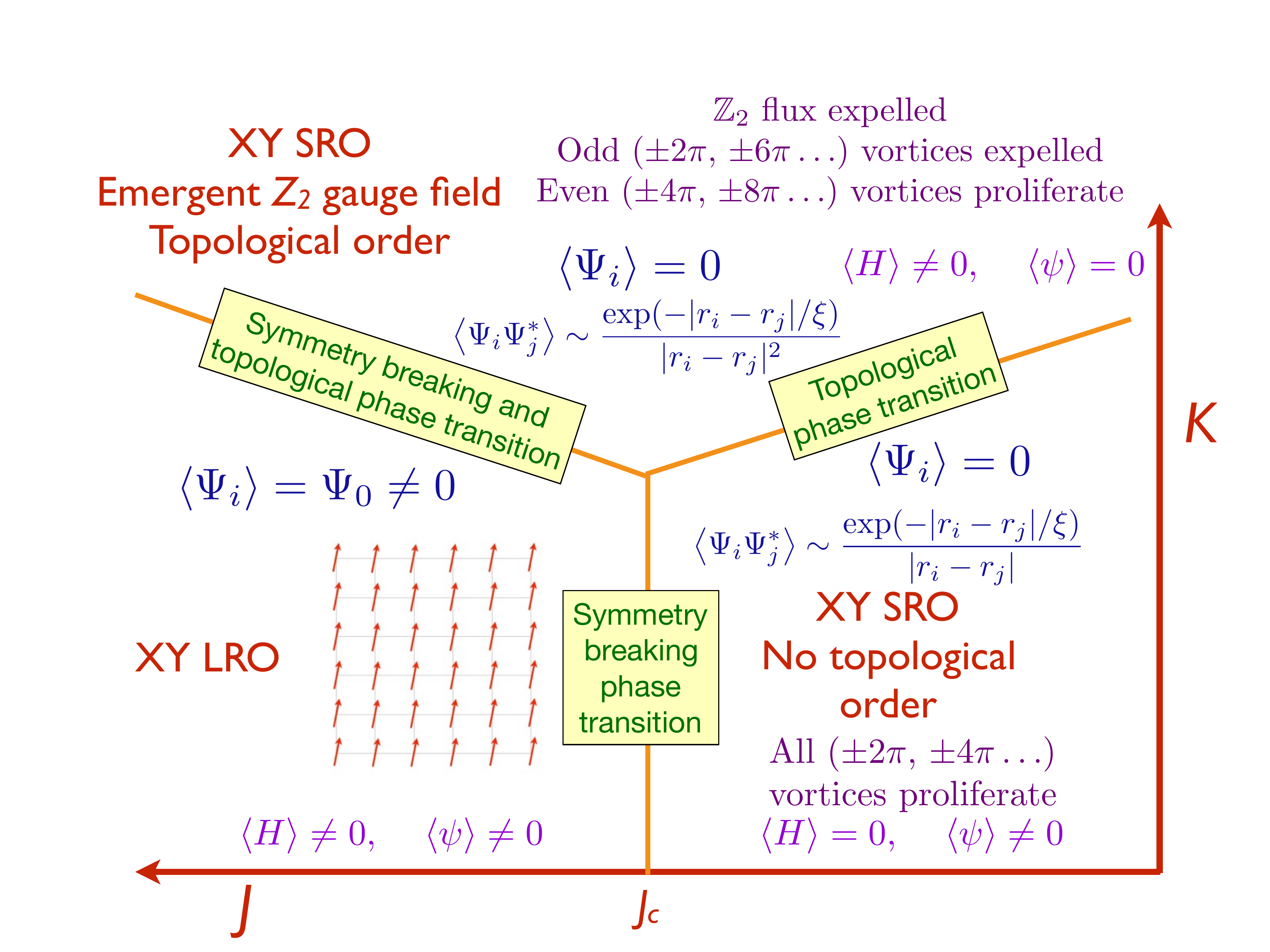}
\end{center}
\caption{Schematic phase diagram of the classical $D=3$ XY model at non-zero temperature in Eq.~(\ref{tzxy}), or the quantum $D=2+1$
XY model at zero temperature in Eq.~(\ref{bhxy}) along with the constraint in Eq.~(\ref{GXY1}). The XY order parameter is $\Psi$ (Eq.~(\ref{defPsi}).
These models correspond to the case of bosons on the square lattice with short-range interactions, and at integer filling.
The meaning of the Higgs field $H$, and half-boson-number field $\psi$ will become clear in Section~\ref{sec:xyeven},
but we note here that $\Psi = H \psi^2$ (Eq.~(\ref{PsiHpsi})).
The two SRO phases differ in the prefactor of exponential decay of correlations of 
the order parameter. But more importantly, the large $K$ phase has  topological order 
associated with the expulsion of odd vortices: this topological order is associated with an emergent $\mathbb{Z}_2$ gauge field, and is the same
as that in the $\mathbb{Z}_2$ gauge theory of Section~\ref{sec:ising} at small $g$. The transition between the SRO phases is
also in the same university class as the confinement-deconfinement transition of the $\mathbb{Z}_2$ gauge theory of Section~\ref{sec:ising}.
A numerical simulation of a model with the same phase diagram is in Ref.~\cite{SM02}.}
\label{fig:xytopo}
\end{figure}

Section~\ref{sec:xytopo} turns to XY models in $D=3$, where we show that topological order, similar to that found by KT in $D=2$, 
is possible also in three (and higher) dimensions. 
We examine $D=3$ classical XY
models at non-zero temperature with suitable short-range couplings between the XY spins; these 
models are connected to $d=2$ quantum models at zero temperature of 
bosons with short-range interactions 
on the square lattice at integer filling \cite{RJSS91,SSMV99,SM02,SM02PRL}. 
The situation is however more subtle than in $D=2$: the topologically
ordered phase in $D=3$ has exponentially decaying XY correlations, unlike the power-law correlations in $D=2$. There is also a 
`trivial' disordered phase with exponentially decaying correlations, as shown in Fig.~\ref{fig:xytopo}; but
the two phases with short-range order (SRO) in Fig.~\ref{fig:xytopo} are distinguished by the power-law prefactor of the 
exponential decay. 
More importantly, the topological phase only expels vortices with a winding number which is an {\it odd multiple of $2\pi$}; 
the latter should be compared with the expulsion of {\it all} vortices in the KT topological phase of the XY model in $D=2$.
The topological phase also has an emergent $\mathbb{Z}_2$ gauge field, and the topological order is the same as that in the $\mathbb{Z}_2$ gauge theory of Section~\ref{sec:ising} at small $g$.
Including the phase with XY 
long-range order (LRO), we have 3 possible states, arranged schematically as in Fig.~\ref{fig:xytopo}. This phase diagram will form a template
for subsequent phase diagrams of more complex models that are presented in this review; in particular the interplay between topological
and symmetry-breaking phase transitions will be similar to that in Fig.~\ref{fig:xytopo}.

Section~\ref{sec:z2even} introduces a powerful technical tool in the analysis of topological states and their phase transitions.
We embed the model into a related theory with a large local gauge invariance, and then use the Higgs mechanism
to reduce the residual gauge invariance: this leads to states with the topological order of interest. The full gauge theory
allows one to more easily incorporate matter fields, including gapless matter, and to account for global symmetries of the Hamiltonian.
In Section~\ref{sec:xyeven}, we apply this method to the $D=3$ XY models of Section~\ref{sec:xytopo}; after incorporating the methods of 
particle-vortex duality, we will obtain a field-theoretic description of all the phases in Fig.~\ref{fig:xytopo}, and potentially also
of the phase transitions. 

Sections~\ref{sec:z2odd} and~\ref{sec:xyodd} are a detour from the main presentation, and may be skipped on an
initial reading. Here we consider the influence of static background electric charges on the gauge theories of topological phases.
These charges introduce Berry phases, and we describe the
subtle interplay between these Berry phases and the manner in which the square lattice space group symmetry is realized.
Such background charges are needed to describe the boson/XY models of Section~\ref{sec:xytopo} for the cases when the boson
density is half-integer; these models also correspond to easy-plane $S=1/2$ antiferromagnets on the square lattice, and to quantum dimer models
\cite{DRSK88,EFSK90,NRSS90,RJSS91,SSMV99}, and so are of
considerable physical importance. We find several new phenomena: the presence of `symmetry-enriched' topological (SET) phases \cite{EH13,SPTSET} with a projective symmetry group $D_8$ (the 16 element non-abelian dihedral group) \cite{HPS11},
the necessity of broken translational symmetry (with valence bond solid (VBS) order) in the 
confining phase \cite{NRSS89,NRSS90,RJSS91,SSMV99}, and the presence of
deconfined critical points \cite{RJSS91,SSMV99,senthil1,senthil2}. 
In particular, the larger gauge groups of Section~\ref{sec:higgs} are not optional at the deconfined
critical points, and remain unbroken in the deconfined critical theory.
\begin{figure}[htb]
\begin{center}
\includegraphics[height=4.1in]{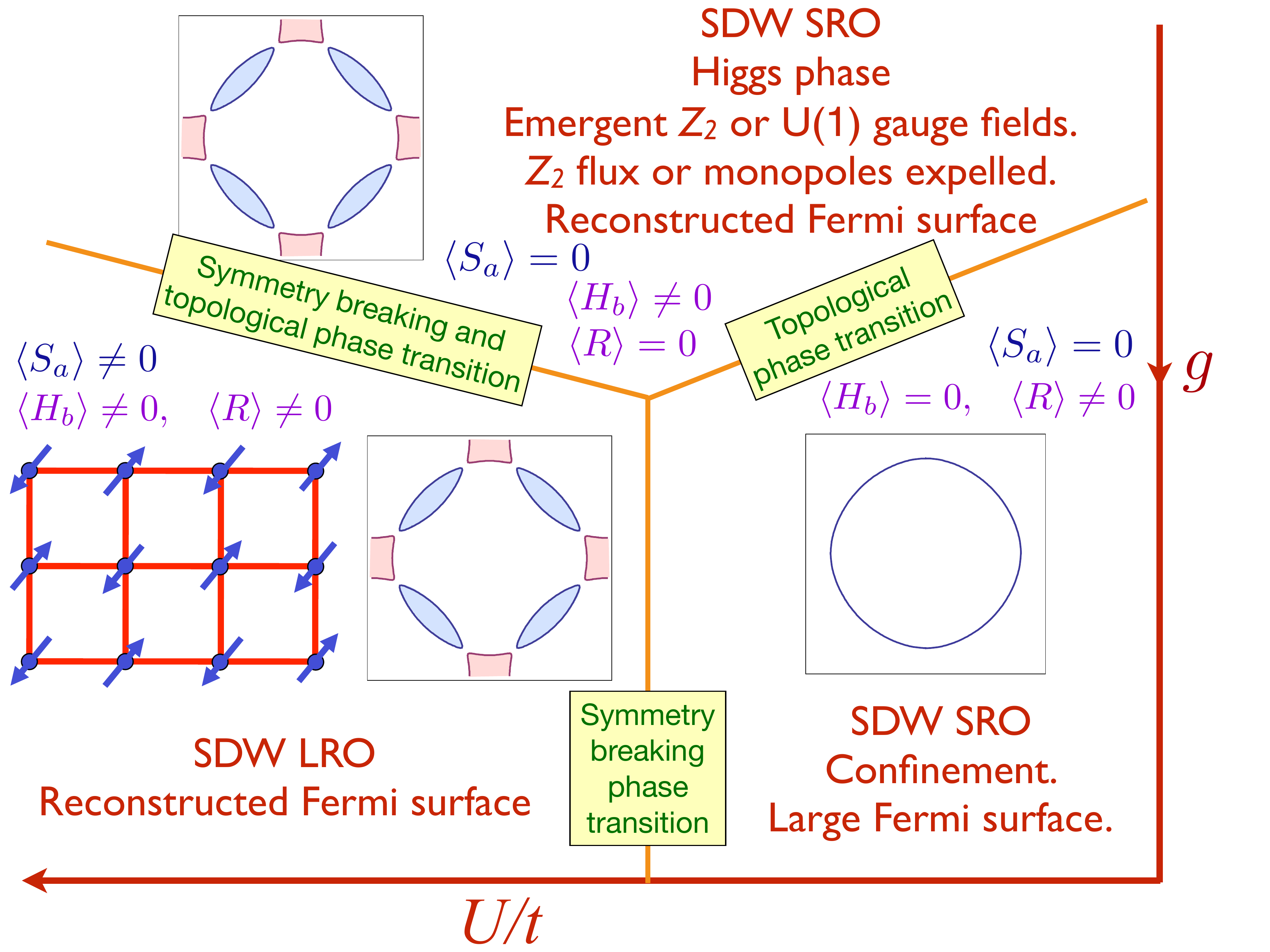}
\end{center}
\caption{Schematic phase diagram of the electronic Hubbard model at generic density, to be discussed in Section~\ref{sec:hubbard}. 
Note the similarity to Fig.~\ref{fig:xytopo}. These phases are realized in a formulation with an emergent
SU$_s$(2) gauge field. The condensation of the Higgs field, $H_b$, can break the gauge invariance down to smaller
groups. The `spinon' field, $R$ carries charges under both the gauge SU$_s$(2), and the global SU(2) spin, and it is the analog
of $\psi$ in Fig.~\ref{fig:xytopo}. The SDW order $S_a$ is related to $H_b$ and $\psi$ via Eq.~(\ref{SH}), which is the analog
of Eq.~(\ref{PsiHpsi}) for the XY model.
The emergent gauge fields and topological order are associated with the expulsion of defects in the SDW order. 
The reconstructed Fermi surface in the state with topological order can have `chargon' ($f_p$) or electron-like quasiparticles.
At half-filling, the states with reconstructed Fermi surfaces can become insulators without Fermi surfaces (in this case, the insulator
with U(1) topological order is unstable to confinement and valence bond solid (VBS) order).}
\label{fig:sdwtopo}
\end{figure}

Section~\ref{sec:hubbard} will turn finally to the important case of electronic Hubbard models
on the square lattice. Here, we will present a SU$_s$(2) gauge theory \cite{SS09} which contains phases closely analogous
to those of the $D=3$ XY model in Fig.~\ref{fig:xytopo}, as is clear from Fig.~\ref{fig:sdwtopo}. 
We use the subscript $s$ in the gauge theory to distinguish from the global SU(2)
spin rotation symmetry (which will have no subscript).
Note the similarity between Figs.~\ref{fig:xytopo} and~\ref{fig:sdwtopo}
in the placement of the topological and symmetry-breaking phase transitions.
The simplest state of the Hubbard model is the one adiabatically connected to the free electron limit.
This has no broken symmetries, and has a `large' Fermi surface which obeys the Luttinger theorem.
The Hubbard model also
has states with conventional broken symmetry, and we focus on the case with spin-density wave (SDW)
order: the SDW order breaks translational symmetry, and so the Fermi surface can reconstruct
in the conventional theory, as we review in Section~\ref{sec:sdw}. But the state of greatest interest in the present
paper is the one with topological order and no broken symmetries, shown at the top of Fig.~\ref{fig:sdwtopo}.
We will show that this state is also characterized by the expulsion of topological defects, and a deconfined emergent $\mathbb{Z}_2$ or U(1)
gauge field. The expulsion of defects will be shown to
allow reconstruction of the Fermi surface into small pocket Fermi surfaces with `chargon' ($f_p$) or electron-like
quasiparticles.

Finally, Section~\ref{sec:conc} will briefly note application of these results on fluctuating SDW order to the pseudogap
phase of the cuprate superconductors \cite{SSRoyal,SSNambu,SCSS17,CSS17,WSCSGF,SCWFGS,MSSS18}. Experimental connections  \cite{ZXtopo,LTCP15,SP15,Greven16,Boebinger11} were already noted above.
We will also mention extensions \cite{XS10} which incorporate 
pairing fluctuations into more general theories of fluctuating order for the pseudogap.

\section{$\mathbb{Z}_2$ gauge theory in $D=2+1$}
\label{sec:ising}

Wegner defined the $\mathbb{Z}_2$ gauge theory as a classical statistical mechanics partition function on the
cubic lattice. We consider the partition function \cite{wegner71}
\begin{eqnarray}
\widetilde{\mathcal{Z}}_{\mathbb{Z}_2} &=& \sum_{\{\sigma_{ij}\} = \pm 1}  
\exp \left( - \widetilde{\mathcal{H}}_{\mathbb{Z}_2} /T \right) \nonumber \\
\widetilde{\mathcal{H}}_{\mathbb{Z}_2} &=&  - K \sum_{\square}
\prod_{(ij) \in \square} \sigma_{ij}\,, \label{tz2}
\end{eqnarray}
The degrees of freedom in this partition function are the binary variables $\sigma_{ij} = \pm 1$ on the links $\ell \equiv (ij)$ of the cubic lattice.
The $\square$ indicates the elementary plaquettes of the cubic lattice.

\begin{figure}[htb]
\begin{center}
\includegraphics[height=1.7in]{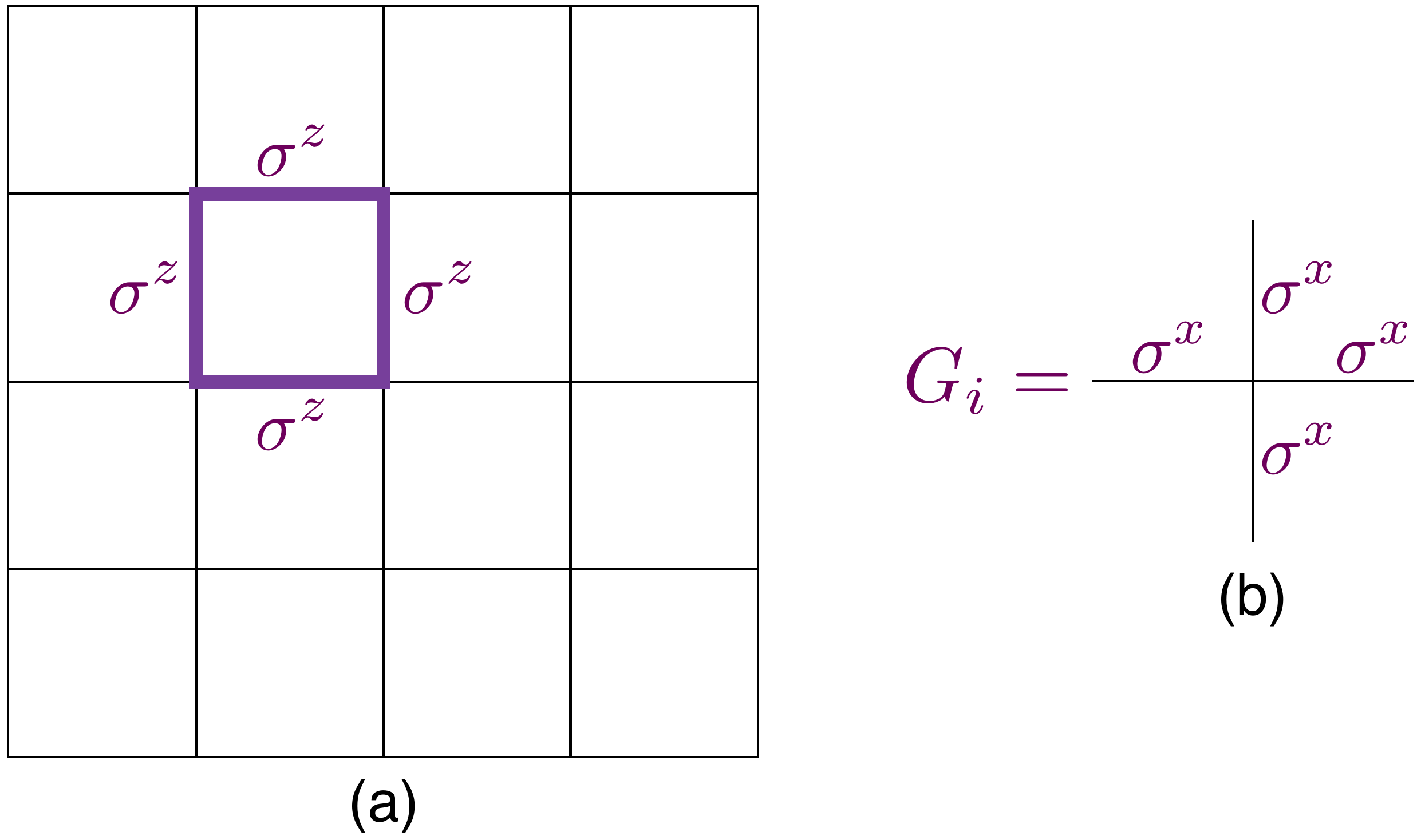}
\end{center}
\caption{(a) The plaquette term of the $\mathbb{Z}_2$ lattice gauge theory. (b) The operators $G_i$ which generate
$\mathbb{Z}_2$ gauge transformations.}
\label{fig:plaq}
\end{figure}

We will present our discussion in this section entirely in terms of the corresponding quantum model on the square lattice.
This degrees of freedom of the quantum model are
qubits on the links, $\ell$, of a square lattice. The Pauli operators $\sigma^{\alpha}_\ell$ ($\alpha = x,y,z$)
act on these qubits, and $\sigma_{ij}$ variables in Eq.~(\ref{tz2}) are promoted to the operators $\sigma^z_\ell$ on the spatial links.
We set $\sigma_{ij}=1$ on the temporal links as a gauge choice.
The Hamiltonian of the quantum $\mathbb{Z}_2$ gauge theory is \cite{wegner71,kogut79}
\beq
\mathcal{H}_{\mathbb{Z}_2} = - K \sum_{\square}\, \prod_{\ell \, \in \, \square} \sigma^z_\ell - g \sum_{\ell} \sigma^x_{\ell} \,,
\label{Hz2}
\eeq
where $\square$ indicates the elementary plaquettes on the square lattice, as indicated in Fig.~\ref{fig:plaq}a.

On the infinite square lattice, we can define operators on each site, $i$, of the lattice which commute with $\mathcal{H}_{\mathbb{Z}_2}$
(see Fig.~\ref{fig:plaq}b)
\beq
G_i = \prod_{\ell \, \in \, +} \sigma^x_{\ell} \,, \label{defGi}
\eeq
which clearly obey $G_i^2 = 1$.
We have $G_i \sigma^z_\ell G_i = \varrho_i \sigma^z_\ell$, where $\varrho_i = -1$ only if the site $i$ is at the end of link $\ell$,
and $\varrho_i = 1$ otherwise: the $G_i$ generates a space-dependent $\mathbb{Z}_2$ gauge transformation on the site $i$.
There are an even number of $\sigma^z_\ell$ emanating from each site in the $K$ term in $\mathcal{H}_{\mathbb{Z}_2}$, and so 
\beq
[\mathcal{H}_{\mathbb{Z}_2}, G_i] = 0 \,.
\eeq
The spectrum of $\mathcal{H}_{\mathbb{Z}_2}$ depends upon the values of the conserved $G_i$, and here we will take 
\beq
G_i = 1 \,; \label{G1}
\eeq
this corresponds
to a `pure' $\mathbb{Z}_2$ gauge theory with no matter fields. We will consider matter fields later.

\begin{figure}[htb]
\begin{center}
\includegraphics[height=2.5in]{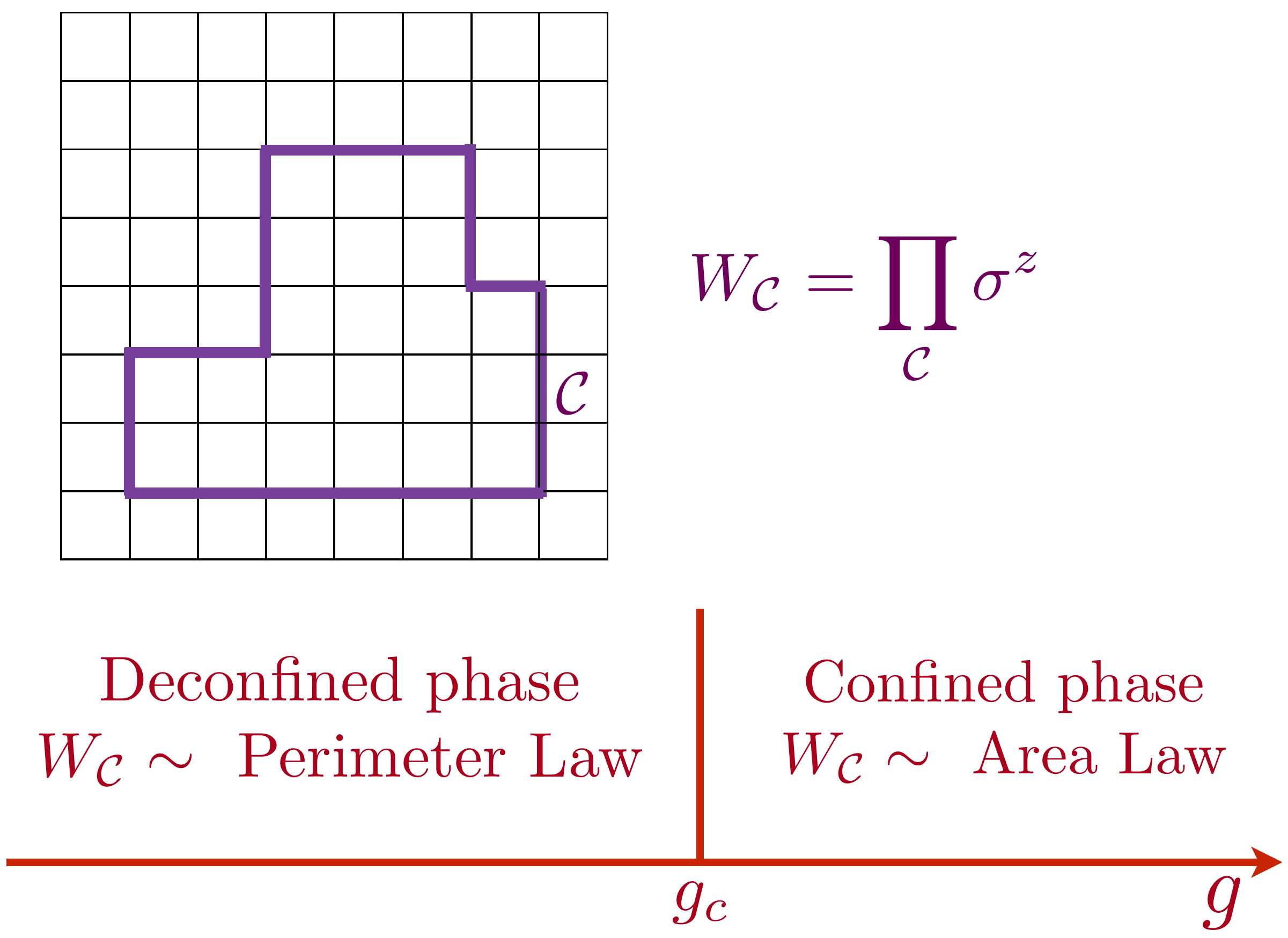}
\end{center}
\caption{The Wegner-Wilson loop operator $W_\mathcal{C}$ on the closed loop $\mathcal{C}$. Shown below is a schematic ground state phase diagram of $\mathcal{H}_{\mathbb{Z}_2}$, with the distinct behaviors of $W_\mathcal{C}$ in the 
deconfined and confined phases.}
\label{fig:wilson}
\end{figure}

Wegner \cite{wegner71} showed that there were two gapped phases in the theory, which are necessarily separated by
a phase transition. Remarkably, unlike all previously known cases, this phase transition was not required
by the presence of a broken symmetry in one of the phases: there was no local order parameter characterizing
the phase transition. Instead, Wegner argued for the presence of a phase transition using the behavior 
of the Wegner-Wilson loop operator $W_{\mathcal{C}}$, which is the product of $\sigma^z$ on the links of any
closed contour $\mathcal{C}$ on the direct square lattice, as  illustrated in Fig.~\ref{fig:wilson}. ($W_\mathcal{C}$ is usually, and improperly, referred to just 
as a Wilson loop.)
The two phases are:\\
({\em i\/}) At $g \gg K$ we have the `confining' phase. 
In this phase $W_\mathcal{C}$ obeys the area law: $\left\langle W_{\mathcal{C}} \right\rangle \sim \exp( - \alpha A_{\mathcal{C}})$ for large contours $\mathcal{C}$, 
where $A_{\mathcal{C}}$ is the area enclosed by the contour $\mathcal{C}$ and $\alpha$ is a constant. 
This behavior can easily be seen by a small $K$ expansion of $\left\langle W_{\mathcal{C}} \right\rangle$: one power of $K$ is needed for every plaquette enclosed by $\mathcal{C}$ for the first non-vanishing contribution to $\mathcal{W}_C$.
The rapid decay of $\left\langle W_{\mathcal{C}} \right\rangle$ is a consequence of the large fluctuations in the
$\mathbb{Z}_2$ flux, $\prod_{\ell \, \in \, \square} \sigma^z_\ell$, through each plaquette\\
({\em ii\/}) At $K \gg g$ we have the `deconfined' phase. In this phase, the $\mathbb{Z}_2$ flux 
is expelled, and $\prod_{\ell \, \in \, \square} \sigma^z_\ell$ usually equals $+1$ in all plaquettes.
We will see later that the flux expulsion is analogous to the Meissner effect in superconductors.
The small residual fluctuations of the flux lead to a perimeter law decay, 
$\left\langle W_{\mathcal{C}} \right\rangle \sim \exp( - \alpha' P_{\mathcal{C}})$ for large contours $\mathcal{C}$, 
where $P_{\mathcal{C}}$ is the perimeter of the contour $\mathcal{C}$ and $\alpha'$ is a constant. 

Along with establishing the existence of a phase transition using the distinct behaviors of the Wegner-Wilson loop,
Wegner also determined the critical properties of the transition. He performed a Kramers-Wannier duality transformation,
and showed that the $\mathbb{Z}_2$ gauge theory was equivalent to the classical Ising model. This establishes that
the confinement-deconfinement transition is in the universality class of the the Ising
Wilson-Fisher \cite{wilsonfisher72} conformal field theory in 3 spacetime dimensions (a CFT3).
The phase with the dual Ising order is the confining phase, and the phase with Ising `disorder' is the deconfined phase.
We will derive the this Ising criticality in Section~\ref{sec:z2even} by a different method. For now, we note that 
the critical theory is not precisely the Wilson-Fisher Ising CFT, but 
what we call the Ising* theory. In the Ising* theory, the only allowed
operators are those which are invariant under $\phi  \rightarrow -\phi$, where $\phi$ is the Ising primary 
field \cite{2016PhRvL.117u0401S,2016PhRvB..94h5134W}.

\subsection{Topological order}
\label{sec:z2topo}

While Wegner's analysis yields a satisfactory description of the pure $\mathbb{Z}_2$ gauge theory, the Wegner-Wilson
loop is, in general, not a useful diagnostic for the existence of a phase transition. Once we add dynamical matter fields
(as we will do below), $W_{\mathcal{C}}$ invariably has a perimeter law decay, although the confinement-deconfinement phase transition can persist.

The modern interpretation of the existence of the phase transition in the $\mathbb{Z}_2$ lattice gauge theory is that
it is present because the deconfined phase has $\mathbb{Z}_2$ `topological' order \cite{NRSS91,Wen91,Bais92,MMS01,Kitaev03,Nayak04,Hansson04}, while the confined phase is `trivial'. We now describe two characteristics of this topological order: both characteristics can survive the introduction of additional degrees of freedom; but we will see that the first is more robust, and is present even in cases with gapless excitations carrying $\mathbb{Z}_2$ charges.

\begin{figure}[htb]
\begin{center}
\includegraphics[height=2.3in]{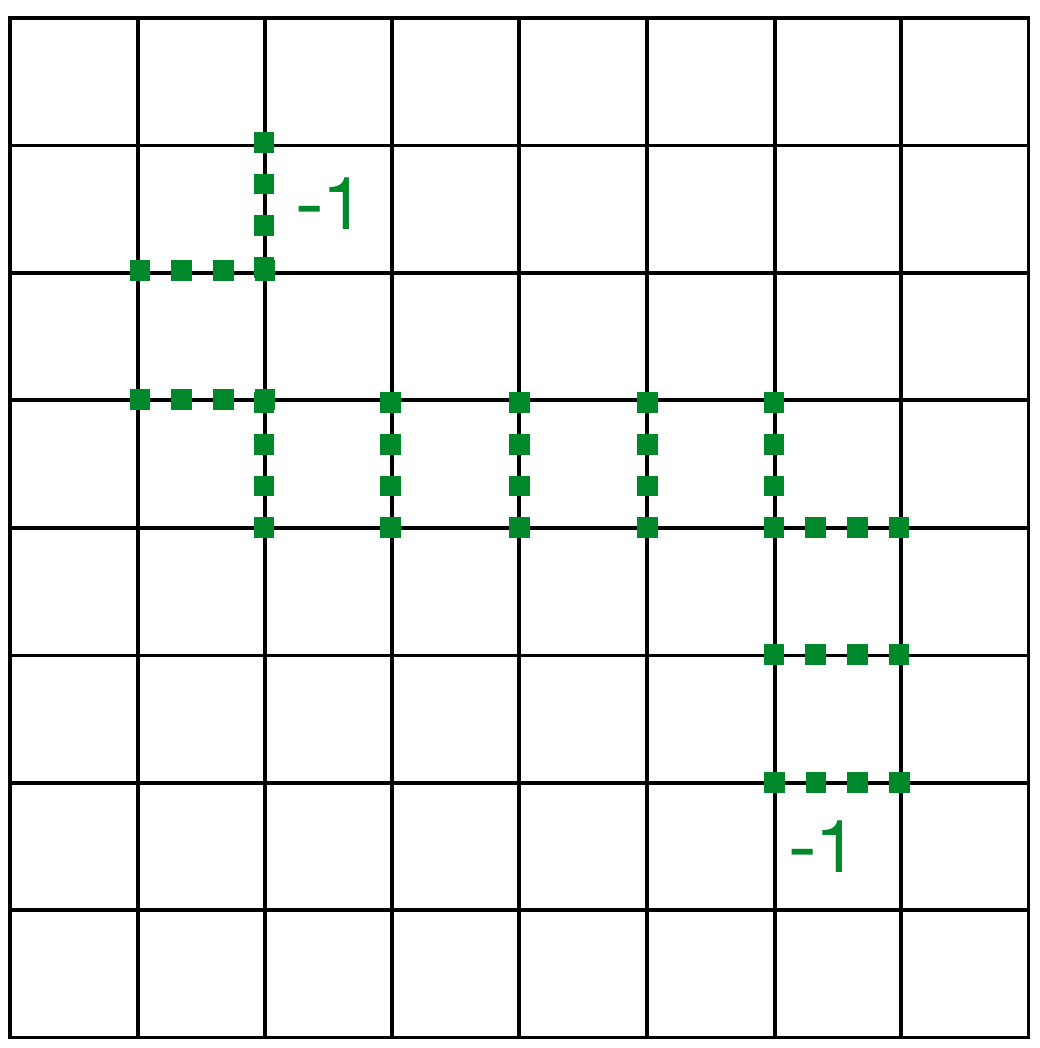}
\end{center}
\caption{Two visons (indicated by the $-1$'s in the plaquettes) connected by an invisible string. The dashed lines
indicate the links, $\ell$, on which the $\sigma^x_\ell$ operators acted on $\left|0 \right\rangle$ to create a pair
of separated visons. The plaquettes with an even number of dashed lines on their edges carry no $\mathbb{Z}_2$ fluxes,
and so are `invisible'.}
\label{fig:vison}
\end{figure}

The first characteristic is that there are stable low-lying excitations of the topological phase in the infinite lattice model which
cannot be created by the action of any local operator on the ground state ({\it i.e.\/} there are `superselection' sectors \cite{Kitaev03}). 
This excitation is a particle, often called a 
`vison', which carries $\mathbb{Z}_2$ flux of -1 \cite{Kivelson89,RC89,SenthilFisher}. Recall that the ground state of the deconfined phase expelled the $\mathbb{Z}_2$ flux: at $g = 0$ the state with all spins up, $\left|\Uparrow \right\rangle$, ({\it i.e.\/} eigenstates of $\sigma^z_\ell$ with eigenvalue $+1$) is a ground state, and this has no $\mathbb{Z}_2$ flux. This state is not an eigenstate of the $G_i$, but this is easily remedied by a gauge transformation:
\beq
\left| 0 \right\rangle = \prod_i (1 + G_i) \left| \Uparrow \right\rangle \label{eq:gs0}
\eeq
is an eigenstate of all the $G_i$.
Now we apply the $\sigma^x_\ell$ operator on a link $\ell$, the neighboring plaquettes acquire
$\mathbb{Z}_2$ flux of -1. We need a non-local `string' of $\sigma^x$ operators to separate these $\mathbb{Z}_2$ fluxes so that we obtain 2 well separated vison excitations; see Fig.~\ref{fig:vison}. 
Each vison is stable in its own region, and it can only be annihilated 
when it encounters another vison. Such a vison particle is present only in the deconfined phase: all excitations in the confined phase can be created by local operators, as is easily verified in a small $K$ expansion.

\begin{figure}[htb]
\begin{center}
\includegraphics[height=2in]{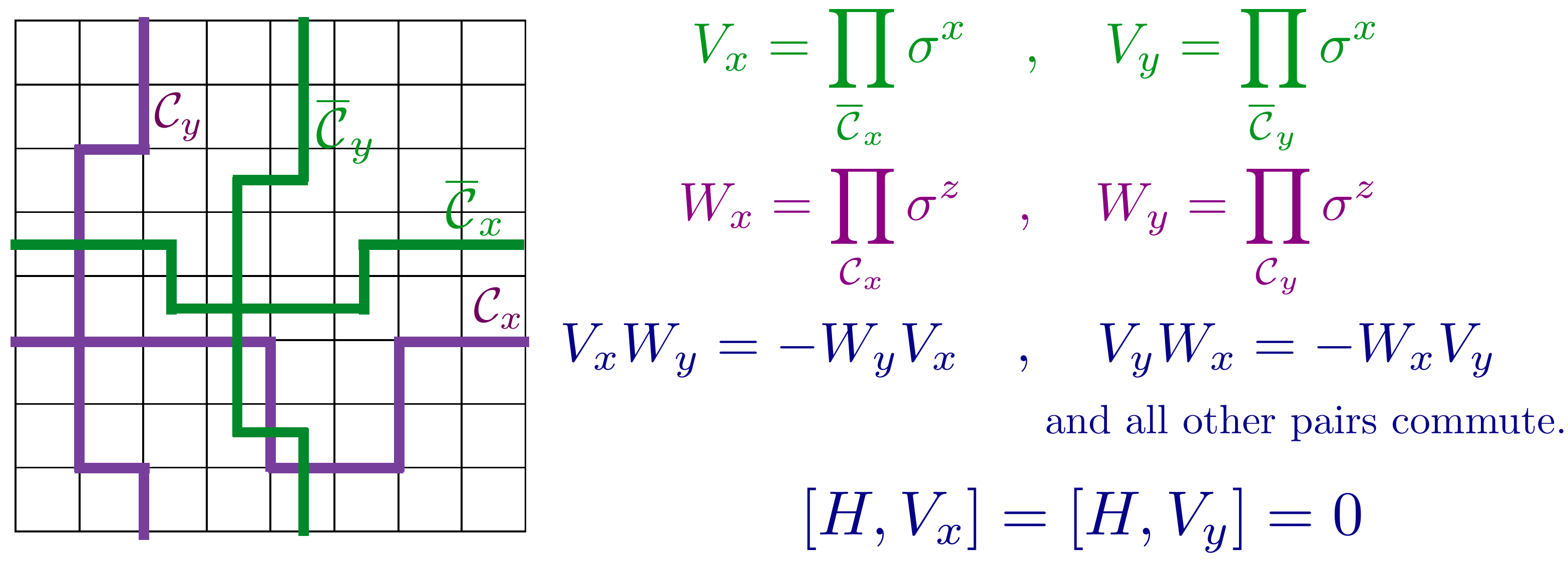}
\end{center}
\caption{Operators in a torus geometry: periodic boundary conditions are implied on the lattice.}
\label{fig:thooft}
\end{figure}

The second topological characteristic emerges upon considering the low-lying states of $\mathcal{H}_{\mathbb{Z}_2}$ on a topologically non-trivial geometry, like
the torus. A key observation in such geometries is that the $G_i$ (and their products) do not exhaust the set of operators which
commute with $\mathcal{H}_{\mathbb{Z}_2}$. On a torus, there are 2 additional independent operators which commute with $\mathcal{H}_{\mathbb{Z}_2}$: these operators, $V_x$, $V_y$, are illustrated
in Fig.~\ref{fig:thooft} (these are analogs of 'tHooft loops).
The operators are defined on contours, $\overline{\mathcal{C}}_{x,y}$ which reside on the dual square lattice, and encircle the 
two independent cycles of the torus. The specific contours do not matter, because we can deform the contours locally by multiplying them with the $G_i$. It is also useful to define Wegner-Wilson loop operators $W_{x,y}$ on direct lattice contours $\mathcal{C}_{x,y}$ which encircle the cycles of the torus; note that the $W_{x,y}$ do not commute with $\mathcal{H}_{\mathbb{Z}_2}$, while the $V_{x,y}$ do commute. 
Because the contour $\mathcal{C}_x$ intersects the contour $\overline{\mathcal{C}}_{y}$
an odd number of times (and similarly with $\mathcal{C}_y$ and $\overline{\mathcal{C}}_{x}$) we obtain the anti-commutation relations
\beq
W_x V_y = - V_y W_x \quad , \quad W_y V_x = - V_x W_y\,, \label{eq:vwwv}
\eeq
while all other pairs commute.

With this algebra of topologically non-trivial operators at hand, we can now identify the distinct signatures of the phases
without and with topological order. All eigenstates of $\mathcal{H}_{\mathbb{Z}_2}$ must also be eigenstates of $V_x$ and $V_y$.
First, consider the non-topological confining phase at large $g$. At $g = \infty$, the ground state, $\left| \Rightarrow \right\rangle$, has
all spins pointing 
to the right ({\it i.e.\/} all qubits are eigenstates of $\sigma^x_\ell$ with eigenvalue $+1$). This state clearly has eigenvalues $V_x = V_y = +1$. States with $V_x = -1$ or $V_y = -1$ must have at least one spin pointing to the left, and so cost a large energy $g$: such states cannot be degenerate with the ground state, even in the limit of an infinite volume for the torus.

\begin{figure}[htb]
\begin{center}
\includegraphics[height=2.3in]{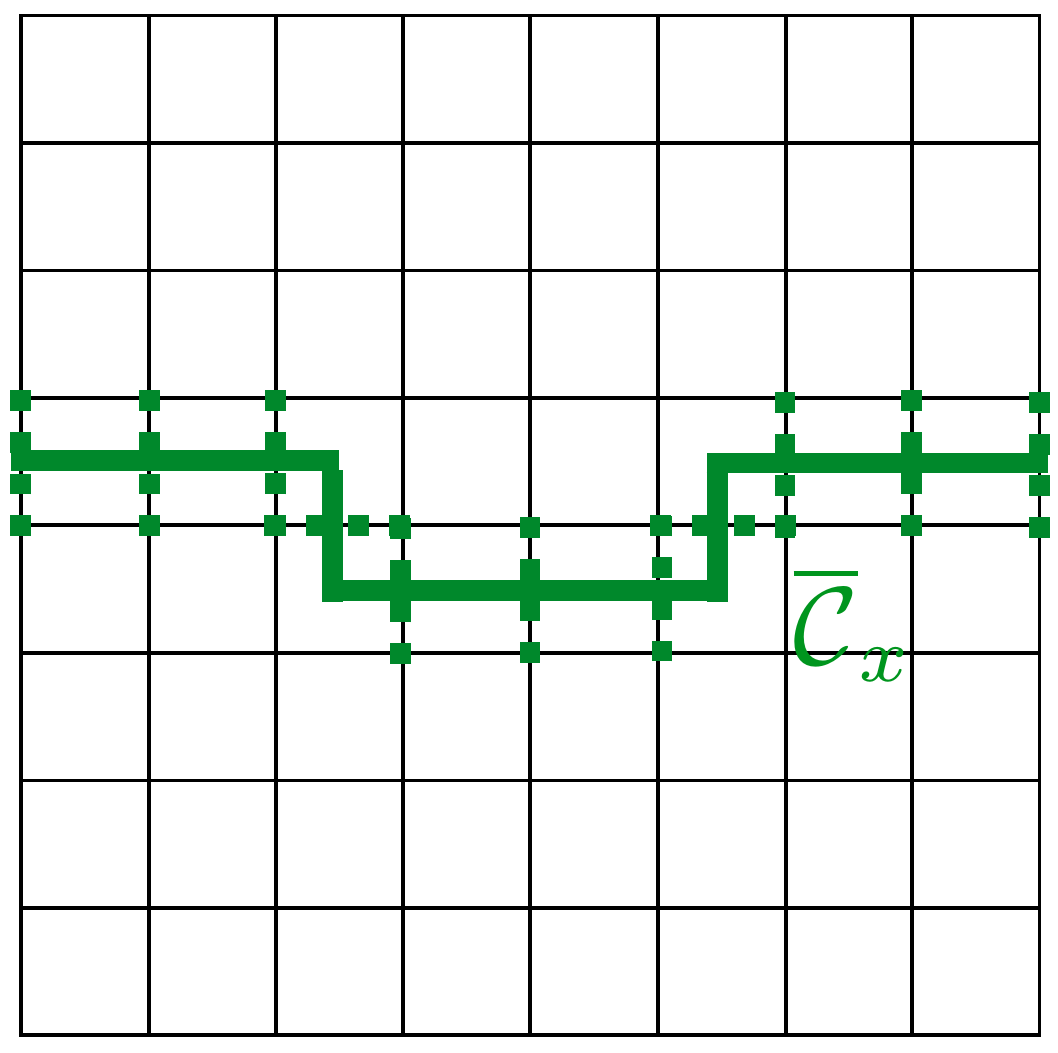}
\end{center}
\caption{The state $V_x \left | 0 \right\rangle$ (the dashed lines indicate $\sigma^x$ operators on the ground state: 
periodic boundary conditions are implied on the lattice.
Notice that every plaquette has $\mathbb{Z}_2$ flux $+1$, and so this is a ground state at $g=0$. This state
has $W_x = 1$ and $W_y = -1$. At small non-zero $g$, there is a non-zero tunnelling amplitude between 
$\left | 0 \right\rangle$ and $V_x \left | 0 \right\rangle$ of order $g^{L_x}$, where $L_x$ is the length of $\overline{\mathcal{C}}_x$.}
\label{fig:gstate}
\end{figure}

Next, consider the topological deconfined phase at small $g$. The ground state $\left| 0 \right\rangle$ is not an eigenstate of $V_{x,y}$,
but is instead an eigenstate of $W_{x,y}$ with $W_x = W_y = 1$. The state $V_x \left| 0 \right\rangle$ is easily seen to be
an eigenstate of $W_{x,y}$ with $W_x = 1$ and $W_y = -1$: so this state has $\mathbb{Z}_2$ flux of $-1$ through one of the 
holes of the torus. At $g=0$, the state $V_x \left| 0 \right\rangle$ is also a
ground state of $\mathcal{H}_{\mathbb{Z}_2}$, degenerate with $\left| 0 \right\rangle$: see Fig.~\ref{fig:gstate}. 
Similarly, we can create two other ground states, 
$V_y \left| 0 \right\rangle$ and $V_y V_x \left| 0 \right\rangle$, which are also eigenstates of $W_{x,y}$ with distinct eigenvalues.
So at $g=0$, we have a 4-fold degeneracy in the ground state, and all other states are separated by an energy gap.
When we turn on a non-zero $g$, the ground states will no longer be eigenstates of $W_{x,y}$ because these operators do not
commute with $\mathcal{H}_{\mathbb{Z}_2}$. Instead the ground states will become eigenstates of $V_{x,y}$; at $g=0$ we can take the linear
combinations $(1 \pm V_x) (1 \pm V_y) \left| 0 \right\rangle$ to obtain degenerate states with eigenvalues $V_x = \pm 1$
and $V_y = \pm 1$. At non-zero $g$, these 4 states will no longer be degenerate, but will acquire an exponentially small splitting of order
$g(g/K)^{L}$, where $L$ is a linear dimension of the torus: this is due to a non-zero tunneling amplitude between states
with distinct $\mathbb{Z}_2$ fluxes through the holes of the torus.

\begin{figure}[htb]
\begin{center}
\includegraphics[height=1.85in]{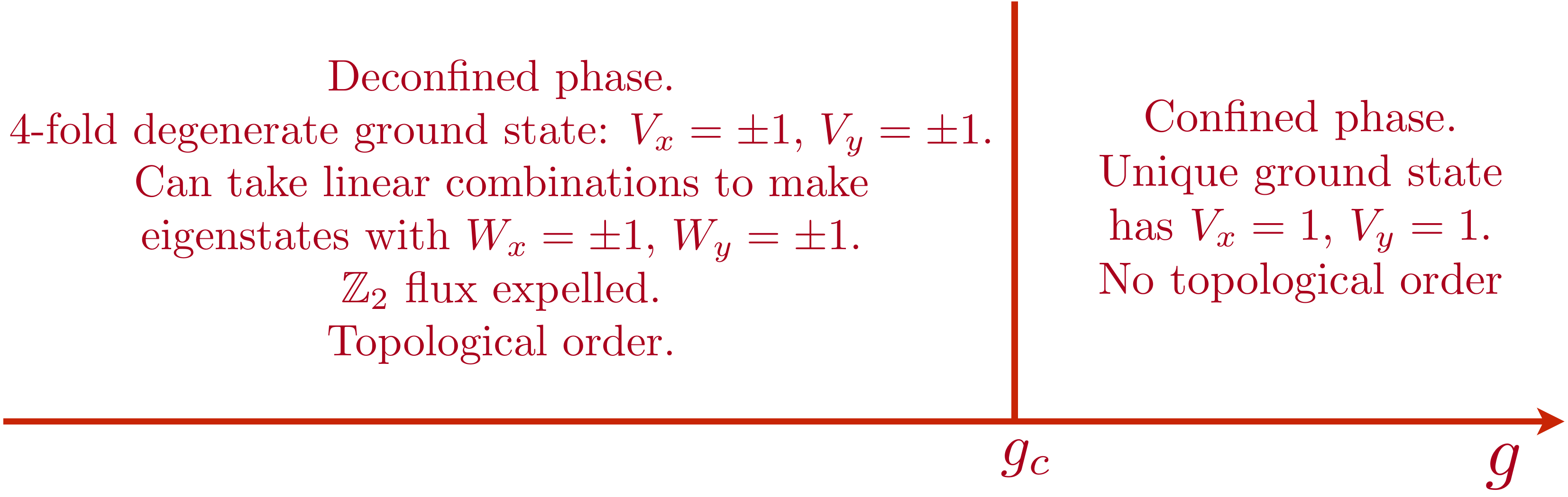}
\end{center}
\caption{An updated version of the phase diagram of $\mathcal{H}_{\mathbb{Z}_2}$ in Fig.~\ref{fig:wilson}.
The confinement-deconfinement phase transition is described by the Ising$^\ast$ Wilson-Fisher CFT, as is described
in Fig.~\ref{fig:u1}.}
\label{fig:z2topo}
\end{figure}

The presence of these 4 lowest energy states, which are separated by an energy splitting which vanishes exponentially with the linear
size of the torus, is one of the defining characteristics of $\mathbb{Z}_2$ topological order. We can take linear combinations of these 4 states to obtain distinct states with eigenvalues $W_x = \pm 1$, $W_y = \pm 1$
of the $\mathbb{Z}_2$ flux through the holes of the torus; or we can take energy eigenvalues, which are also eigenstates
of $V_{x,y}$ with $V_x = \pm 1$, $V_y = \pm 1$.
These feature are present throughout the entire 
deconfined phase, while the confining state has a unique ground state with $V_x = V_y = 1$. See Fig.~\ref{fig:z2topo}.

We close this section by noting that the $\mathbb{Z}_2$ topological order described above can also be realized in
a U(1)$\times$U(1) Chern-Simons gauge theory \cite{Nayak04,MMS01}. This is the theory with the 2+1 dimensional Lagrangian
\beq
\mathcal{L}_{\rm cs} = \frac{i}{\pi} \int d^3 x \, \epsilon_{\mu\nu\lambda} A_\mu \partial_\nu b_{\lambda}\,,
\label{eq:lcs}
\eeq
where $A_\mu$ and $b_\mu$ are the 2 U(1) gauge fields. The Wilson loop operators of these gauge fields
\beq
W_i = \exp \left( i \int_{\mathcal{C}_i} A_\mu dx_\mu \right) \quad , \quad V_i = \exp \left( i \int_{\overline{\mathcal{C}}_i} b_\mu dx_\mu \right)\,,
\eeq
are precisely the operators $W_{x,y}$ and $V_{x,y}$ when the contours $\mathcal{C}_i$ and $\overline{\mathcal{C}}_i$
encircle the cycles of the torus. This can be verified by reproducing the commutation relations in Eq.~(\ref{eq:vwwv})
from Eq.~(\ref{eq:lcs}). We will present an explicit derivation of $\mathcal{L}_{\rm cs}$ in Section~\ref{sec:xyeven}.

\section{The classical XY models}
\label{sec:xy}

This section recalls some well-established results on the classical statistical mechanics of the XY model at non-zero temperature
 in dimensions
$D=2$ and $D=3$. Later, we will extend these models to studies of topological order in quantum models at zero temperature.

The degrees of freedom of the XY model are angles $0 \leq \theta_i < 2 \pi$ on the sites $i$ of a square or cubic lattice.
The partition function is
\begin{eqnarray}
\mathcal{Z}_{XY} &=& \prod_i \int_{0}^{2 \pi} \frac{d \theta_i}{2 \pi} \exp \left( - \mathcal{H}_{XY} /T \right) \nonumber \\
\mathcal{H}_{XY} &=& -J \sum_{\langle i j \rangle} \cos(\theta_i - \theta_j) \,, \label{ZXY}
\end{eqnarray}
where the coupling $J >0$ is ferromagnetic and so the $\theta_i$ prefer to align at low temperature. 
A key property of the model is that the $\mathcal{H}_{XY}$ is invariant under $\theta_i \rightarrow \theta_i + 2\pi n_i$,
where the $n_i$ are arbitrary integers.

\subsection{Symmetry breaking in $D=3$}
\label{sec:xyd3}

\begin{figure}[htb]
\begin{center}
\includegraphics[height=1.8in]{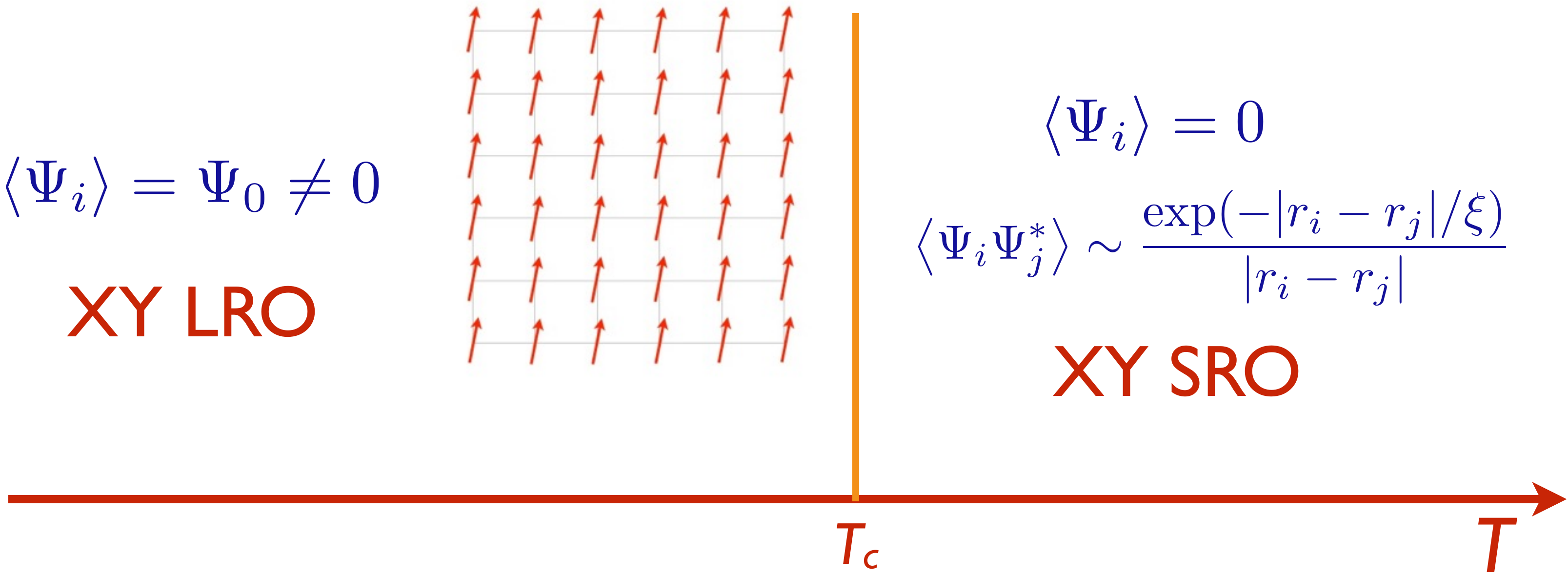}
\end{center}
\caption{Phase diagram of the classical XY model in Eq.~(\ref{ZXY}) in $D=3$ dimensions.
The low $T$ phase has long-range order (LRO) in $\Psi$, while the high $T$ has
only short-range order (SRO).}
\label{fig:xyd3}
\end{figure}

There is a well-studied phase transition in $D=3$, associated with the breaking of the symmetry
$\theta_i \rightarrow \theta_i + c$, where $c$ is any $i$-independent real number. As shown in Fig.~\ref{fig:xyd3},
below a critical temperature $T_c$, the symmetry is broken and there are long-range correlations in the complex
order parameter 
\beq
\Psi_j \equiv e^{i \theta_j} \label{defPsi}
\eeq
with
\beq
\lim_{|r_i - r_j| \rightarrow \infty} \left\langle \Psi_i \Psi_j^\ast \right\rangle = |\Psi_0|^2 \neq 0\,.
\eeq 
For $T>T_c$, the symmetry is restored and there are exponentially decaying correlations, along with a power-law prefactor, as indicated
in Fig.~\ref{fig:xyd3}. This prefactor is the Ornstein-Zernike form \cite{Fisher67}, and arises from the three-dimensional Fourier transform of $(\vec{p}^2 + \xi^{-2})^{-1}$, where $\vec{p}$ is a three-dimensional momentum.
The critical theory at $T=T_c$ is described by the XY Wilson-Fisher CFT \cite{wilsonfisher72}.

\subsection{Topological phase transition in $D=2$}

\begin{figure}[htb]
\begin{center}
\includegraphics[height=2.5in]{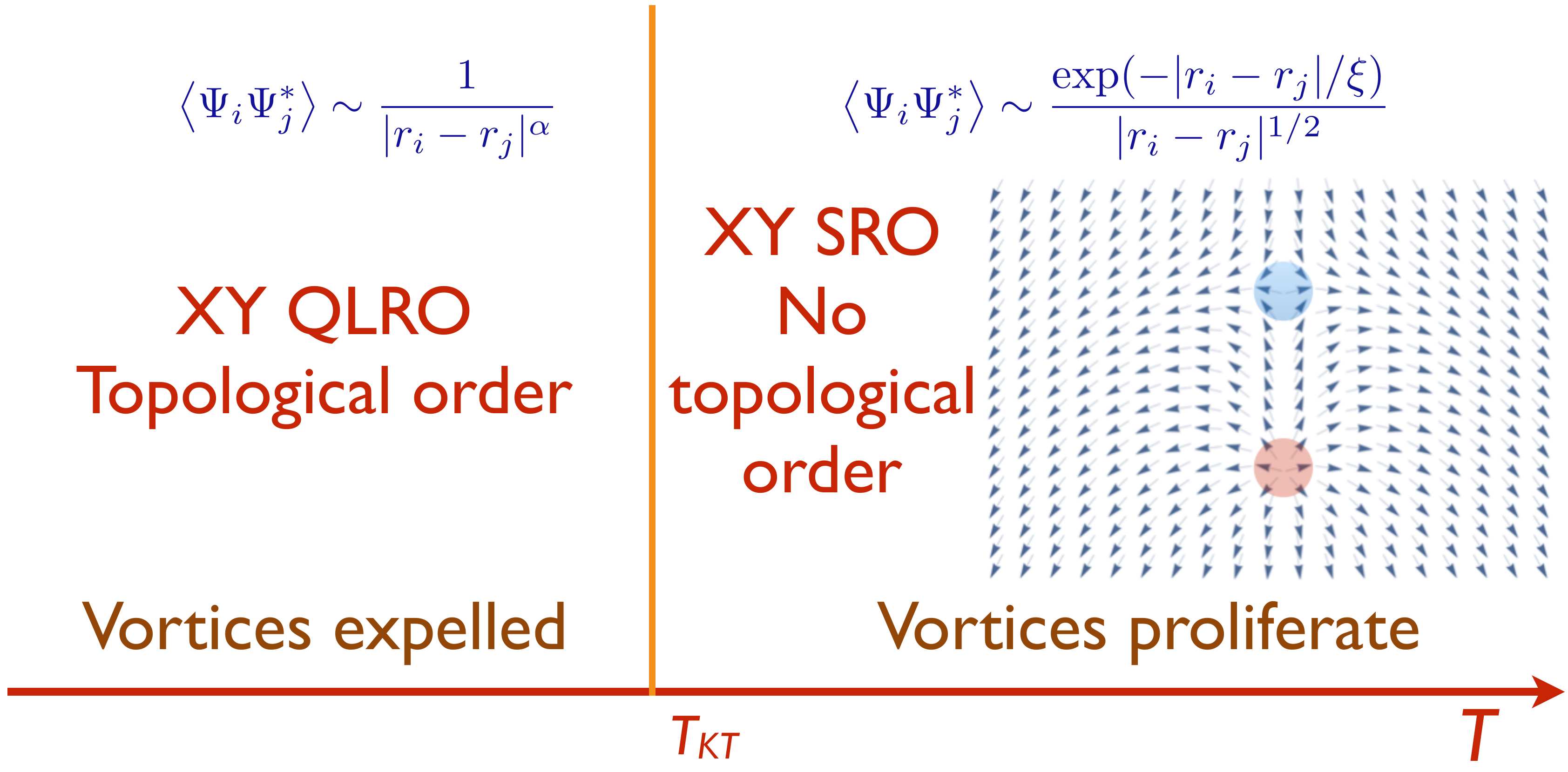}
\end{center}
\caption{Phase diagram of the classical XY model in Eq.~(\ref{ZXY}) in $D=3$ dimensions.
There is no LRO at any $T$.
The low $T$ phase has quasi long-range order (QLRO) in $\Psi$, while the high $T$ has
SRO. The KT transition is associated with the proliferation of vortices, and also a change in the form of the correlations
of the XY order parameter from power-law to exponential.}
\label{fig:xyd2}
\end{figure}

In dimension $D=2$, the symmetry $\theta_i \rightarrow \theta_i + c$
is preserved at all non-zero $T$. There is no LRO, and 
\begin{displaymath} 
\left\langle \Psi_i \right\rangle = 0~ \mbox{for all $T>0$.}
\end{displaymath}
Nevertheless, as illustrated in Fig.~\ref{fig:xyd2}, there is a Kosterlitz-Thouless (KT) phase transition at $T=T_{KT}$ \cite{Berezinskii1,Berezinskii2,KT73,KT74}, where the nature of the correlations changes from a power-law decay at $T < T_{KT}$, to an exponential decay (with an Ornstein-Zernike prefactor) for $T>T_{KT}$.
At low $T$, long-wavelength spin-wave fluctuations in the $\theta_i$ are sufficient to destroy the LRO and turn it into quasi-LRO (QLRO)
with a power-law decay of fluctuations. At high $T$, there is SRO with exponential decay of correlations. KT showed that the transition 
between these phases occurs as a consequences of the proliferations of point-like vortex and anti-vortex defects, illustrated
in Fig.~\ref{fig:xyd2}. Each defect is associated with a winding in the phase gradient far from the core of the defect:
\beq
\oint dx_i \, \partial_i \theta = 2 \pi n_v \,, \label{defvortex}
\eeq
where the integer $n_v$ is a topological invariant characterizing the vorticity. In the QLRO phase, the vortices occur only in 
tightly bound pairs of $n_v =\pm 1$ so that there is no net vorticity at large scales; and in the SRO phase, these pairs
undergo a deconfinement transition to a free plasma. So the QLRO phase is characterized by the suppression of
the topological vortex defects. By analogy with the suppression of $\mathbb{Z}_2$ flux defects in the 
topological-ordered phase of the $\mathbb{Z}_2$ gauge theory discussed in Section~\ref{sec:z2topo}, 
we conclude that the low $T$ phase of the $D=2$ XY model has {\it topological order}, and the KT transition
is a topological phase transition \cite{KT73}. Of course, in the present case, the phase transition can also be identified by the two-point
correlator of $\Psi_i$ changing from the QLRO to the SRO form, but KT showed that the underlying mechanism is the
proliferation of vortices and so it is appropriate to identify the KT transition as a topological phase transition.

\section{Topological order in XY models in $D=2+1$}
\label{sec:xytopo}

In the study of classical XY models in Section~\ref{sec:xy}, we found only a symmetry breaking phase transition
in $D=3$ dimensions. In contrast, the $D=2$ case exhibited a topological phase transition without a symmetry breaking
order parameter. This section shows that modified XY models can also exhibit a topological phase transition in $D=3$ dimensions.

\begin{figure}[htb]
\begin{center}
\includegraphics[height=3in]{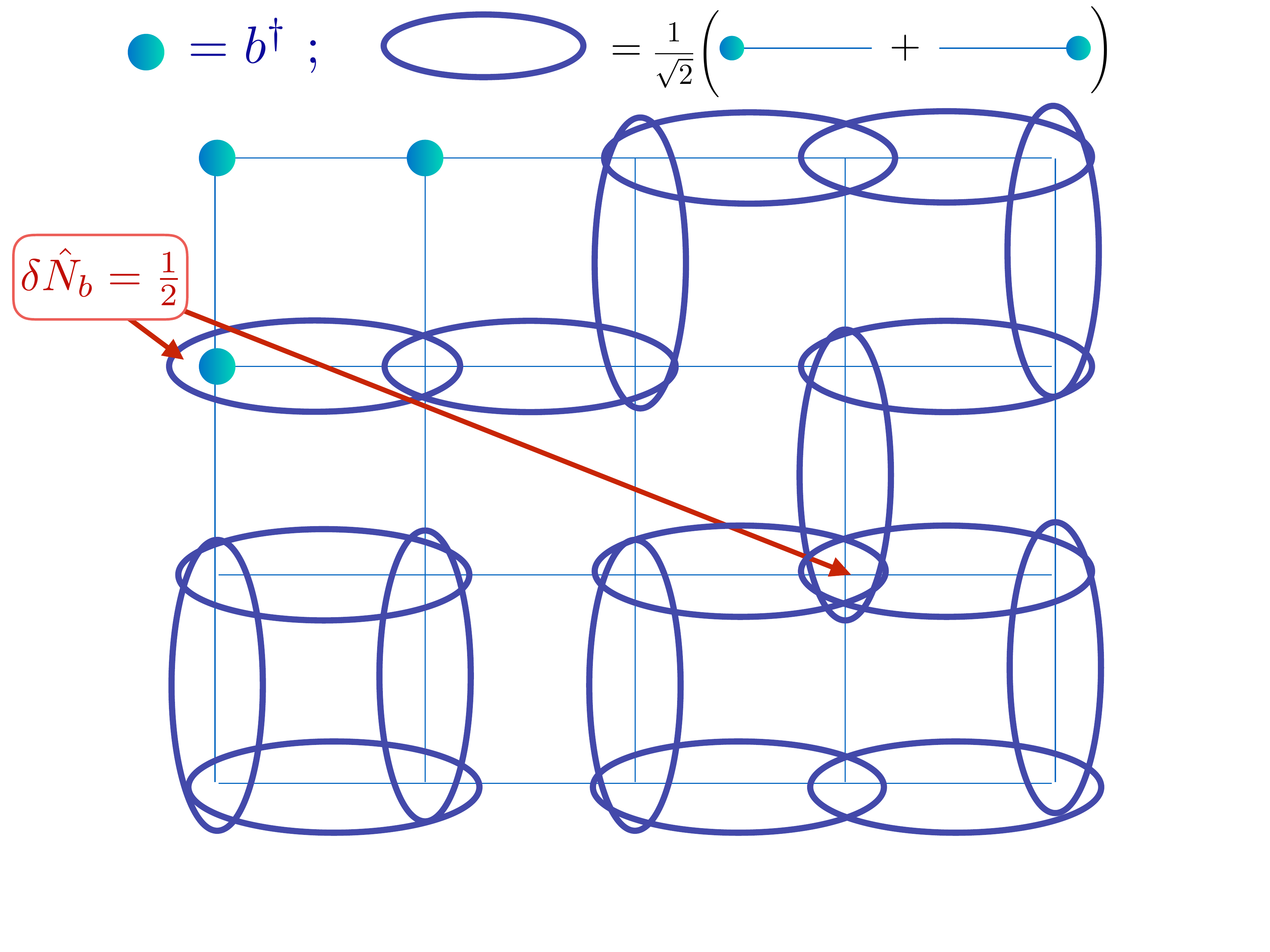}
\end{center}
\caption{Schematic representation of a topologically ordered, `resonating valence bond' state in the boson models of Refs.~\cite{RJSS91,SSMV99,SM02,SM02PRL}. The boson $b^\dagger$ can reside
either on sites (indicated by the filled circles) or in a bonding orbital  (a`valence bond') between sites (indicated by the ellipses). The average boson density 
of the ground state is 1. A single additional boson has been added above, and it has fractionalized into 2 excitations carrying boson number 
$\delta\hat{N}_b = 1/2$ (this becomes clear when we consider each bonding orbital as contributing a density of $1/2$ to each of the two sites it connects).}
\label{fig:dimer}
\end{figure}

Classical XY models also have an interpretation as quantum XY models at zero temperature 
in spatial dimensionality $d=D-1$, where one of the classical dimensions is interpreted as the imaginary time of the quantum model. And the quantum XY models have the same phases and phase transitions
as models of lattice bosons with short-range interactions. Specifically, the classical $D=3$ XY models we study below
map onto previously studied models of bosons on the square lattice at an average boson number density, $\langle \hat{N}_b \rangle$, 
which is an integer \cite{RJSS91,SSMV99,SM02,SM02PRL}.
These boson models are illustrated in Fig.~\ref{fig:dimer}.
As indicated in Fig.~\ref{fig:dimer}, it is possible for such boson models to have topologically ordered phases which have excitations
with a fractional boson number $\delta \hat{N}_b = 1/2$. 
We will also describe the physics for the case of half-integer boson density later in Section~\ref{sec:xyodd}; this case is also related to 
quantum dimer models \cite{DRSK88,EFSK90,NRSS90,RJSS91,SSMV99}.

We now return to the discussion of classical XY models in $D=3$ because they offer a transparent and intuitive route
to describing the nature of topological order in $D=2+1$ dimensions.
The quantum extension of the discussion below will appear in Section~\ref{sec:quantumxy}.
We consider an XY model which augments the Hamiltonian in Eq.~(\ref{ZXY}) by longer-range couplings between
the $\theta_i$, {\it e.g.\/}:
\beq
\widetilde{\mathcal{H}}_{XY} = -J \sum_{\langle i j \rangle} \cos(\theta_i - \theta_j) + 
\sum_{ijk\ell} K_{ijk\ell} \cos(\theta_i + \theta_j - \theta_k - \theta_\ell) + 
\ldots\ldots \label{HKij}
\eeq
The additional couplings $K_{ijk\ell}$ preserve the 
basic properties of the XY model: invariance under the global U(1) symmetry $\theta_i \rightarrow \theta_i + c$, 
and periodicity in $\theta_i \rightarrow \theta_i + 2 \pi n_i$. 
We will not work out the specific forms of the $K_{ijk\ell}$ needed for our purposes, but instead use an alternative form in Eq.~(\ref{tzxy}) below,
in which these couplings are decoupled by an auxiliary Ising variable, and they all depend upon a single additional coupling $K$.
At small $K$, the model will have the same phase diagram
as that in Fig.~\ref{fig:xyd3}. But at larger $K$, we will obtain an additional phase with topological order, as shown in
Fig.~\ref{fig:xytopo}.
We will design the additional couplings so that the topological phase proliferates only {\it even\/} line vortex defects {\it i.e.\/} vortex lines for which the integer $n_v$ in Eq.~(\ref{defvortex})
is even. So the transition to topological order from the non-topological SRO phase occurs via the expulsion
of odd vortex defects, including the elementary vortices with $n_v=\pm 1$. The additional $K$-dependent couplings in the XY model
will be designed to suppress vortices with $n_v=\pm 1$.
This transition should be compared to the KT 
transition in $D=2$, where both even and odd vortices are suppressed as the temperature is lowered into the topological phase.
Note that the new topological phase only has SRO with exponentially decaying correlations of the order parameter, 
unlike the QLRO phase of the $D=2$ XY model. But, there is a subtle difference between the two-point correlators
of $\Psi_i$ in the two SRO phases in Fig.~\ref{fig:xytopo}: the power-law prefactors of the exponential are different between the topological
and non-topological phases. 

We now present the partition function of the XY model of Fig.~\ref{fig:xytopo}, related to models in several previous studies \cite{SSNR91,RJSS91,LRT93,CSS93,LRT95a,LRT95b,SSMV99,SenthilFisher,SP01,SM02,SM02PRL,SSS02,PS02}:
\begin{eqnarray}
\widetilde{\mathcal{Z}}_{XY} &=& \sum_{\{\sigma_{ij}\} = \pm 1} \prod_i \int_{0}^{2 \pi} \frac{d \theta_i}{2 \pi} \exp \left( - \widetilde{\mathcal{H}}_{XY} /T \right) \nonumber \\
\widetilde{\mathcal{H}}_{XY} &=& -J \sum_{\langle i j \rangle} \sigma_{ij} \cos\left[(\theta_i - \theta_j)/2\right] - K \sum_{\square}
\prod_{(ij) \in \square} \sigma_{ij}\,, \label{tzxy}
\end{eqnarray}
where sites $i$ reside on the $D=3$ cubic lattice. This partition function is the basis for the schematic phase
diagram in Fig.~\ref{fig:xytopo}, and numerical results for such a phase diagram appear in Ref.~\cite{SM02}.

As written, the partition function has an additional degree of freedom $\sigma_{ij} = \pm 1$ on the links
$\ell \equiv (ij)$ of the cubic lattice: these are Ising gauge fields 
similar to those in Eq.~(\ref{tz2}). It is not difficult to sum over the $\sigma_{ij}$ explicitly order-by-order
in $K$, and then the resulting effective action for $\theta_i$ has all the properties required of a XY model:
periodicity in $\theta \rightarrow \theta + 2 \pi$ and global U(1) symmetry. We can view the $\sigma_{ij}$ as a 
discrete Hubbard-Stratanovich variable which has been used to decouple the $K_{ijk\ell}$ term in Eq.~(\ref{HKij}). 
So we are justified in describing
$\widetilde{\mathcal{Z}}_{XY}$ as a modified XY model. However, for our purposes, it will be useful to keep
the $\sigma_{ij}$ explicit.

In the form in Eq.~(\ref{tzxy}), a crucial property of $\widetilde{\mathcal{Z}}_{XY}$ is its invariance under $\mathbb{Z}_2$ gauge transformations generated by $\varrho_i = \pm 1$:
\beq
\theta_i \rightarrow \theta_i + \pi (1-\varrho_i) \quad, \quad \sigma_{ij} \rightarrow \varrho_i \sigma_{ij} \varrho_j \,.
\label{thetagauge}
\eeq
It will turn out that $\sigma_{ij}$ is the advertized emergent $\mathbb{Z}_2$ gauge field of the topological phase. 
Note that the XY order parameter, $\Psi_i$, is gauge-invariant.

\begin{figure}[htb]
\begin{center}
\includegraphics[height=2.3in]{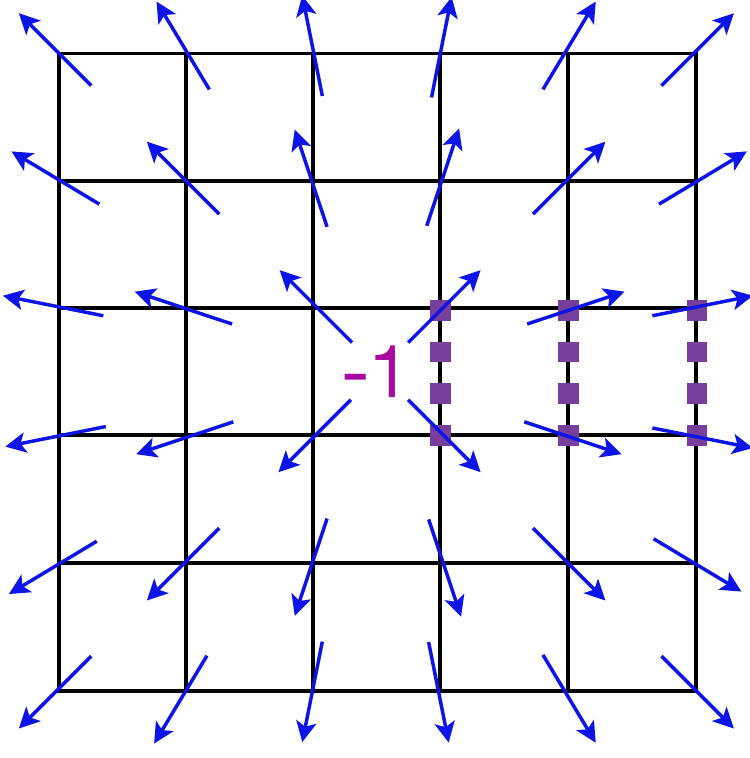}
\end{center}
\caption{A $2 \pi$ vortex in $\theta_i$. The $\mathbb{Z}_2$ gauge field $\sigma_{ij} = -1$ on links (indicated by 
thick dashed lines)
across which $\theta_i$ has a branch cut, and $\sigma_{ij}=1$ otherwise.
The $\mathbb{Z}_2$ flux of -1 is present only in the central plaquette, and so a vison is present
at the vortex core. If the contour of $\sigma_{ij}=-1$ deviates from the branch cut in $\theta_i$, there is an energy cost proportional
to the length of the deviation. Consequently, the vison is confined to the vortex core.}
\label{fig:xyvortex}
\end{figure}

The rationale for our choice of $\widetilde{\mathcal{H}}_{XY}$ becomes evident upon considering the
structure of a $2 \pi$ vortex in $\theta_i$, sketched in Fig.~\ref{fig:xyvortex}. 
Let us choose the values of $\theta_i$ around the central plaquette
of this vortex as (say) $\theta_i = \pi/4, 3 \pi/4, 5 \pi/4, 7 \pi/4$. Then we find that the values of 
$  \cos \left[ (\theta_i - \theta_j)/2 \right] > 0$ on all links except for that across the branch cut between $\pi/4$
and $7 \pi/4$. For $J>0$, such a vortex will have $\sigma_{ij} = -1$ only for the link across the branch cut.
So a $2 \pi$ vortex will prefer $\prod_{(ij)\in \square} \sigma_{ij} = -1$, {\it i.e.\/} a $2 \pi$ vortex has $\mathbb{Z}_2$ flux $=-1$ in its core, and then a large $K>0$ will suppress (odd) $ 2 \pi$ vortices.
Note that there is no analogous suppression of (even) $4 \pi$ vortices. This explains why it is possible for
$\widetilde{\mathcal{H}}_{XY}$ to have large $K$ phase with odd vortices suppressed, as indicated
in Fig.~\ref{fig:xytopo}.

The existence of a phase transition between the two SRO phases of Fig.~\ref{fig:xytopo} can be established by explicitly
performing the integral over the $\theta_i$ in $\widetilde{\mathcal{Z}}_{XY}$ order-by-order in $J$. Such a procedure
should be valid because correlations in $\theta_i$ decay exponentially. Then, it is easy to see that the resulting effective action
for the $\sigma_{ij}$ is just the $\mathbb{Z}_2$ gauge theory of Section~\ref{sec:ising}, in Wegner's classical cubic lattice formulation;
this is evident from the requirements imposed by the gauge invariance in Eq.~(\ref{thetagauge}). 
To leading order, the main effect of the $\theta_i$ integral is a renormalization in the coupling $K$. 
The $\mathbb{Z}_2$ gauge theory has a confinement-to-deconfinement transition with increasing $K$, 
and this is just the transition for the onset of topological order in the SRO regime.

\subsection{Quantum XY models}
\label{sec:quantumxy}

Further discussions on the nature of topological phase are more easily carried out in the language of the corresponding
quantum model in $d=2$ spatial dimensions. The quantum language will also enable us to connect with the discussion
on the $\mathbb{Z}_2$ gauge theory in Section~\ref{sec:ising}.

The quantum form of $\widetilde{\mathcal{H}}_{XY}$ in Eq.~(\ref{tzxy}) is obtained by 
transforming the temporal direction of the partition function
into a `kinetic energy' expressed in terms of canonically conjugate quantum variables. We introduce the half-angle: 
\beq
\vartheta_i \equiv \theta_i/2\,, 
\eeq
and a canonically conjugate number variable $\hat{n}_i$ with integer eigenvalues. 
Just as in Eq.~(\ref{Hz2}), the $\sigma_{ij}$ are promoted to the Pauli matrices $\sigma^z_{ij}$,
and we will also need the Pauli matrix $\sigma^x_{ij}$. So we obtain 
\begin{eqnarray}
\overline{\mathcal{H}}_{XY} &=& -J \sum_{\langle i j \rangle} \sigma^z_{ij} \cos (\vartheta_i - \vartheta_j) - K \sum_{\square}
\prod_{(ij) \in \square} \sigma^z_{ij} \nonumber \\
&~&~~~ + U \sum_{i} (\hat{n}_i)^2 - g \sum_{\langle ij \rangle} \sigma^x_{ij}\,;  \nonumber \\
 &&~~~~ [ \vartheta_i, \hat{n}_j ] = i \delta_{ij} \,.
\label{bhxy}
\end{eqnarray}
The set of operators which commute with $\overline{\mathcal{H}}_{XY}$ are now modified from Eq.~(\ref{defGi}) to 
\beq
G_i^{XY} = e^{i \pi \hat{n}_i } \, \prod_{\ell \, \in \, +} \sigma^x_{\ell} \,. \label{defGXY}
\eeq
Each $e^{i \vartheta}$ boson carries unit $\mathbb{Z}_2$ electric charge, and so the Gauss law
has been modified by the total electric charge on site $i$. The Gauss law constraint in Eq.~(\ref{G1}) now becomes
\beq
G_i^{XY} = 1\,. \label{GXY1}
\eeq

The properties of the large $K$ topological phase of $\overline{\mathcal{H}}_{XY}$ are closely connected to those of the deconfined
phase of the $\mathbb{Z}_2$ gauge theory in Section~\ref{sec:ising}. There is four-fold degeneracy on the torus, and a stable
`vison' excitations carrying magnetic $\mathbb{Z}_2$ flux of -1. In the present context, the `vison' can also be interpreted as gapped
odd vortex in the $\theta_i$; because of the condensation of even vortices, there is only a single independent odd vortex excitation.

A significant new property is the presence of fractionalized bosonic excitations which carry `electric' charges under the $\mathbb{Z}_2$ gauge field.
These are the particles created by the 
\beq
\psi = e^{i \vartheta} \label{defpsi}
\eeq 
operator, and the anti-particles created by $\psi^\ast = e^{-i \vartheta}$.
These are the excitations illustrated in the boson models of Fig.~\ref{fig:dimer}, and they carry boson number $\hat{N} = 1/2$.
Note that the XY order parameter, $\Psi$, and correspondingly the XY boson number, $\hat{N}_b$, obey
\beq
\Psi = \psi^2 \quad , \quad \hat{N}_b = \hat{n}/2. \label{Psipsi}
\eeq
It is clear from Eq.~(\ref{thetagauge}) that the $\psi$ particles carry $\mathbb{Z}_2$ charges. Also, parallel transporting an electric charge
around a vison leads to a Berry phase of $-1$, and hence the $\psi$ and the visons are mutual semions. This structure of 
electric and magnetic excitations, and of the degeneracy on the torus, is that found in the solvable `toric code' model \cite{Kitaev03}.

The presence of the $\psi$ excitations also helps us understand the nature of the XY order parameter 
correlations in the topological SRO phase, as indicated in Fig.~\ref{fig:xytopo}. The $\psi$ are deconfined, gapped, bosonic excitations,
and the Hamiltonian has a charge conjugation symmetry under $\psi \rightarrow \psi^\ast$: so the $\psi$ are described at low energies
as massive relativistic charge particles, and this implies that the 2-point $\psi$ correlator has a Ornstein-Zernike form, with a $1/r$ prefactor.
Then using $\Psi = \psi^2$, we find the exponential decay of the XY order, with the $1/r^2$ prefactor, as shown in Fig.~\ref{fig:xytopo}.

\section{Embedding into Higgs phases of larger gauge groups}
\label{sec:higgs}

Sections~\ref{sec:ising} and~\ref{sec:xytopo} 
have so far described states with $\mathbb{Z}_2$ topological order using a $\mathbb{Z}_2$ gauge theory.
However, as we will be amply demonstrated below, it is often useful to consider the topological state arising as a phase
of a theory with a larger gauge group, in which condensation of a Higgs field breaks the gauge group back down to $\mathbb{Z}_2$.
Such an approach yields a powerful method of analyzing the influence of additional matter fields in the topological state, and also
of describing `deconfined' critical points at which the topological order is lost: often, the larger gauge group emerges as unbroken
in the theory of deconfined criticality \cite{senthil1,senthil2}.

We will begin in Section~\ref{sec:z2even} 
by recasting the $\mathbb{Z}_2$ gauge theory of Section~\ref{sec:ising} as a U(1) gauge theory \cite{FradkinShenker}.
This does not immediately offer advantages over the $\mathbb{Z}_2$ formulation, but does allow us to address the 
nature of the confinement-deconfinement phase transition using the well-studied methods of particle-vortex duality.
Section~\ref{sec:z2odd} will then consider an extension of the $\mathbb{Z}_2$ gauge theory to include static matter
with a net density of one electric charge per site: this is the so-called `odd' $\mathbb{Z}_2$ gauge theory \cite{MSF01} (correspondingly,
the original $\mathbb{Z}_2$ gauge theory of Section~\ref{sec:ising} is called an `even' gauge theory). We will show
that the deconfinement-confinement transition in the odd $\mathbb{Z}_2$ gauge theory is described by a deconfined
critical U(1) gauge theory. 

\subsection{Even $\mathbb{Z}_2$ gauge theory}
\label{sec:z2even}

This is section will reconsider the $\mathbb{Z}_2$ gauge theory $\mathcal{H}_{\mathbb{Z}_2}$ in Eq.~(\ref{Hz2})
for the case with no background $\mathbb{Z}_2$ gauge charges, as specified by Eq.~(\ref{G1}).

We introduce a U(1) gauge field $A_{i\alpha}$ on the link of the square lattice between the sites $i$ and $i + \hat{e}_\alpha$,
where $\alpha = \pm x, \pm y$ and $\hat{e}_\alpha$ are the unit vectors to the nearest neighbors of site $i$. Unlike the $\mathbb{Z}_2$
gauge field, the U(1) gauge field is oriented, and so $A_{i + \hat{e}_\alpha, -\alpha} = - A_{i \alpha}$. The $A_{i \alpha}$ are compact variables
with period $2 \pi$. We will reduce them to nearly discrete variables by applying a potential $\sim - \cos(2 A_{i \alpha})$ so that the values $A_{i \alpha} = 0, \pi$ are preferred. Then we choose the mapping between the gauge fields of the $\mathbb{Z}_2$ and U(1) gauge theories
\beq
\sigma^{z}_{ix} \rightarrow \exp \left( i \eta_i A_{ix} \right) \quad, \quad \sigma^{z}_{iy} \rightarrow \exp \left(- i \eta_i A_{iy} \right) \label{mapz}
\eeq
where
\beq
\eta_i = (-1)^{i_x + i_y} \,, \label{defeta}
\eeq
takes opposite signs on the two sublattices of the square lattice. 

We also introduce a canonically conjugate `electric field', $E_{i \alpha}$, on each link of the lattice,
\beq
[A_{i\alpha}, E_{j\beta}] = i \delta_{ij} \delta_{\alpha\beta}\,, \label{AE}
\eeq
so that the $E_{i \alpha}$ have integer eigenvalues. As the $\sigma^x_{\ell}$ flip the eigenvalues of $\sigma^{z}_\ell$, the corresponding
operator in the U(1) gauge field should shift $A_{i \alpha}$ by $\pi$. So we have the mapping
\beq
\sigma^{x}_{i\alpha} \rightarrow \exp \left( i \pi \eta_i E_{i\alpha} \right)\,. \label{mapx}
\eeq

We apply the mapping in Eq.~(\ref{mapz}) to the plaquette term in Eq.~(\ref{Hz2}), and include the potential to favor gauge fields at $0, \pi$, to obtain the Hamiltonian
\beq
\mathcal{H}_{U(1)} = - K \sum_{\square} \cos \left( \epsilon_{\alpha\beta} \Delta_\alpha A_{i \beta} \right) + h \sum_{i,\alpha} E_{i \alpha}^2
- L \sum_{i \alpha} \cos(2 A_{i \alpha}) \,, \label{Hu1a}
\eeq
where $\Delta_\alpha$ is a discrete lattice derivative ({\it i.e.\/} $\Delta_\alpha f(i) \equiv f(i+ \hat{e}_\alpha) - f(i)$), and $\epsilon_{\alpha\beta}$ is the unit antisymmetric tensor.  

Eq.~(\ref{Hu1a}) also contains a `kinetic energy' term for the U(1) gauge field $\sim E_{i \alpha}^2$. With this term included, the 
Hamiltonian on each link becomes $h \, E^2  - L \cos(2 A)$, where (via Eq.~(\ref{AE})) $A$ and $E$ are canonically conjugate variables. This Hamiltonian describes a `particle' moving on a circle with periodic co-ordinate $0 \leq A \leq 2 \pi$ in a potential with degenerate minima at $A = 0, \pi$. Such a particle will have 2 low-lying states in its spectrum, and we map these two states 
to the two eigenstates of the $\sigma^x_\ell$ operator on each link of the $\mathbb{Z}_2$ gauge
theory in Eq.~(\ref{Hz2}). 

While the form of $\mathcal{H}_{U(1)}$ in Eq.~(\ref{Hu1a}) is, in principle, adequate for our purposes, it is inconvenient to work
with because the $L$ term is not invariant under U(1) gauge transformations (it is invariant only under $\mathbb{Z}_2$ gauge transformations).
However, it is possible to make it U(1) gauge invariant: we generate a U(1) gauge transformation of $A_{i \alpha}$ by the angular variable $\Theta_i$, and make $\Theta_i$ a dynamical degree of freedom. This introduces redundant degrees of freedom which allow for full U(1) gauge invariance. Explicitly, the modified Hamiltonian is
\bea
\mathcal{H}_{U(1)} &=& - K \sum_{\square} \cos \left( \epsilon_{\alpha\beta} \Delta_\alpha A_{i \beta} \right) + h \sum_{i,\alpha} E_{i \alpha}^2
\nonumber \\
&~&- L \sum_{i \alpha} \cos(\Delta_\alpha \Theta_i - 2 A_{i \alpha}) + \widetilde{h} \sum_{i} \hat{N}_i^2 \,, \label{Hu1}
\eea
where $\hat{N}_i$ is the conjugate integer-valued number operator to $\Theta_i$
\beq
[\Theta_i, \hat{N}_j] = i \delta_{ij}\,. \label{TN}
\eeq
The spectrum of Eq.~(\ref{Hu1}) at $\widetilde{h}=0$ is identical to that of Eq.~(\ref{Hu1a}). The form in Eq.~(\ref{Hu1a}) is invariant under U(1) gauge transformations generated by the arbitrary field $f_i$, where
\beq
A_{i \alpha} \rightarrow A_{i \alpha} + \Delta_\alpha f_i \quad, \quad \Theta_{i} \rightarrow \Theta_i + 2 f_i \,, \label{u1gauge}
\eeq
and so
\beq
H_i \equiv e^{i \Theta_i}
\eeq
transforms as a charge 2 scalar field. We will refer to $H_i$ as a `Higgs' field, for reasons that will become clear below.

To complete our description of our U(1) gauge theory, we need to present the fate of the site constraints in Eq.~(\ref{G1}).
Just like the $\mathbb{Z}_2$ gauge theory, there are an infinite number of operators that commute with $\mathcal{H}_{U(1)}$, associated with
the gauge invariance in Eq.~(\ref{u1gauge}). If we use the mapping in Eq.~(\ref{mapx}), the constraint transforms simply to 
the Gauss Law $\Delta_\alpha E_{i \alpha} = 0$. However this constraint does not commute with $\mathcal{H}_{U(1)}$ because of the contribution of the Higgs field. The proper Gauss law constraint is
\beq
\Delta_\alpha E_{i \alpha} - 2 \hat{N}_i = 0\,, \label{Gu1}
\eeq
which is expected, given the presence of a charge 2 matter field. It can be verified that Eq.~(\ref{Gu1}) commutes with Eq.~(\ref{Hu1}).

We can now state the main result of this subsection, obtained by Fradkin and Shenker \cite{FradkinShenker}: the phases and phase transitions of the 
U(1) gauge theory with a charge 2 Higgs field, defined by Eqs.~(\ref{AE},\ref{Hu1},\ref{TN},\ref{Gu1}), are the same as those
of the $\mathbb{Z}_2$ gauge theory, defined by Eqs.~(\ref{Hz2},\ref{G1}).
The U(1) formulation allows easy access to a continuum limit, which then allows us to use the powerful methods of field theory and particle-vortex
duality. 

Let us analyze the properties of the U(1) gauge theory in such a continuum theory. We impose the constraint in Eq.~(\ref{Gu1}) by a Lagrange
multiplier $A_{i \tau}$, which will serve as a time component of the gauge field. The continuum limit is expressed in terms of a 
U(1) gauge field $A_\mu$ ($\mu=x,y,\tau$) and the Higgs field $H$, and takes the form of a standard relativistic theory of the Higgs field with
the Lagrangian density
\bea
\mathcal{L}_{U(1)} &=& \mathcal{L}_{H}  + \mathcal{L}_{\rm monopole} \nonumber \\
  \mathcal{L}_{H}  &=& |(\partial_\mu - 2 i A_\mu) H|^2 + g |H|^2 + u |H|^4 + K (\epsilon_{\mu\nu\lambda} \partial_\nu A_\lambda)^2  \,. \label{LU1}
\eea
The gauge invariance in Eq.~(\ref{u1gauge}) has now been lifted to the continuum
\beq
A_{\mu} \rightarrow A_{\mu} + \partial_\mu f \quad, \quad H \rightarrow H e^{2if} \,. \label{u1gaugec}
\eeq
This theory is similar to the conventional Landau-Ginzburg theory of a superconductor coupled to an electromagnetic field, but with two important
differences: the fluctuations of the gauge field are not weak, and we have to allow for Dirac monopole instantons in which the U(1) gauge
flux changes by $2 \pi$. The latter are represented schematically by the source term 
$\mathcal{L}_{\rm monopole}$, and such instantons are present because
of the periodicity of the gauge field on the lattice. 

\begin{figure}[htb]
\begin{center}
\includegraphics[height=1.85in]{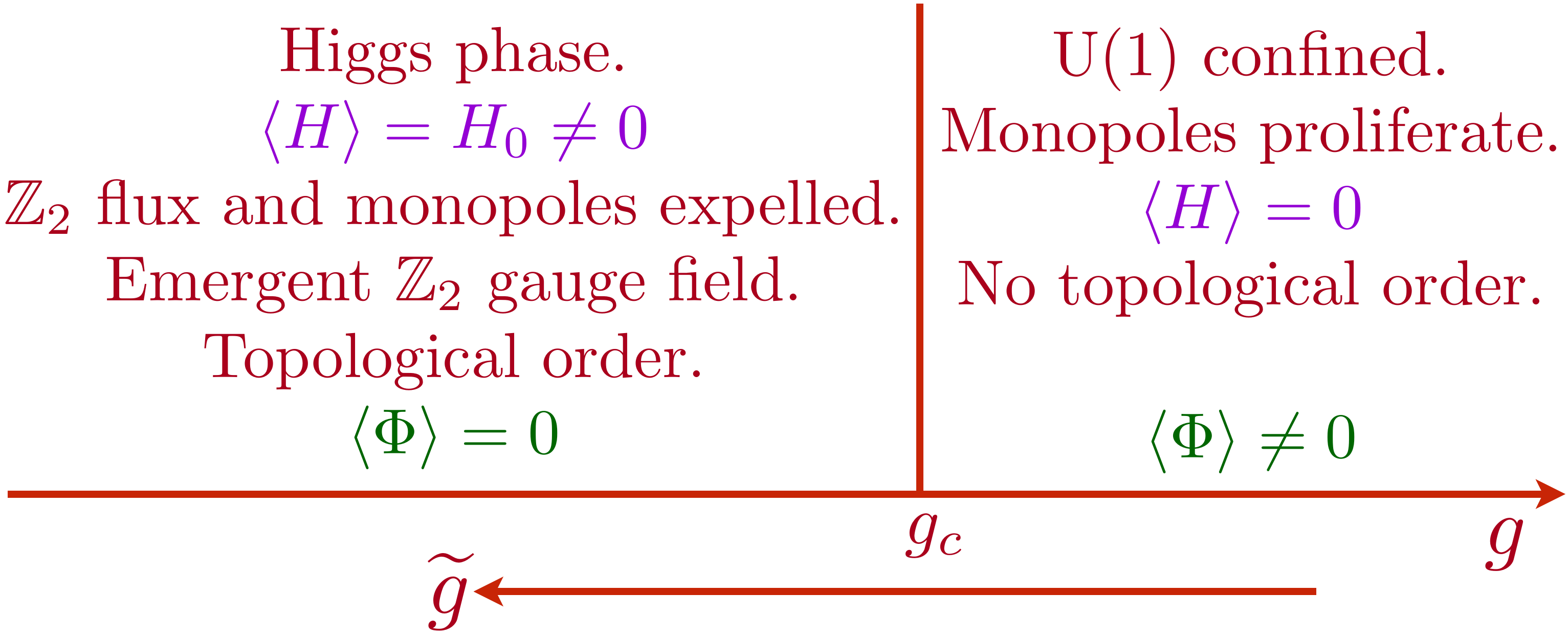}
\end{center}
\caption{Phase diagram of the U(1) gauge theory in Eq.~(\ref{LU1}), which corresponds to the phase diagrams of the $\mathbb{Z}_2$ gauge
theory in Figs.~\ref{fig:wilson} and~\ref{fig:z2topo}. The vison field $\Phi$ represents a $2 \pi$ vortex in $H$, corresponding to $p=\pm 1$
in Fig.~\ref{fig:abrikosov}. The above is also the phase diagram of the theory for the visons in Eq.~(\ref{Lvison}), as a function of $\widetilde{g}$; this vison theory shows that the critical points is described by the Ising$^\ast$ Wilson-Fisher CFT.}
\label{fig:u1}
\end{figure}

\begin{figure}[htb]
\begin{center}
\includegraphics[height=2.7in]{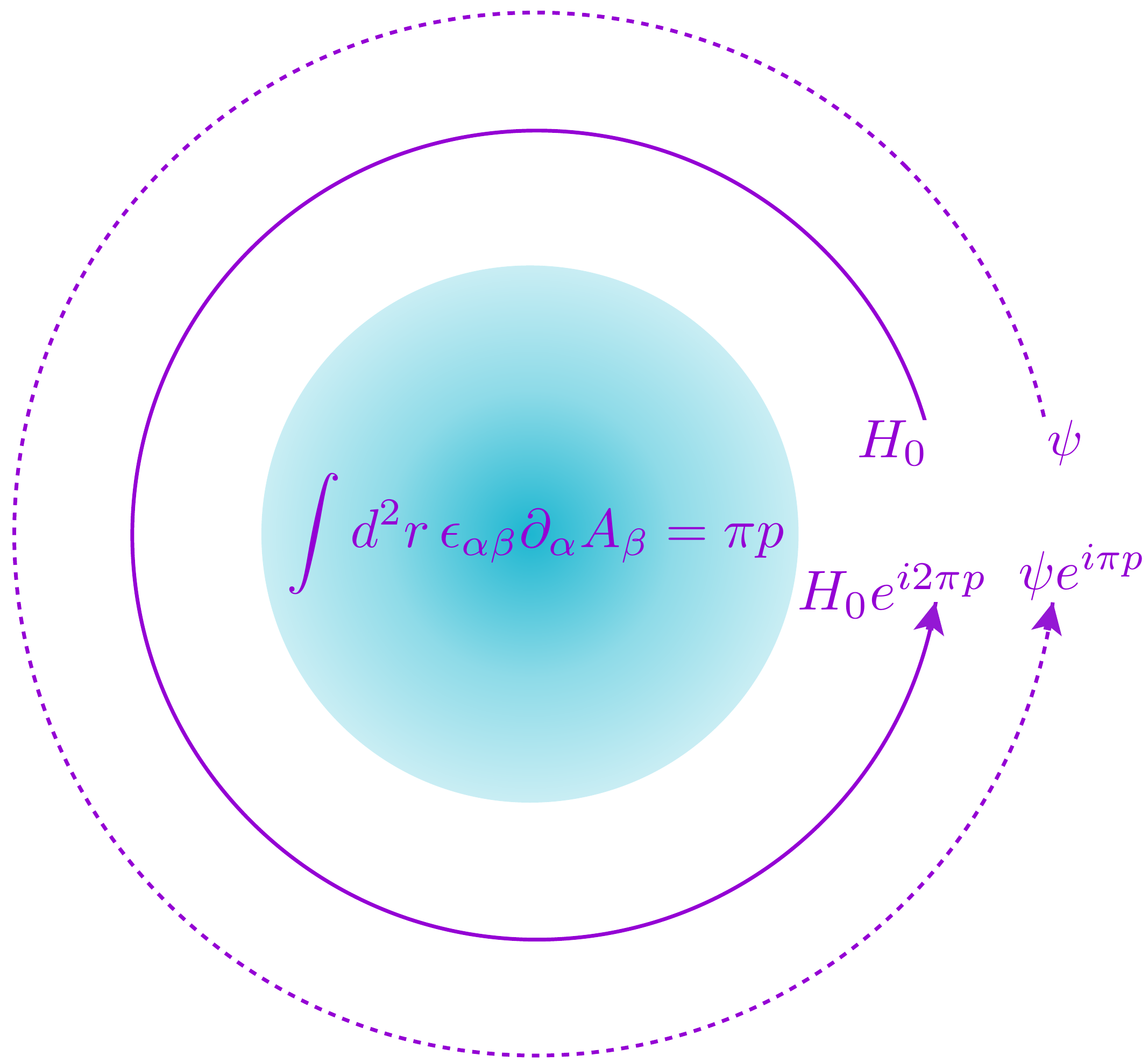}
\end{center}
\caption{Structure of an Abrikosov vortex saddle point of Eq.~(\ref{LU1}). The Higgs field magnitude $\left| \langle H \rangle \right| 
\rightarrow 0$ as $r \rightarrow 0$. Far from the vortex core, $\left| \langle H \rangle \right| \rightarrow |H_0| \neq 0$, and the phase
of $H_0$ winds by $2 \pi p$, where $p$ is an integer. 
There is gauge flux trapped in the vortex core; far from the core, the gauge field screens the Higgs field gradients, and so the energy of
the vortex is finite.
The trapped flux is defined only modulo $2 \pi$ because of the monopole source term, 
and so ultimately all odd values of $p$ map to the same vortex (the vison). The dashed line indicated the Berry phase picked up by a $\psi$ excitation
upon parallel transport around the vortex, as discussed below Eqs.~(\ref{Psipsi}) and (\ref{PsiHpsi}). Because this Berry phase equals $-1$ for a vison,
it is not possible for the $\psi$ field to condense in the phase with topological order.}
\label{fig:abrikosov}
\end{figure}

The two phases of $\mathcal{L}_{U(1)}$ correspond to the two phases of the $\mathbb{Z}_2$ gauge theory in 
Figs.~\ref{fig:wilson} and~\ref{fig:z2topo}, and are sketched in Fig.~\ref{fig:u1}. 
For $g>g_c$, we have no Higgs condensate, $\langle H \rangle =0$,
and then $\mathcal{L}_{U(1)}$ reduces to a pure U(1) gauge theory will monopole sources in the action: such a theory was shown by Polyakov \cite{Polyakov77}
to be confining, and this corresponds to the confining phase of the $\mathbb{Z}_2$ gauge theory. 
For $g<g_c$, we realize the Higgs phase with $\langle H \rangle = H_0 \neq  0$, which corresponds to the deconfined phase of the 
$\mathbb{Z}_2$ gauge theory. Because of the presence of a gauge field, such a condensate does not
correspond to a broken symmetry. But the Higgs phase is topological because there is a stable point-like topological defect, realizing the vison 
of the deconfined phase of the $\mathbb{Z}_2$ gauge theory. This defect is similar to the finite energy 
Abrikosov vortex of the Landau-Ginzburg theory, 
and is sketched in Fig.~\ref{fig:abrikosov}: the phase of $H$ winds by $2 \pi p$ around the core the defect ($p$ is an integer), and this traps a U(1) gauge flux of $\pi p$. However, because of the presence
of monopoles, the flux is conserved only modulo $2 \pi$, and so there is only a single $\pm \pi$ flux defect, which preserves time-reversal
symmetry. 
This $\pi$ flux is clearly the analog
of the $\mathbb{Z}_2$ flux of $-1$ for the vison.  

A dual description of the above Higgs transition provides an elegant route to properly treat the monopole insertion within the theory, rather than as an 
afterthought above. We apply the standard 2+1 dimensional duality \cite{Peskin78,DH81} between a complex scalar ($H$) coupled to a U(1) gauge field ($A_\mu$)
and just a complex scalar ($\Phi$). Here, the field $\Phi$ represents the $\pi$ flux vortex illustrated in Fig.~\ref{fig:abrikosov}. 
A monopole insertion carries flux $2 \pi$, and so turns out to correspond here to the operator $\Phi^2$. In this manner, we obtain
the following theory, which is the particle-vortex dual of Eq.~(\ref{LU1}), including the monopole insertion \cite{RJSS91,SSMV99}
\bea
\mathcal{L}_{d,U(1)} &=& \mathcal{L}_\Phi + \mathcal{L}_{\rm monopole} \nonumber \\
\mathcal{L}_\Phi &=& |\partial_\mu \Phi|^2 + \widetilde{g} |\Phi|^2 + \widetilde{u} |\Phi|^4  \nonumber \\
\mathcal{L}_{\rm monopole} &=& - \lambda \left( \Phi^2 + \Phi^{\ast 2} \right)\,.
\label{Lvison}
\eea
It is also possible \cite{SSMV99} to explicitly derive Eq.~(\ref{Lvison}) by carrying out the duality transformation on the lattice using a Villain form
of the original lattice gauge theory in Eq.~(\ref{Hu1}). 
The interesting feature here is the explicit form of the $\mathcal{L}_{\rm monopole}$ term, which inserts monopoles and anti-monopoles,
and which is always strongly relevant. For $\lambda >0$ (say), $\mathcal{L}_{\rm monopole}$ prefers the real 
part of $\Phi$ over the imaginary part of $\Phi$: so effectively, at low energies, $\mathcal{L}_{d,U(1)}$ is actually the theory of a real (and not complex) scalar. The phase where $\Phi$ is condensed, corresponding to the proliferation of $\mathbb{Z}_2$ flux in the $\mathbb{Z}_2$ gauge theory,
is the confining phase, as illustrated in Fig.~\ref{fig:u1}. And the phase where $\Phi$ is gapped is the deconfined phase: this has is a gapped real particle carrying $\mathbb{Z}_2$ flux, the vison. 

The new result which can be obtained from Eq.~(\ref{Lvison}) is the universality class of the confinement-deconfinement transition. We integrate out the always gapped imaginary part of $\Phi$, and then $\mathcal{L}_{d,U(1)}$ becomes the Wilson-Fisher
theory of the Ising transition in 2+1 dimensions. So the phase transition is in the Ising universality class,
a result already obtained by Wegner \cite{wegner71,TKPS08}, using a Kramers-Wannier duality on the lattice $\mathbb{Z}_2$ gauge theory.
Strictly speaking, as we briefly noted in Section~\ref{sec:ising}, the transition is actually is in the Ising$^\ast$ universality class \cite{2016PhRvL.117u0401S,2016PhRvB..94h5134W}.
This differs from the Ising universality by dropping operators which are odd under $\Phi \rightarrow - \Phi$, because the topological order
prohibits creation of single visons. 

\subsection{Quantum XY model at integer filling}
\label{sec:xyeven}

We can easily extend the U(1) gauge theory mapping of Section~\ref{sec:z2even} to the $D=3$ XY models of Section~\ref{sec:xytopo}.
The key additional feature we need is the presence of the half-boson-number, $\hat{N}=1/2$, excitations $\psi$, defined in Eq.~(\ref{defpsi}). 
These carry $\mathbb{Z}_2$ electric charges, and from the structure of the $J$ term in Eq.~(\ref{bhxy}), we see that they should also carry
a unit U(1) charge. Combined with gauge invariance and symmetry arguments, we conclude that the continuum Lagrangian of the XY models
at integer filling, defined by the Hamiltonian in Eq.~(\ref{bhxy}) and the constraint in Eq.~(\ref{GXY1}),
is obtained by extending Eq.~(\ref{LU1}) to 
\bea
\mathcal{L}_{XY} &=& \mathcal{L}_H + \mathcal{L}_\psi + \mathcal{L}_{\rm monopole} \nonumber \\
  \mathcal{L}_{H}  &=& |(\partial_\mu - 2 i A_\mu) H|^2 + g |H|^2 + u |H|^4 + K (\epsilon_{\mu\nu\lambda} \partial_\nu A_\lambda)^2 \nonumber \\
\mathcal{L}_\psi &=& |(\partial_\mu + i A_\mu) \psi|^2 + s |\psi|^2 + u' |\psi|^4 \,. \label{LXY}
\eea
The gauge invariance in Eq.~(\ref{u1gaugec}) is now extended to
\beq
A_{\mu} \rightarrow A_{\mu} + \partial_\mu f \quad, \quad H \rightarrow H e^{2if} \quad, \quad \psi \rightarrow \psi e^{-if} \,. \label{u1gaugecc}
\eeq
Note that a term of the form $H \, \psi^2$, although gauge invariant under Eq.~(\ref{u1gaugecc}), 
is not allowed in the Lagrangian, because such a term carries a charge
under the global U(1) symmetry of the XY model, which is linked to number conservation in the boson models of Fig.~\ref{fig:dimer}.
Indeed, this combination is the gauge-invariant XY order parameter, which is modified from the
$\mathbb{Z}_2$ gauge theory form in Eq.~(\ref{Psipsi}) to the U(1) gauge theory form
\beq
\Psi = H \, \psi^2 \quad, \quad  \hat{N}_b = \hat{N} + \hat{n}/2 \,. \label{PsiHpsi}
\eeq

\begin{figure}[htb]
\begin{center}
\includegraphics[height=3.5in]{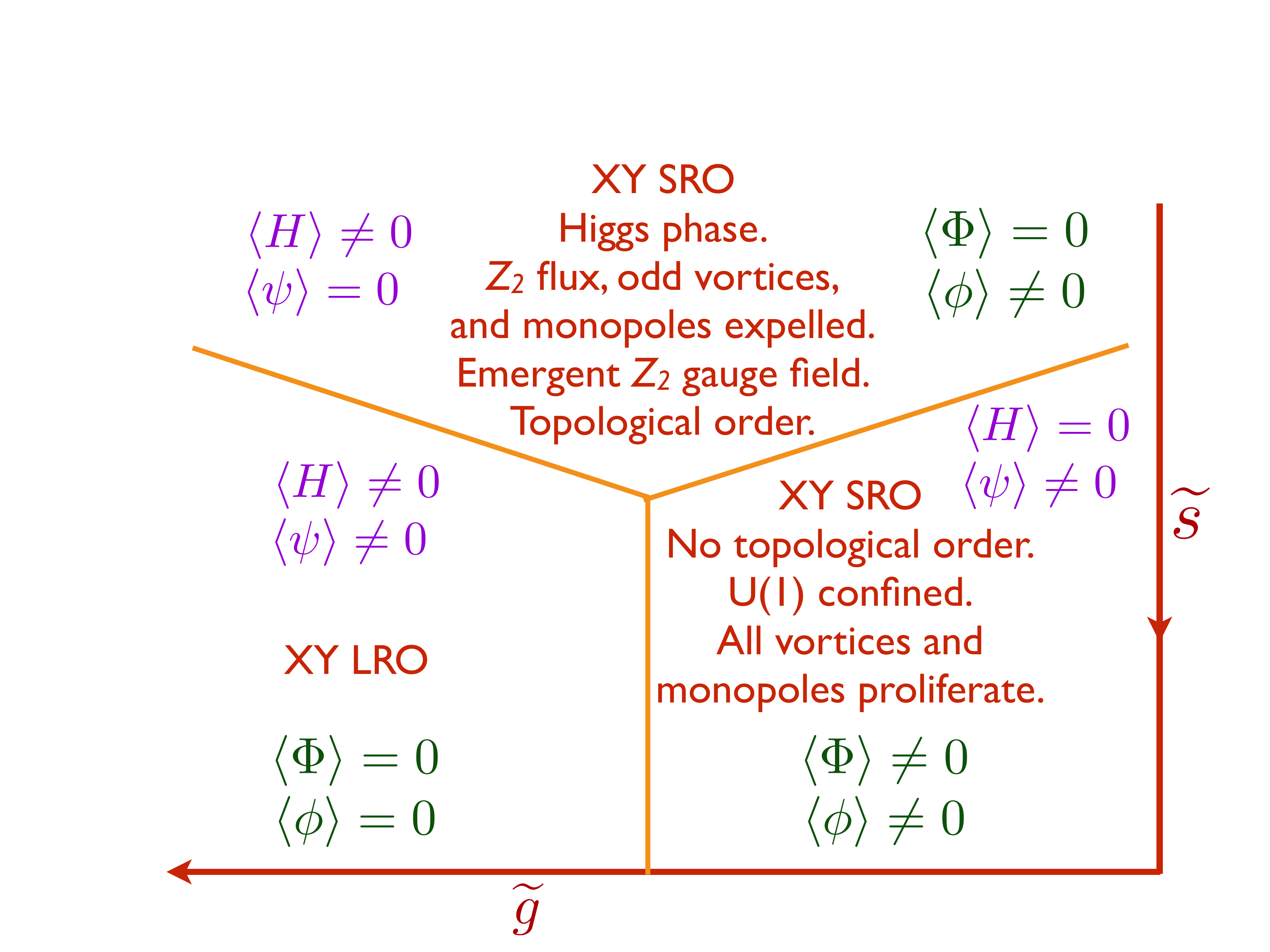}
\end{center}
\caption{The schematic phase diagram of the XY model at integer filling in Fig.~\ref{fig:xytopo}, presented in the language of the 
U(1) gauge theory in Eq.~(\ref{LXY}), and its dual vortex representation in Eq.~(\ref{LdXY}).  $\Phi$ is a vortex in $H$, and $\phi$ is a vortex in $\psi$.
In terms of the gauge-invariant XY order parameter $\Psi$, $\Phi$ is a $2\pi$ vortex in $\Psi$, and $\phi$ is a $4 \pi$ vortex in $\Psi$. Note that the condensates
in $\Phi$ and $\phi$ specify the same vortex proliferations as in Fig.~\ref{fig:xytopo}.}
\label{fig:lxy}
\end{figure}

We can now identify the phases in the phase diagram of Fig.~\ref{fig:xytopo} using the degrees of freedom in the U(1) gauge theory in Eq.~(\ref{LXY});
the updated phase diagram is shown in Fig.~\ref{fig:lxy}.
As in Fig.~\ref{fig:u1}, the topological phase is the `Higgs phase', where the U(1) Higgs field, $H$, is condensed, but the $\psi$ excitations remain gapped.
Because of the unit U(1) charge of $\psi$ in Eq.~(\ref{LXY}), the gapped $\psi$ excitations pick up a Berry phase of -1 around a vison, as indicated in 
Fig.~\ref{fig:abrikosov}.
Also, as in Fig.~\ref{fig:u1}, the confining phase has proliferation of U(1) monopoles; a confining phase is smoothly connected to a Higgs phase where unit 
charges are condensed \cite{FradkinShenker}, and so we have also identified this phase with the presence of a $\psi$ condensate in Fig.~\ref{fig:lxy}. 
The new phase in Fig.~\ref{fig:lxy}, not present in Fig.~\ref{fig:u1}, is the phase with XY LRO: as is clear from Eq.~(\ref{PsiHpsi}), LRO order is only present when both $H$ and $\psi$ are condensed. 

As in Section~\ref{sec:z2even}, for a complete continuum action which can account for $\mathcal{L}_{\rm monopole}$, and 
capture all the phases and phase transitions in Fig.~\ref{fig:lxy}, we need to perform
a duality transform of $\mathcal{L}_{XY}$ to vortices. Such a duality mapping of Eq.~(\ref{LXY}) proceeds as that outlined for Eq.~(\ref{LU1}).
In addition to vortex field $\Phi$, dual to $H$, we need a vortex field, $\phi$, dual to $\psi$. Because the $\psi$ particle carries XY boson number $\hat{N}_b = 1/2$,
the dual vortex $\phi$ will be a $4 \pi$ vortex. Further details of the mapping appear in Refs.~\cite{LFS01,SP01,SS04review}, 
but most features can be deduced from gauge invariance and general arguments.
In particular, the dual theory must have a remnant U(1) gauge field $b_\mu$, so that the flux of $b_\mu$ is the number current of the original bosons:
\beq
J_\mu = \frac{1}{2 \pi} \epsilon_{\mu\nu\lambda} \partial_\nu b_\lambda \quad , \quad J_\tau = \hat{N}_b \,.
\eeq
Because $J_\mu$ is conserved, there can be no source terms for monopoles of $b_\mu$ in the action. This is an important advantage of the dual
formalism, which enables a common continuum limit across the phase diagram of Fig.~\ref{fig:lxy}. In this manner, we deduce the following theory
dual to Eq.~(\ref{LXY}), which generalizes Eq.~(\ref{Lvison}):
\bea
\mathcal{L}_{d,XY} &=& \mathcal{L}_{\Phi} + \mathcal{L}_\phi + \mathcal{L}_{\rm monopole} \nonumber \\
\mathcal{L}_{\Phi} &=& |(\partial_\mu -i b_\mu)\Phi|^2 + \widetilde{g} |\Phi|^2 + \widetilde{u} |\Phi|^4 \nonumber \\
\mathcal{L}_\phi &=& |(\partial_\mu - 2i  b_\mu)\phi|^2 + \widetilde{s} |\phi|^2 + \widetilde{u}' |\phi|^4 \nonumber \\
\mathcal{L}_{\rm monopole} &=&  - \lambda \left( \phi^\ast \Phi^2 + \phi\, \Phi^{\ast 2} \right) 
\label{LdXY}
\eea
As before $\mathcal{L}_{\rm monopole}$ represents the source terms for monopoles and anti-monopoles in the $A_\mu$ gauge field, as in Eq.~(\ref{Lvison});
the additional factors of $\phi$ and $\phi^\ast$ in these terms are required for $b_\mu$ gauge invariance. 
Alllowing for condensates in one or both of $\Phi$ and $\phi$, we obtain all the phases in Fig.~\ref{fig:lxy}. Because of the $\lambda$ term,
a $\phi$ condensate must be present when there is a $\Phi$ condensate, and that is why there are only 3 phases in Fig.~\ref{fig:lxy}: a phase with $\Phi$ condensate
but no $\phi$ condensate is not allowed. 

It is useful to obtain an effective theory for the excitations of the topological phase in this language. The $\Phi$ field is gapped, and its
quanta are evidently the visons. The original $\psi$ particles are also valid gapped excitations (because the dual $\phi$ field is condensed), and
so should be kept `alive' in the effective theory. We can obtain the needed theory by performing a {\it partial\/} duality transform on the original
theory $\mathcal{L}_{XY}$ in Eq.~(\ref{LXY}): we apply the particle-vortex duality \cite{Peskin78,DH81} on $\mathcal{L}_H$ but {\em not\/} on $\mathcal{L}_\psi$, 
while viewing $A_\mu$ as a background gauge field. In this manner we obtain a Lagrangian for the topological phase
\beq
\mathcal{L}_{\rm topo} = \mathcal{L}_\psi + \mathcal{L}_\Phi + \mathcal{L}_{\rm monopole} +  \mathcal{L}_{\rm cs}\,,
\eeq
where $\mathcal{L}_\psi$ (defined in Eq.~(\ref{LXY})) describes the gapped $\psi$ particle, 
$ \mathcal{L}_\Phi + \mathcal{L}_{\rm monopole}$ (defined in Eq.~(\ref{LdXY}) with $\phi$
replaced by its condensate value) describes the gapped $\Phi$ particle, and
$\mathcal{L}_{\rm cs}$ is precisely the U(1)$\times$U(1) Chern-Simons term postulated earlier in Eq.~(\ref{eq:lcs}).
Here $\mathcal{L}_{\rm cs}$ accounts for the mutual semionic statistics between the $\psi$ and $\Phi$ particles.
When we neglect the gapped $\Phi$ and $\psi$ excitations, then $\mathcal{L}_{\rm topo}$ reduces to the purely topological Chern-Simons
theory, which describes the ground state degeneracy on the torus and other manifolds, as in Section~\ref{sec:z2topo}.

We close this subsection by noting the universality classeses of the 3 phase transitions in Fig.~\ref{fig:xytopo} or Fig.~\ref{fig:lxy}.
\begin{itemize}
\item 
The topological transition 
between the two XY SRO phases: $\phi$ is condensed on both sides, and this gaps out the gauge field $b_\mu$. 
Then the theory in Eq.~(\ref{LdXY}) reduces to a theory for $\Phi$ alone with the same Lagrangian as in Eq.~(\ref{Lvison}). This implies
that this confinement transition is just as in the pure $\mathbb{Z}_2$ gauge theory, in the Ising$^\ast$ universality class.
\item
The symmetry breaking transition between XY LRO and the non-topological XY SRO: this is the conventional transition already discussed in Section~\ref{sec:xyd3},
and is in the Wilson-Fisher XY universality class
\item 
The symmetry breaking and topological transition 
between XY LRO and the $\mathbb{Z}_2$ topological order: we return to the undualized description in Eq.~(\ref{LXY}),
and note that $H$ is condensed on both sides of the transition, gapping out $A_\mu$. Then we have a theory for $\psi$ alone,
and this is in the XY$^\ast$ universality class \cite{CSS93}, because only operators even in $\psi$ are observable.
\end{itemize}

\section{Half-filling, Berry phases, and deconfined criticality}
\label{sec:half}

This section will consider a new set of models, with properties distinct from those we have considered so far. 
These models are ultimately related to the square lattice boson models illustrated in Fig.~\ref{fig:dimer}, at an average
boson density, $\langle \hat{N}_b \rangle$, which is an half-integer \cite{RJSS91,SSMV99,SenthilFisher,MSF01,SM02,SM02PRL},
and also to quantum dimer models \cite{DRSK88,EFSK90,NRSS90,RJSS91,SSMV99}. Readers may skip ahead to Section~\ref{sec:hubbard}
without significant loss of continuity.

Section~\ref{sec:z2odd} will generalize the $\mathbb{Z}_2$ gauge theory of Section~\ref{sec:ising}, and Section~\ref{sec:xyodd}
will extend the analysis of Section~\ref{sec:xyeven} to the XY model at half-integer filling.

\subsection{Odd $\mathbb{Z}_2$ gauge theory}
\label{sec:z2odd}

For the simplest of these models, we return to the square lattice $\mathbb{Z}_2$ gauge theory in Eq.~(\ref{Hz2}), 
and replace the Gauss law constraint in Eq.~(\ref{G1}) by
\beq
G_i = -1 \label{oddc}
\eeq
on all sites, $i$. This corresponds to placing a static background $\mathbb{Z}_2$ electric charge on each lattice site. 
The system has to be globally neutral, and so on a torus of size $L_x \times L_y$, the number of sites, $L_x L_y$, has to be even for there to be any states which satisfy Eq.~(\ref{oddc}). 

Such an `odd' $\mathbb{Z}_2$ gauge theory was not considered by Wegner \cite{wegner71}. As with the even gauge theory in
Section~\ref{sec:ising}, performing the Kramers-Wannier duality with the condition in 
Eq.~(\ref{oddc}) leads to an Ising model in a transverse field on the dual lattice; however in the odd gauge theory, the 
signs of the couplings in each spatial plaquette are frustrated \cite{SenthilFisher,RJSS91}. 
Such a fully frustrated Ising model has been investigated in recent experiments on superconducting qubits \cite{dwave}.

The seemingly innocuous change between Eq.~(\ref{G1}) and Eq.~(\ref{oddc}) turns out to have very significant consequences
when combined with lattice space group symmetries:
\begin{itemize}
\item The topological phase with no broken symmetries is present, but its fractionalized excitations are endowed with additional
degeneracies and transform non-trivially under lattice symmetries. This is a `symmetry enriched' topological (SET) phase with a
$D_8$ symmetry.
\item The confining phase must spontaneously break square lattice symmetries: we will find valence bond solid (VBS) order
in the confining phase. A `trivial' confining is not possible.
\item The phase transition between the topological and confining phase exhibits deconfined criticality. The critical theory is described
by a U(1) gauge theory with an emergent critical photon. In a dual representation, the critical theory is the XY${^\ast}$ Wilson-Fisher CFT, to be 
contrasted from the Ising$^\ast$ Wilson-Fisher CFT criticality for the even gauge theory.
\end{itemize}

\begin{figure}[htb]
\begin{center}
\includegraphics[height=2.8in]{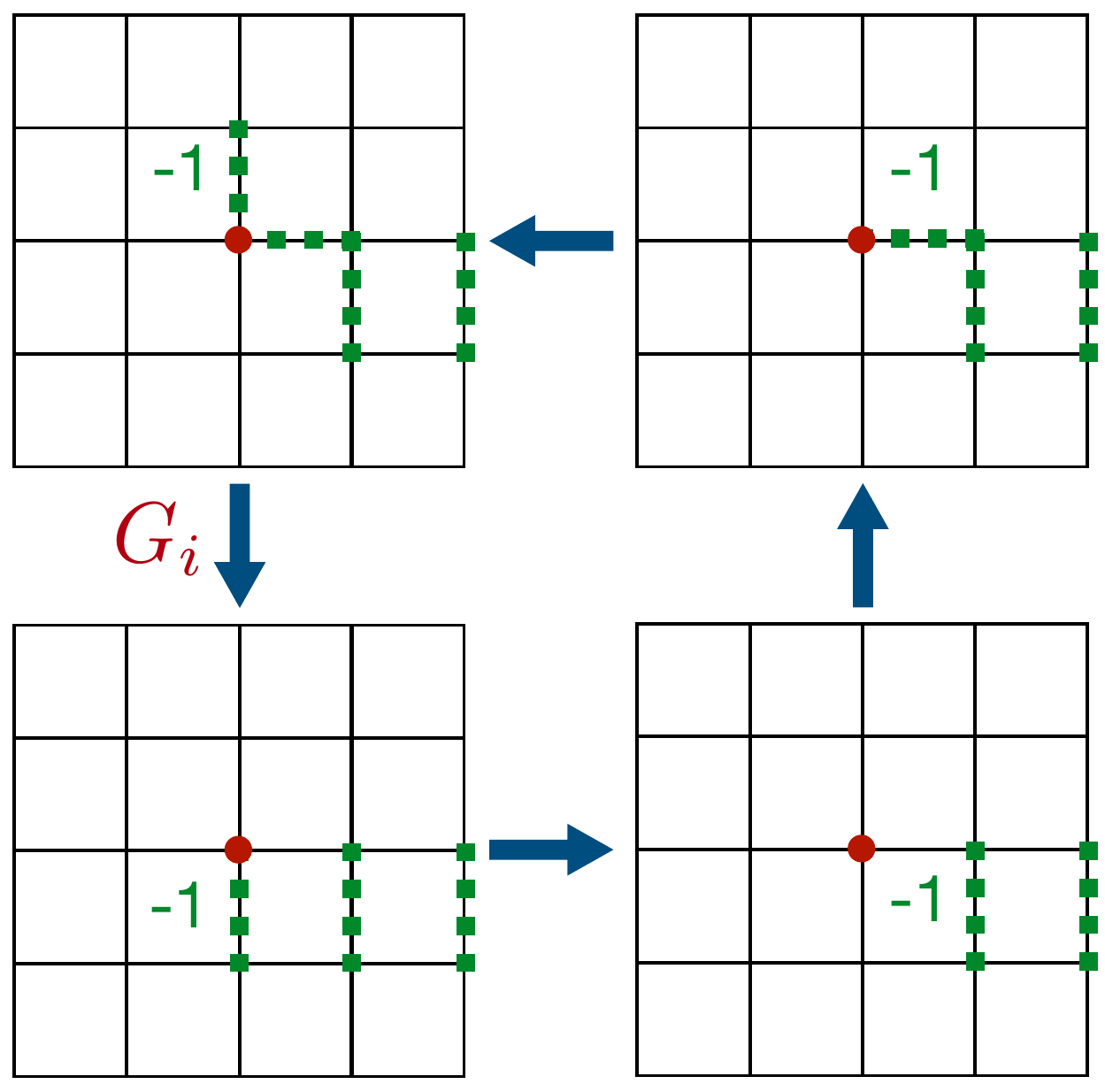}
\end{center}
\caption{Starting from the lower-left, we illustrate a vison undergoing the operations $T_x$, $T_y$, $T_x^{-1}$, $T_y^{-1}$.
The final state differs from the initial state by the action of $G_i$ on the single encircled site. Using Eq.~(\ref{oddc}), we then obtain
Eq.~(\ref{TT}), the $\pi$ Berry phase of a vison moving on the path shown.}
\label{fig:txty}
\end{figure}

First, let us return to the $\mathbb{Z}_2$ gauge theory Hamiltonian in Eq.~(\ref{Hz2}), and deduce some exact consequences of the
odd constraint in Eq.~(\ref{oddc}):\\
({\it i\/}) 
Let $T_x$ ($T_y$) be the operator which translates the system by one lattice spacing along the $x$ ($y$) direction.
Clearly, the operators $T_{x,y}$ commute with the Hamiltonian. Now consider the operators $V_x$ and $V_y$ defined as in Fig.~\ref{fig:thooft}, 
on a $L_x \times L_y$ torus, for convenience on contours $\overline{\mathcal{C}}_x$ and $\overline{\mathcal{C}}_y$ which are straight {\it i.e.\/}
of lengths $L_x$ and $L_y$ respectively. These operators $V_x$ and $V_y$ also commute with the Hamiltonian.
But, as illustrated in Fig.~\ref{fig:vxty}, $V_{x,y}$ and $T_{x,y}$ don't always commute with each other:
\begin{figure}[htb]
\begin{center}
\includegraphics[height=1.3in]{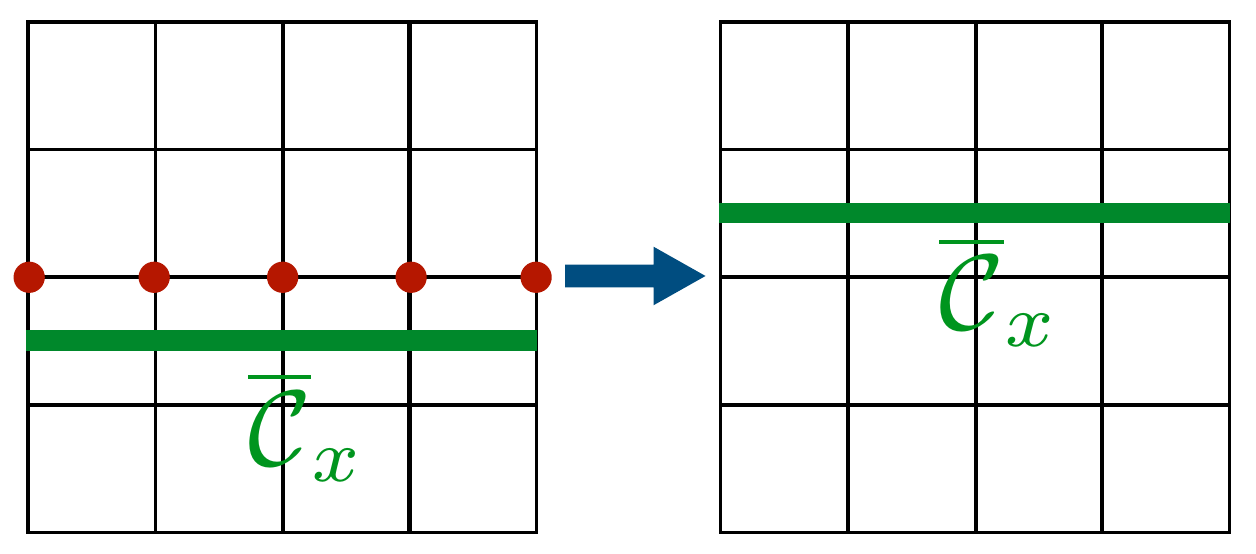}
\end{center}
\caption{The operator $V_x$ on the contour $\overline{\mathcal{C}}_x$ is translated by $T_y$ upon the action of $G_i$ on the encircled sites.
Then using Eq.~(\ref{oddc}), we obtain Eq.~(\ref{TV}).}
\label{fig:vxty}
\end{figure}
\beq
T_x V_y = (-1)^{L_y} V_y T_x \quad \quad, \quad \quad T_y V_x = (-1)^{L_x} V_x T_y 
\label{TV}
\eeq
The relations in Eq.~(\ref{TV}) are valid on any state obeying Eq.~(\ref{oddc}), and they imply that there is no trivial non-degenerate ground state of $\mathcal{H}_{\mathbb{Z}_2}$.\\
({\it ii\/}) In the small $g$ limit, the topological state is modified from Eq.~(\ref{eq:gs0}) to
\beq
\left| 0 \right\rangle = \prod_i (1 - G_i) \left| \Uparrow \right\rangle \label{eq:gs1}
\eeq
({\it iii\/}) $T_x$ and $T_y$ do not commute when acting on a vison state $| v \rangle$:
\beq
T_x T_y |v \rangle = - T_y T_x |v \rangle \,. \label{TT}
\eeq
The proof of this relation is presented in Fig.~\ref{fig:txty}.
This implies that the vison accumulates a Berry phase of $\pi$ when transported around a single square lattice site.

Next, we will deduce the consequences of these properties of the $\mathbb{Z}_2$ gauge theory by proceeding to the embedding
in a U(1) gauge theory. The lattice gauge theory Hamiltonian, $\mathcal{H}_{U(1)}$ remains the same as in Eq.~(\ref{Hu1}). But now
the constraint in Eq.~(\ref{oddc}) changes the local constraint in Eq.~(\ref{Gu1}) to 
\beq
\Delta_\alpha E_{i \alpha} - 2 \hat{N}_i = \eta_i \,, \label{Gu1o}
\eeq
where $\eta_i$ was defined in Eq.~(\ref{defeta}). So there is a background unit U(1) electric charge on each lattice site, but its sign is staggered.
The staggering is a consequence of that in Eq.~(\ref{mapz}), where it was required to enable the flux term to have the form of a lattice curl.
We proceed as in Section~\ref{sec:z2even} to the continuum limit of the U(1) gauge theory: the form in Eq.~(\ref{LU1}) is changed to 
the action \cite{RJSS91,SSMV99}
\bea
\mathcal{S}_{o,U(1)} &=& \int d^3 x \, \mathcal{L}_H + \mathcal{S}_B + \mathcal{S}_{\rm monopole}  \nonumber \\
\mathcal{L}_H &=&
|(\partial_\mu - 2 i A_\mu) H|^2 + g |H|^2 + u |H|^4 + K (\epsilon_{\mu\nu\lambda} \partial_\nu A_\lambda)^2 
\nonumber \\
\mathcal{S}_B &=&  i \sum_i \eta_i \int d \tau A_{i \tau} \nonumber \\
\mathcal{S}_{\rm monopole} &=&   \sum_i \int d\tau \, \mathcal{L}_{\rm monopole}
\,. \label{LU1o}
\eea
The terms in the $\mathcal{S}_B + \mathcal{S}_{\rm monopole}$ are required to be evaluated on the lattice, so they have not been absorbed into the continuum theory.
The Berry phase term, $\mathcal{S}_B$, descends from the right-hand-side of Eq.~(\ref{Gu1o}), after $A_{i \tau}$ is used as a Lagrange multiplier
to impose Eq.~(\ref{Gu1o}).

Finally, following Section~\ref{sec:z2even}, we apply particle-vortex duality to Eq.~(\ref{LU1o}) to obtain an effective theory for the vison
excitations. The explicit computation is presented in Ref.~\cite{SSMV99}. Here we will obtain the result by a general argument. 
As indicated in Fig.~\ref{fig:txty}, each vison moves in a background $\pi$ flux per plaquette of the dual lattice
due to the presence of the electric charges on the sites of the direct lattice. It is a simple matter to diagonalize the dispersion
of a particle moving in $\pi$ flux on the square lattice, and we obtain a doubly-degenerate spectrum: 
specifically the lowest energy states are doubly degenerate,
and we denote the real vison particles at these energy minima by $\varphi_{1,2}$. For a suitable choice of gauge, the transformation of these real particles under square lattice symmetries is specified by \cite{BBBSS05,HPS11,PCAS16}
\bea
T_x &:& \varphi_1 \rightarrow \varphi_2 \quad; \quad \varphi_2 \rightarrow \varphi_1 \nonumber \\
T_y &:& \varphi_1 \rightarrow \varphi_1 \quad; \quad \varphi_2 \rightarrow -\varphi_2 \nonumber \\
R_{\pi/2} &:& \varphi_1 \rightarrow \frac{1}{\sqrt{2}}(\varphi_1 + \varphi_2) \quad; \quad \varphi_2 \rightarrow \frac{1}{\sqrt{2}}(\varphi_1 - \varphi_2), \label{TTR}
\eea
where $R_{\pi/2}$ is the symmetry of rotations about a dual lattice point. The transformations in Eq.~(\ref{TTR}), and their 
compositions, form the projective symmetry group which constrains the theory of the topological phase and of its phase transitions.
Direct computation shows that the group generated by Eq.~(\ref{TTR}) is the 16 element non-abelian dihedral group $D_8$ \cite{HPS11}. This 
$D_8$ symmetry plays a central role in the phenomena described in Sections~\ref{sec:z2odd} and \ref{sec:xyodd}.
Now we combine these real particles into a
single complex field
\beq
\Phi = e^{-i \pi/8} \left( \varphi_1 + i \varphi_2 \right) \label{Phivarphi}
\eeq
With these phase factors, $\Phi$ transforms under $D_8$ as
\beq
T_x : \Phi \rightarrow e^{i  \pi/4}  \Phi^\ast \quad; \quad T_y:  \Phi \rightarrow e^{- i \pi/4} \Phi^\ast \quad ; \quad 
R_{\pi/2}: \Phi \rightarrow \Phi^\ast \,.\label{TTRP}
\eeq
These transformations 
are chosen so that the monopole operator $m = \Phi^2$ transforms as 
\beq
T_x: m \rightarrow i m^\ast \quad; \quad T_y: m \rightarrow -i m^\ast \quad; \quad R_{\pi/2}: m \rightarrow m^\ast \,, \label{TTRm}
\eeq
which are also the transformations implied by the monopole Berry phases in Refs.~\cite{Haldane88,NRSS90,senthil1,senthil2}. 
Note that under the vison $D_8$ operations in Eq.~(\ref{TTRP}), $T_x$ and $T_y$ anticommute (as required by Eq.~(\ref{TT})),
while they commute under the monopole operations in Eq.~(\ref{TTRm}).
Then the effective theory for $\Phi$, which is the
new form of $\mathcal{L}_{d,U(1)}$ in Eq.~(\ref{Lvison}), is the simplest Lagrangian invariant under the $D_8$ symmetry
\bea
\mathcal{L}_{od,U(1)} &=& \mathcal{L}_\Phi + \mathcal{L}_{\rm monopole} \nonumber \\
\mathcal{L}_H &=& |\partial_\mu \Phi|^2 + \widetilde{g} |\Phi|^2 + \widetilde{u} |\Phi|^4 \nonumber \\
\mathcal{L}_{\rm monopole} &=& - \overline{\lambda} \left( \Phi^8 + \Phi^{\ast 8} \right)\,.
\label{Lvisono}
\eea
The important new feature of Eq.~(\ref{Lvisono}) is that $\mathcal{L}_{\rm monopole}$ now involves
8 powers of the vison field operator! This implies that only quadrupled monopoles are permitted in the action, in contrast to single
monopoles in Eq.~(\ref{Lvison}). All smaller monopoles cancel
out of the action due to quantum interference arising from Berry phases from $\mathcal{S}_B$ in Eq.~(\ref{LU1o}).

\begin{figure}[htb]
\begin{center}
\includegraphics[height=3.35in]{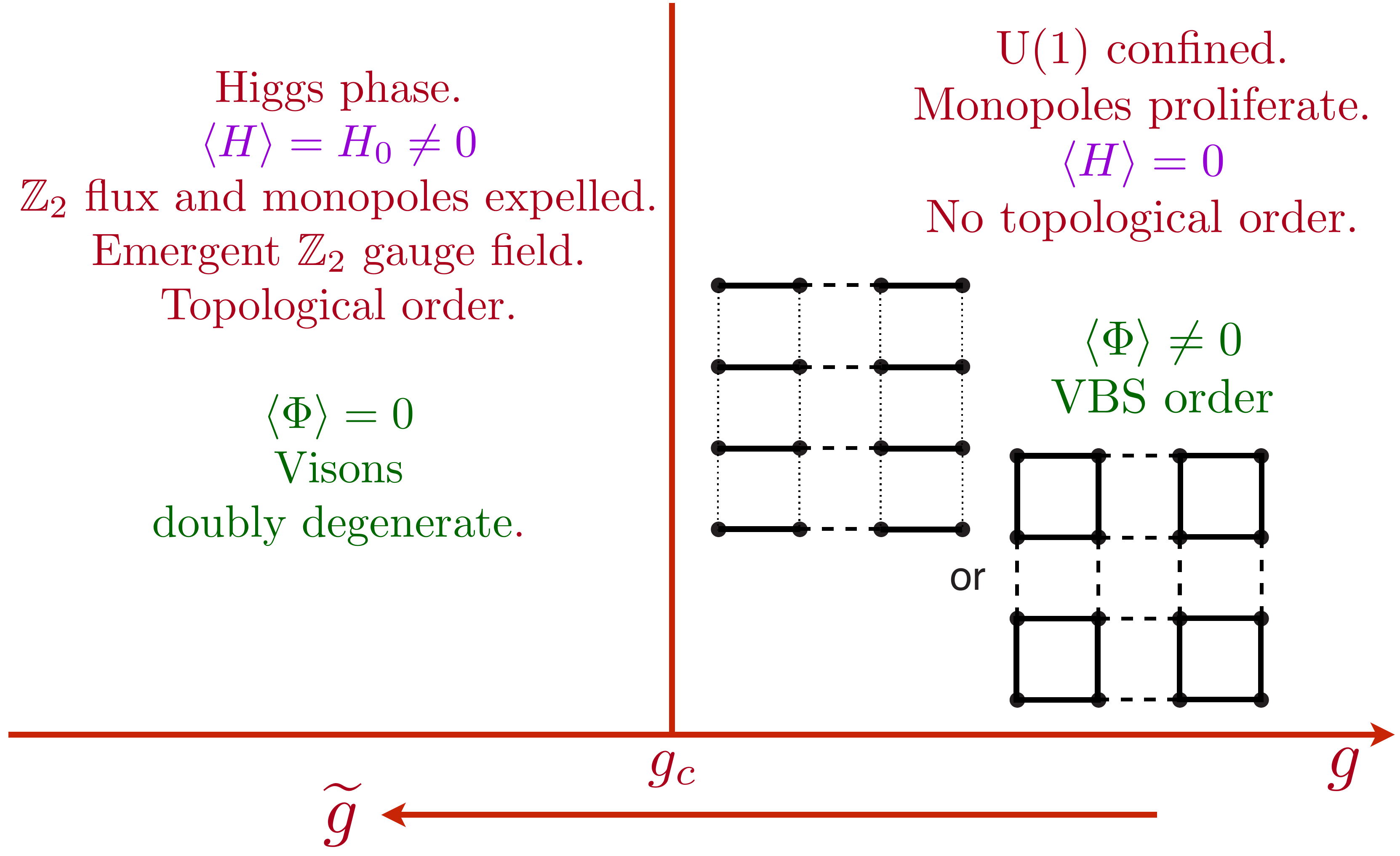}
\end{center}
\caption{Phase diagram of the U(1) gauge theory in Eq.~(\ref{LU1o}) which describes the physics of the odd $\mathbb{Z}_2$ gauge theory defined by Eqs.~(\ref{Hz2}) and (\ref{oddc}). 
Compare to Fig.~\ref{fig:u1} for the even $\mathbb{Z}_2$ gauge theory.
The vison field $\Phi$ represents a $2 \pi$ vortex in $H$. The theory for the visons is Eq.~(\ref{Lvisono}), with the tuning 
parameter $\widetilde{g}$. Monopoles are suppressed at the deconfined critical point at $g=g_c$ above, and consequently 
there is an emergent critical U(1) photon described by the deconfined critical theory $\mathcal{L}_H$ in Eq.~(\ref{LU1o}); in the dual representation of the doubly-degenerate visons, the critical theory is the
XY$^\ast$ Wilson-Fisher CFT, described by $\mathcal{L}_\Phi$ in Eq.~(\ref{Lvisono}). In contrast, monopoles are not suppressed
at $g=g_c$ in the even $\mathbb{Z}_2$ gauge theory phase diagram of Fig.~\ref{fig:u1}. This phase diagram is the earliest example of deconfined criticality, and a numerical study appeared in Ref.~\cite{RJSS91}.}
\label{fig:u1a}
\end{figure}
The phase diagram of $\mathcal{L}_{od,U(1)}$ is modified from Fig.~\ref{fig:u1} to Fig.~\ref{fig:u1a}.
The topological phase has a gapped $\Phi$ excitation. A crucial difference from the even $\mathbb{Z}_2$ gauge theory is that this excitation
is doubly degenerate: $\mathcal{L}_{\rm monopole}$ is sufficiently high order that the degeneracy between the real and imaginary
parts of $\Phi$ is no longer broken (unlike in Eq.~(\ref{Lvison})). So the vison is a complex relativistic particle, unlike the real particle
in Section~\ref{sec:z2even}. This double degeneracy in the vison states is a feature of the symmetry-enriched topological 
order \cite{EH13,SPTSET}, and is intimately linked to the $D_8$ symmetry and 
to the anti-commutation relation \cite{BBBSS05} in Eq.~(\ref{TT}):
it is not possible obtain vison states which form a representation of the algebra of $T_x$ and $T_y$ without this degeneracy.

Turning to the confined phase where $\Phi$ is condensed, the non-trivial transformations in Eq.~(\ref{TTRP}) imply that lattice symmetries
must be broken. The precise pattern of the broken symmetry depends upon the sign of $\overline{\lambda}$, and the two
possibilities are shown in Fig.~\ref{fig:u1a}.

Finally, we address the confinement-deconfinement transition in Fig.~\ref{fig:u1a}. 
In Eq.~(\ref{Lvisono}), $\mathcal{L}_{\rm monopole}$ is an irrelevant perturbation to 
$\mathcal{L}_\Phi$, and the critical point of $\mathcal{L}_\Phi$ is the XY$^\ast$ Wilson-Fisher CFT \cite{RJSS91,SSMV99,senthil1,senthil2} (contrast this with the Ising$^\ast$ Wilson-Fisher CFT
in Fig.~\ref{fig:u1}). Undoing the duality mapping back to Eq.~(\ref{LU1o}), we note that the XY$^\ast$ Wilson-Fisher CFT 
undualizes precisely to $\mathcal{L}_H$. So $\mathcal{S}_B$ and $\mathcal{S}_{\rm monopole}$ in 
Eq.~(\ref{LU1o}) combine to render to each other
irrelevant in the critical theory: the Berry phases in $\mathcal{S}_B$ suppress the monopole tunneling events. 
Consequently, the resulting U(1) gauge theory, $\mathcal{L}_H$, retains a critical photon: 
this is the phenomenon of deconfined criticality \cite{RJSS91,SSMV99,senthil1,senthil2}. The embedding of the $\mathbb{Z}_2$ 
gauge theory into the U(1) gauge theory is now not optional: it is necessary to obtain a complete description of the critical
theory of the phase transition in Fig.~\ref{fig:u1a}. And the critical theory is $\mathcal{L}_H$ in Eq.~(\ref{LU1o}), 
the abelian Higgs model in 2+1 dimensions, which describes a
critical scalar coupled to a U(1) gauge field {\it i.e.\/} the naive continuum limit of
the lattice U(1) gauge theory yields the correct answer for the critical theory, and monopoles and Berry phases can be ignored.
This should be contrasted with the even $\mathbb{Z}_2$ gauge theory case in Section~\ref{sec:z2even}, where monopoles were relevant.

In closing, we note that the above phase diagram also applies to quantum dimer models on the square lattice \cite{RJSS91,SSMV99}. 
The extension to quantum dimer
models on other lattices have also been considered \cite{MSC99,RMSLS01,MSC01,VBS04,FHMOS04,HPS11}.

\subsection{Quantum XY model at half-integer filling}
\label{sec:xyodd}

In this final subsection, we briefly address the case of bosons with short-range interactions on the square lattice
at half-integer filling. The results apply also to easy-plane $S=1/2$ antiferromagnets on the square lattice, which were
the focus of attention in the studies of Refs.~\cite{NRSS89,NRSS90,senthil1,senthil2}.
The analysis involves some rather subtle interplay between Berry phases and particle-vortex duality,
and readers may skip this section without loss of continuity.

We describe here the properties of the Hamiltonian in Eq.~(\ref{bhxy}), but the constraint in Eq.~(\ref{GXY1}) is now modified to
`odd' constraint appropriate to half-integer filling.
\beq
G_i^{XY} = -1\,. \label{GXYm1}
\eeq
Our results will be obtained by combining the U(1) gauge theories of Sections.~\ref{sec:xyeven} and~\ref{sec:z2odd}.
We begin with the integer-filling XY model theory of Eq.~(\ref{LXY}), and add to it the odd $\mathbb{Z}_2$ gauge theory
Berry phases in Eq.~(\ref{LU1o}) to obtain the action
\bea
\mathcal{S}_{o,XY} &=& \int d^3 x \Bigl[ \mathcal{L}_H + \mathcal{L}_\psi \bigr] + \mathcal{S}_B + \mathcal{S}_{\rm monopole}  \nonumber \\
\mathcal{L}_H &=&
|(\partial_\mu - 2 i A_\mu) H|^2 + g |H|^2 + u |H|^4 + K (\epsilon_{\mu\nu\lambda} \partial_\nu A_\lambda)^2 
\nonumber \\
\mathcal{L}_\psi &=& |(\partial_\mu + i A_\mu) \psi|^2 + s |\psi|^2 + u' |\psi|^4 \nonumber \\
\mathcal{S}_B &=&  i \sum_i \eta_i \int d \tau A_{i \tau} \nonumber \\
\mathcal{S}_{\rm monopole} &=&   \sum_i \int d\tau \, \mathcal{L}_{\rm monopole}
\,. \label{SXYo}
\eea

\begin{figure}[htb]
\begin{center}
\includegraphics[height=4.5in]{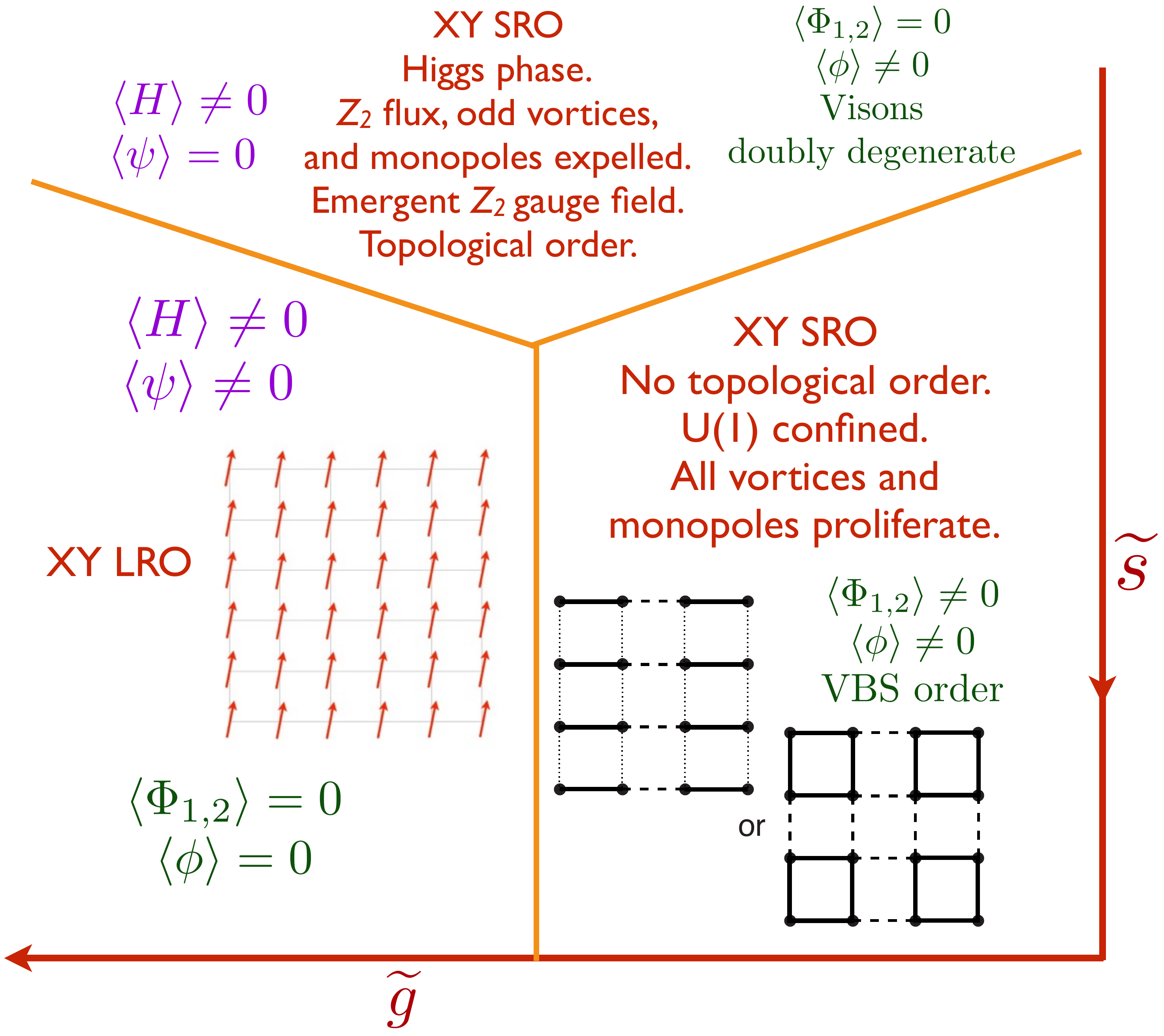}
\end{center}
\caption{Schematic phase diagram of the square lattice quantum XY model at half-integer filling, defined by Eqs.~(\ref{bhxy}) and (\ref{GXYm1}). Compare to the phase diagrams at integer filling in Fig.~\ref{fig:xytopo}
and Fig.~\ref{fig:lxy}. Now there is VBS order in the confining phase, and the $\mathbb{Z}_2$ topological order is symmetry enriched.
The phase transition between VBS and $\mathbb{Z}_2$ topological order (which is the same as that in Fig.~\ref{fig:u1a}), and that
between XY LRO and VBS, are both examples of deconfined criticality. Numerical results on such a phase diagram appear
in Ref.~\cite{PS02}, and a mean-field phase diagram was computed from Eq.~(\ref{LdXYo}) in Ref.~\cite{SP01}.}
\label{fig:lxyo}
\end{figure}

The ground states of Eq.~(\ref{SXYo}) are similar to those of Eq.~(\ref{LXY}) in Fig.~\ref{fig:lxy}, and are now shown in 
Fig.~\ref{fig:lxyo}. The main change from Fig.~\ref{fig:lxy} is that we no longer expect a trivial confining phase: instead, the Berry
phases are expected to introduce VBS order.
This phase diagram is supported by a quantum Monte Carlo study of a suitable sign-problem-free lattice realization which was
presented in Ref.~\cite{PS02}.

As in Section~\ref{sec:xyeven}, evaluating the consequences of $\mathcal{S}_B + \mathcal{S}_{\rm monopole}$ requires a duality 
transform. This was carried out in Refs.~\cite{LFS01,SP01,SS04review}. 
We will not present the general derivation, but present a simple argument which is similar
to that in Section~\ref{sec:z2odd} for the odd $\mathbb{Z}_2$ gauge theory. As in Section~\ref{sec:z2odd}, the main consequence
of the background electric charges is that the vison move in a background $\pi$ flux. However, as illustrated in Fig.~\ref{fig:xyvortex}, 
in the presence of XY degrees of freedom, each vison is attached to a vortex, or anti-vortex, in the XY order. So each vison is 
microscopically a complex particle. We then account for the $\pi$ flux just as in Section~\ref{sec:z2odd}, with the result that
we obtain two {\it complex} visons, $\varphi_{1,2}$, which transform just as in Eq.~(\ref{TTR}). But now we can combine these vison
fields into not one complex field $\Phi$ (as in Eq.~(\ref{Phivarphi})), but two complex fields $\Phi_{1,2}$ which we choose as
(compare to Eq.~(\ref{Phivarphi}))
\beq
\Phi_1 = e^{-i \pi/8} \left( \varphi_1 + i \varphi_2 \right) \quad , \quad \Phi_2 = e^{i \pi/8} \left( \varphi_1 - i \varphi_2 \right) \,. \label{Phi12}
\eeq
From Eqs.~(\ref{TTR}) and (\ref{Phi12}) we then obtain a representation of the 
$D_8$ symmetry transformations (compare to Eq.~(\ref{TTRP}))
\bea
T_x &:& \Phi_1 \rightarrow e^{i \pi/4} \Phi_2 \quad; \quad \Phi_2 \rightarrow e^{-i  \pi/4} \Phi_1 \nonumber \\
T_y &:& \Phi_1 \rightarrow e^{-i \pi/4} \Phi_2 \quad; \quad \Phi_2 \rightarrow e^{i  \pi/4} \Phi_1 \nonumber \\
R_{\pi/2} &:& \Phi_1 \rightarrow  \Phi_2 \quad; \quad \Phi_2 \rightarrow  \Phi_1 \,. \label{TTR12}
\eea
The field definitions in Eq.~(\ref{Phi12}) were chosen 
so that ({\it i\/}) the product $\Phi_1 \Phi_2$ is invariant under all $D_8$ symmetries,
and ({\it ii\/}) the product $m = \Phi_1 \Phi_2^\ast$ transforms as the monopole operator in Eq.~(\ref{TTRm}).

We now proceed with the same arguments as those leading to Eq.~(\ref{LdXY}). The main change is that the field $\Phi$
has been replaced by a two fields $\Phi_{1,2}$, and we have to choose a Lagrangian that is invariant under Eq.~(\ref{TTR12}).
As in $\mathcal{L}_{d,XY}$, we also introduce a $4 \pi$ vortex field $\phi$; this is assumed here to transform trivially under
all the space group operations because the $\phi$ field does not observe any background flux. 
This leads to the half-integer boson density version of $\mathcal{L}_{d,XY}$ \cite{LFS01,SP01}, now invariant under the $D_8$
projective symmetry group:
\bea
\mathcal{L}_{od,XY} &=& \mathcal{L}_\Phi + \mathcal{L}_\phi + \mathcal{L}_{\rm monopole} \nonumber \\
\mathcal{L}_\Phi &=& |(\partial_\mu -i b_\mu)\Phi_1|^2 + |(\partial_\mu -i b_\mu)\Phi_2|^2 + \widetilde{g} \left( |\Phi_1|^2 + |\Phi_2|^2 \right)  \nonumber\\
&~&~~+ \widetilde{u} \left(|\Phi_1|^4 + |\Phi_2|^4 \right)
+ \widetilde{v} |\Phi_1|^2 |\Phi_2|^2 \nonumber \\
\mathcal{L}_\phi &=& |(\partial_\mu - 2i  b_\mu)\phi|^2 + \widetilde{s} |\phi|^2 + \widetilde{u}' |\phi|^4  \nonumber \\
\mathcal{L}_{\rm monopole} &=& - \overline{\lambda} \left( (\Phi_1^\ast \Phi_2)^4 + (\Phi_2^\ast \Phi_1)^4 \right) 
- \lambda \left( \phi^\ast \Phi_1 \Phi_2 + \phi\, \Phi_1^{\ast} \Phi_2^{\ast} \right) \,.
\label{LdXYo}
\eea
Now simple considerations of condensates of $\Phi_{1,2}$ and $\phi$ lead to the phase diagram in Fig.~\ref{fig:lxyo}---a
mean-field phase diagram was computed in Ref.~\cite{SP01}.
The main change from Fig.~\ref{fig:lxy} is that the non-trivial symmetry transformations in Eq.~(\ref{TTR12}) imply
that the presence of $\Phi_{1,2}$ condensates leads to VBS order in the confining phase.

As in Section~\ref{sec:xyeven}, we close this subsection by noting the universality classeses of the three phase transitions in Fig.~\ref{fig:lxyo}:
\begin{itemize}
\item
The transition between the two XY SRO phases: $\phi$ is condensed on both sides, and this gaps out the gauge field $b_\mu$. 
From the $\lambda$ term in Eq.~(\ref{LdXYo}), we may set $\Phi_2 \sim \Phi_1^\ast$. 
Then the theory in Eq.~(\ref{LdXYo}) reduces to a theory for $\Phi_1$ alone with the same Lagrangian as 
that for $\Phi$ 
in Eq.~(\ref{Lvisono}). This implies
that this confinement transition is just as in the odd $\mathbb{Z}_2$ gauge theory in Section~\ref{sec:z2odd}, 
in the XY$^\ast$ universality class.
In the undualized variables, this transition is described by the U(1) gauge theory $\mathcal{L}_H$ in Eq.~(\ref{SXYo}), 
which makes it the earliest example of deconfined criticality \cite{RJSS91,SSMV99,senthil1,senthil2}.
\item
The transition between the XY LRO and VBS states is a prominent example of deconfined criticality \cite{senthil1,senthil2}. For this transition, we can assume
that the $\phi$ field is gapped, and then the Lagrangian reduces to $\mathcal{L}_\Phi$ and the $\overline{\lambda}$ term in Eq.~(\ref{LdXYo}).
The Lagrangian $\mathcal{L}_\Phi$ in Eq.~(\ref{LdXYo}) describes the easy-plane $\mathbb{C}\mathbb{P}^1$ model
in the complex fields $\Phi_{1,2}$. It is assumed that the $\overline{\lambda}$ monopoles are irrelevant at the XY-VBS transition, and then
the critical theory is the critical easy-plane $\mathbb{C}\mathbb{P}^1$ theory. This theory is self-dual \cite{OMAV04}, and it undualizes
to another $\mathbb{C} \mathbb{P}^1$ theory of a pair of relativistic bosons (`spinons') carrying XY boson number $\hat{N}_b = 1/2$.
\item 
The transition between XY LRO and the $\mathbb{Z}_2$ topological order: this the same as that in the XY model at integer filling
in Section~\ref{sec:xyeven}.
We have a theory for $\psi$ alone, in the XY$^\ast$ universality class \cite{CSS93}.
\end{itemize}

\section{Electron Hubbard model on the square lattice}
\label{sec:hubbard}

We are now ready to describe possible states with topological order in Hubbard-like models, relevant for the cuprate superconductors.
Unlike previous sections, the topological states described below can be gapless: they can contain Fermi surfaces of gapless fermions and exhibit
metallic conduction. Nevertheless, the Higgs field approach developed in Section~\ref{sec:higgs} can be deployed on the Hubbard model
largely unchanged.

We consider fermions (electrons), $c_{i\alpha}$, on the sites, $i$, of the square lattice, with spin index $s = \uparrow,\downarrow$.
They are described by the Hubbard Hamiltonian
\beq
\mathcal{H}_U = - \sum_{i< j} t_{ij} c_{i s}^\dagger c_{j s}  + U \sum_{i} \left( n_{i \uparrow} - \frac{1}{2} \right)
\left( n_{i \downarrow} - \frac{1}{2} \right) - \mu \sum_{i} c_{i s}^\dagger c_{i s}
\eeq
where the number operator $n_{i s} \equiv c_{i s}^\dagger c_{i s}$, $t_{ij}$ is the `hopping' matrix element between near-neighbors,
$U$ is the on-site repulsion, and $\mu$ is the chemical potential. 

First, in Section~\ref{sec:sdw}, we will review the mean-field theory of spin density wave order in the Hubbard model.
This will be the analog of the discussion of Section~\ref{sec:xyd3} of LRO in the $D=3$ XY model.
Then, in Sections~\ref{sec:su2} and \ref{sec:su2theory}, we will add a topological phase, as in Section~\ref{sec:xytopo} for the XY model.
Section~\ref{sec:su2} will introduce the argument based upon transformation to a rotating reference frame, showing that topological order
is required (in the absence of translational symmetry breaking) for Fermi surface reconstruction. A more formal argument, based
upon Higgs phases of a SU$_s$(2) gauge theory, appears in Section~\ref{sec:su2theory}.

\subsection{Spin density wave mean-field theory and Fermi surface reconstruction}
\label{sec:sdw}

\begin{figure}[htb]
\begin{center}
\includegraphics[height=2.65in]{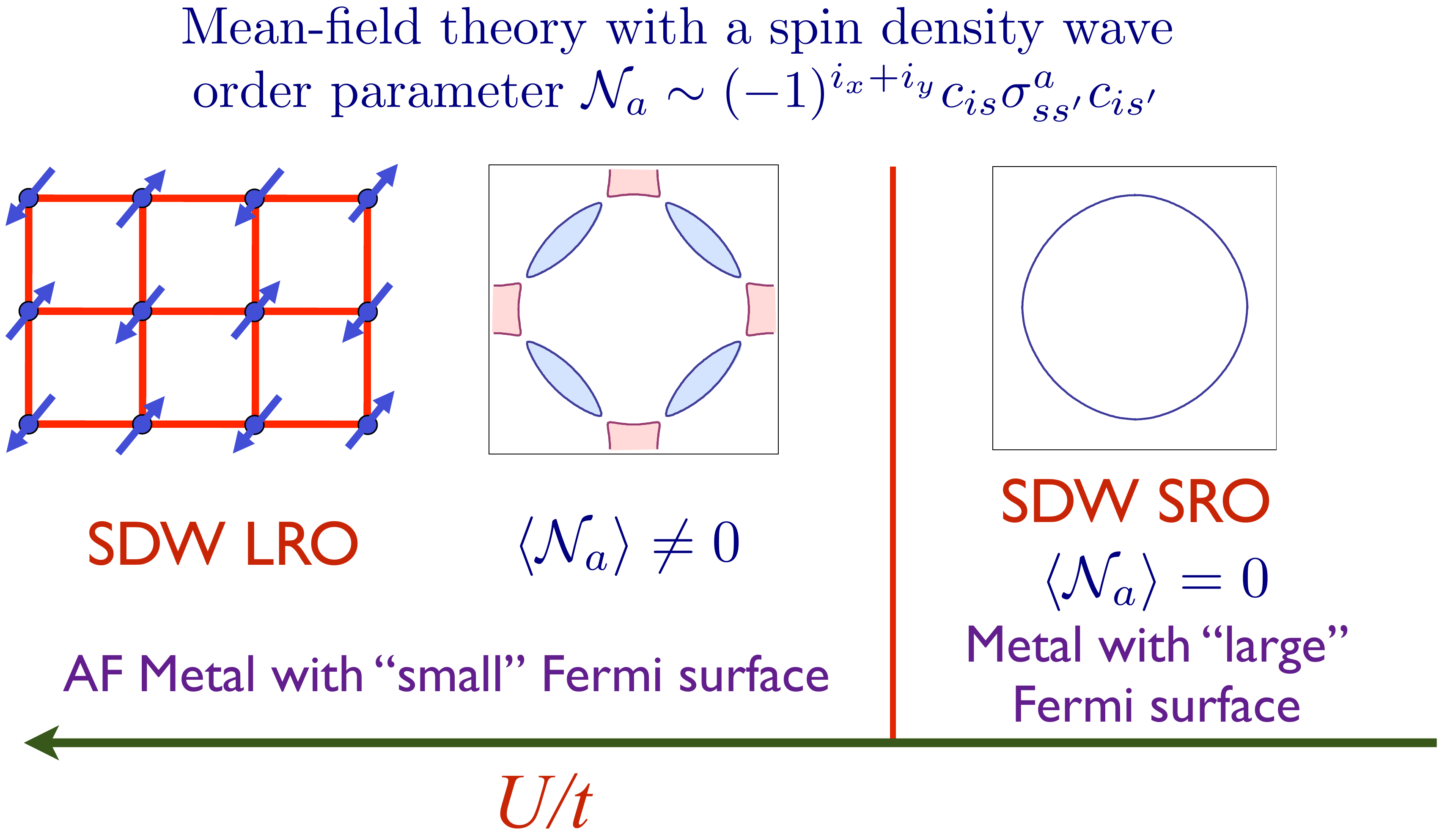}
\end{center}
\caption{Mean field phase diagram of $\mathcal{H}_U$. This is the analog Fig.~\ref{fig:xyd3}, with the XY order parameter, $\Psi$, replaced
by the spin density wave order parameter $S_a$. The new feature is the reconstruction of the large Fermi surface to small
pockets: this reconstruction coincides with the onset of a non-zero $\langle S_a \rangle$, and the associated breaking of translational symmetry.
}
\label{fig:sdw1}
\end{figure}

The traditional mean-field treatment of the Hubbard model proceeds by decoupling the on-site interaction term, $U$,
into fermion bilinears, and optimizing the spin and space dependence of the bilinear condensate. 
For simplicity, we work here with 
a spin density wave (SDW) order parameter; then the effective Hamiltonian for the electrons in the phase with
SDW order has the form
\beq
\mathcal{H}_{\rm sdw} = - \sum_{i< j} t_{ij} c_{i s}^\dagger c_{j s} - \sum_i  \, S_{ia} \, c_{i s}^\dagger \sigma^a_{ss'} c_{i s'}
 - \mu \sum_{i} c_{i s}^\dagger c_{i s} \,, \label{Hsdw}
\eeq 
where $S_{ia}$ is the effective field conjugate to the SDW order on site $i$. One important case is antiferromagnetic SDW order
at wavevector ${\bm K} = (\pi, \pi)$, in which case we write
\beq
S_{ia} = \eta_i \, \mathcal{N}_{a} \label{AFMSDW}
\eeq
where $\mathcal{N}_a$ is the N\'eel order, and $\eta_i$ was defined in Eq.~(\ref{defeta}). Several other spatial configurations of $S_{ia}$ are possible, and have
been discussed elsewhere \cite{CSS17,MSSS18}. Here, we will also consider the case of {\it canted} antiferromagnetism,  with
\beq
S_{ia} = \eta_i \, \mathcal{N}_a + \mathcal{M}_a \quad, \quad \mathcal{N}_a \mathcal{M}_a = 0\,, \label{cantedSDW}
\eeq
where $\mathcal{M}_a$ is a ferromagnetic moment orthogonal to the antiferromagnetic moment.
For our purposes, the key point
to note is that the $\eta_i$ breaks translational symmetry and doubles the unit cell. Consequently $\mathcal{N}_a$ mixes electron states
between momenta ${\bm k}$ and ${\bm k} + {\bm K}$. Diagonalizing the $2 \times 2 $ Hamiltonian at $\mathcal{M}_a=0$, 
we obtain the energy eigenvalues for the antiferromagnetic case
\beq
E_{\bm k} = \frac{\varepsilon_{\bm k} + \varepsilon_{{\bm k} + {\bm K}}}{2} \pm \left[
\left( \frac{\varepsilon_{\bm k} - \varepsilon_{{\bm k} + {\bm K}}}{2} \right)^2 + |\mathcal{N}_a|^2 \right]^{1/2} \,, \label{sdwdisp}
\eeq
where $\varepsilon_{\bm k}$ is the bare electronic dispersion due to the $t_{ij}$. Filling the lowest energy states with such a dispersion,
we obtain the pocket Fermi surfaces shown in Fig.~\ref{fig:sdw1}.
The phase transition in Fig.~\ref{fig:sdw1} involves the breaking of symmetry, described the by the SDW 
order parameter $S_a$, and in this respect it is 
similar XY model transition in $D=3$ shown in Fig.~\ref{fig:xyd3}. However, here the reconstruction of the Fermi surface also accompanies the onset
of SDW order. Luttinger's theorem guarantees that the small Fermi surfaces cannot appear without the breaking of translational symmetry,
and the latter is linked to a non-zero $\langle S_a \rangle$.  

\subsection{Transforming to a rotating reference frame}
\label{sec:su2}

In this section, we wish to move beyond the conventional phases of the Hubbard model 
in Fig.~\ref{fig:sdw1}, and describe also phases with topological order, as was shown already in Fig.~\ref{fig:sdwtopo}. 
This will
be analogous to extending the phase diagram of the $D=3$ XY model from Fig.~\ref{fig:xyd3} to Fig.~\ref{fig:xytopo}.

A diverse set of methods have been employed to describe conducting states on the square lattice with topological order. 
Here we shall follow a strategy similar to that employed in Section~\ref{sec:xytopo}: we will describe a state with fluctuating SDW order,
{\it i.e.\/} a state with SDW SRO, in which certain defects have been suppressed. We will show that the defect suppression
leads to topological order with emergent gauge fields (as was the case with the XY model), 
and also to Fermi surface reconstruction, as is indicated in Fig.~\ref{fig:sdwtopo}.

The approach presented here was proposed in Ref.~\cite{SS09}, and some of the discussion below is adapted from the
review in Ref.~\cite{SSNambu}. The key idea is to transform the electron spin state to a rotating reference
frame. We now show that this leads to a SU$_s$(2) gauge theory, along with a Higgs field, with a structure very similar to that of the 
U(1) gauge theory for the XY model in Section~\ref{sec:xyeven}.
The transformation to a rotating reference frame is defined by
a SU(2) rotation $R_i$ and rotated fermions 
$f_{i,p}$ ($p= \pm$):
\beq
\left( \begin{array}{c} c_{i\uparrow} \\ c_{i\downarrow} \end{array} \right) = R_i \left( \begin{array}{c} f_{i,+} \\ f_{i,-} \end{array} \right),
\label{R}
\eeq
where $R_i^\dagger R_i = R_i R_i^\dagger = 1$. Note that this representation immediately introduces a SU$_s$(2) gauge invariance (distinct from the global SU(2) spin rotation)
\beq
\left( \begin{array}{c} f_{i,+} \\ f_{i,-} \end{array} \right)\rightarrow U_i (\tau) \left( \begin{array}{c} f_{i,+} \\ f_{i,-} \end{array} \right) \quad, \quad R_i \rightarrow R_i U_i^\dagger (\tau), \label{gauge}
\eeq
under which the original electronic operators remain invariant, $c_{is}\rightarrow c_{is}$; here $U_i (\tau) $ is a  SU$_s$(2) gauge-transformation acting on the $p=\pm$ index. 
As noted earlier, we use the subscript $s$ in the gauge theory to distinguish from the global SU(2)
spin rotation symmetry (which has no subscript).
So the $f_p$ fermions are SU$_s$(2) gauge fundamentals, carrying 
the physical electromagnetic global U(1) charge, but not the SU(2) spin of the electron: they are the fermionic  
`chargons' of this theory, and  the density of the $f_p$ is the same as that of the electrons.
The bosonic $R$ fields transform as SU$_s$(2) fundamentals under {\it right\/} multiplication, but they
also carry the global SU(2) spin under {\it left\/} multiplication, and are electrically neutral:
they are bosonic `spinons', and are related, but not identical, to Schwinger bosons \cite{NRSS89,SSNR91,SS09,CSS17,SCWFGS,MSSS18}.
(The Schwinger bosons are canonical bosons, whereas $R$ is initially defined as a SU(2) matrix with no independent dynamics. The Schwinger bosons, and the `rotating reference frame' method used here, ultimately lead to the same results in the undoped antiferromagnet, but the 
latter is far more convenient in the doped antiferromagnet. Also, the latter 
approach is essential for reaching the large Fermi surface Fermi liquid.)

Similarly, we can now transform the SDW order parameter $S_a$ to the rotating reference frame. For reasons which will become
evident, we will call the rotated order parameter a Higgs field, $H_b$. Lifting the spinor rotation, $R$, in Eq.~(\ref{R}) to the adjoint
representation of SU$_s$(2) we obtain the defining relation for $H_b$
\beq
\sigma^a S_a = R \, \sigma^b R^\dagger \, H_b \,, \label{SH}
\eeq
where $\sigma^a$ are the Pauli matrices. From this definition and Eq.~(\ref{gauge}), 
we find that the Higgs field does not carry the global SU(2) spin, but it does transform as an adjoint of the SU$_s$(2) gauge transformations
\beq
\sigma^b H_b \rightarrow U \, \sigma^b H_b \, U^\dagger \,.\label{HU}
\eeq
We now pause to note that the definition in Eq.~(\ref{SH}) is the precise analog of the relation $\Psi = H \psi^2$ in Eq.~(\ref{PsiHpsi})
for the quantum XY model. Indeed Eq.~(\ref{SH}) reduces to Eq.~(\ref{PsiHpsi}) when we limit the SU(2) gauge transformations
to a single U(1) rotation in a plane. The gauge-invariant SDW order parameter $S_a$ is the analog of the gauge invariant XY order
parameter $\Psi$, the Higgs fields $H_b$ and $H$ evidently map to each other, and the spinor $R$ maps to the field $\psi$ (both of which 
carry both gauge and global charges). Specifically, if we choose
\bea
S_a &=& \frac{1}{2}(\Psi + \Psi^\ast, -i(\Psi - \Psi^\ast), 0) \nonumber \\
H_a &=& \frac{1}{2}(H + H^\ast, -i (H - H^\ast), 0) \nonumber \\
R &=& \left( \begin{array}{cc} \psi^\ast & 0 \\ 0 & \psi \end{array} \right) \,,
\eea
then Eq.~(\ref{SH}) reduces to the relation $\Psi = H \psi^2$ in Eq.~(\ref{PsiHpsi}).
Finally, we note that the gauge transformations in Eqs.~(\ref{gauge}) and (\ref{HU}) map to those
in Eqs.~(\ref{u1gaugec}) and (\ref{u1gaugecc}). These mappings are also clear from the correspondences between the condensates
in Fig.~\ref{fig:lxy} and Fig.~\ref{fig:sdwtopo}.

\begin{figure}[htb]
\begin{center}
\includegraphics[height=1.3in]{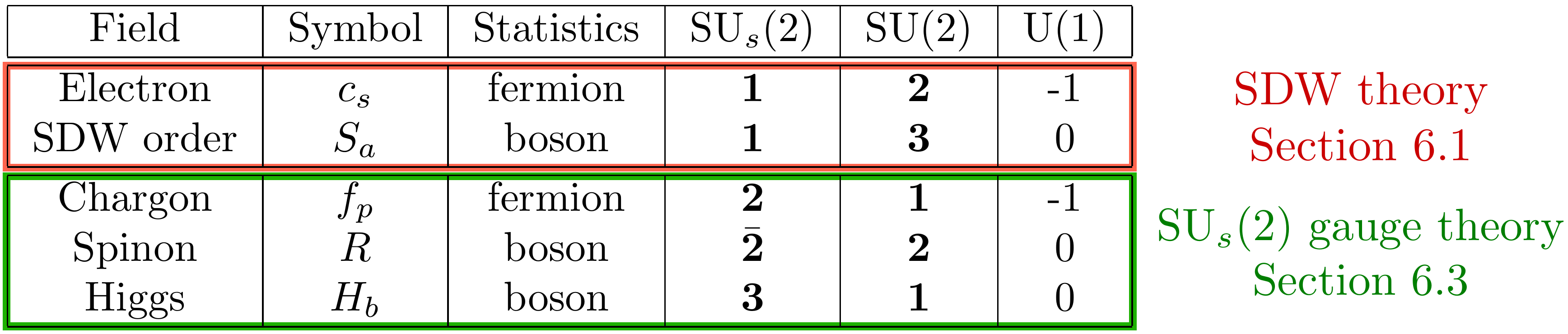}
\end{center}
\caption{Fields and quantum numbers employed in the description of the Hubbard model. 
The transformations under the SU(2)'s are labelled by the dimension
of the SU(2) representation, while those under the electromagnetic U(1) are labeled by the U(1) charge.
The SDW theory can describe only the two conventional phases in Fig.~\ref{fig:sdwtopo}, while the SU$_s$(2) gauge theory
can also describe the third phase with topological order and emergent gauge fields. The two sets of fields are connected via
Eqs.~(\ref{R}) and (\ref{SH}).}
\label{tab:charge}
\end{figure}

A summary of the fields we have introduced so far, and their transformations under the various global symmetries and gauge invariances
are shown in Fig.~\ref{tab:charge}.
These transformations constrain structure of the SU$_s$(2) gauge theory for the $R$, $f$, and $H$ fields, and this theory will be described in Section~\ref{sec:su2theory}.

But for now, we can already present a simple picture of the structure of a possible state with topological order, and how
it allows for small reconstructed Fermi surfaces \cite{SBCS16}. Imagine we are in a state with fluctuating antiferromagnetic SDW order,
where the field $\mathcal{N}_{ia}$ is fluctuating in spacetime (and $\mathcal{M}_a = 0$). 
We want to perform a transformation to a rotating reference
frame in which the corresponding Higgs field has a uniformly staggered spatial arrangement, and is independent of time:
\beq
H_{ib} = \eta_i H_0 e_b \label{HH0}
\eeq
where $e_b$ is a fixed 3-component unit vector. The idea is that the rotated fermions, $f$, will then see a uniform background
antiferromagnetic SDW order. More completely, we can postulate an effective Hamiltonian for the $f$ fermions, which is just
the rotated version of  $\mathcal{H}_{\rm sdw}$ in Eq.~(\ref{Hsdw}):
\beq
\mathcal{H}_{f,{\rm sdw}} = - \sum_{i< j} t_{ij} f_{i p}^\dagger f_{j p} - \sum_i  \, H_{ib} \, f_{i p}^\dagger \sigma^b_{pp'} f_{i p'}
 - \mu \sum_{i} f_{i p}^\dagger f_{i p} \,. \label{Hsdwp}
\eeq 
From the Eqs.~(\ref{HH0}) and (\ref{Hsdwp}) we find that the dispersion of the $f$ fermions is given by Eq.~(\ref{sdwdisp})
with $|\mathcal{N}_a| \rightarrow H_0$. Consequently, it appears that 
the $f$ Fermi surfaces have been reconstructed to small Fermi surfaces.

\begin{figure}[htb]
\begin{center}
\includegraphics[height=2.3in]{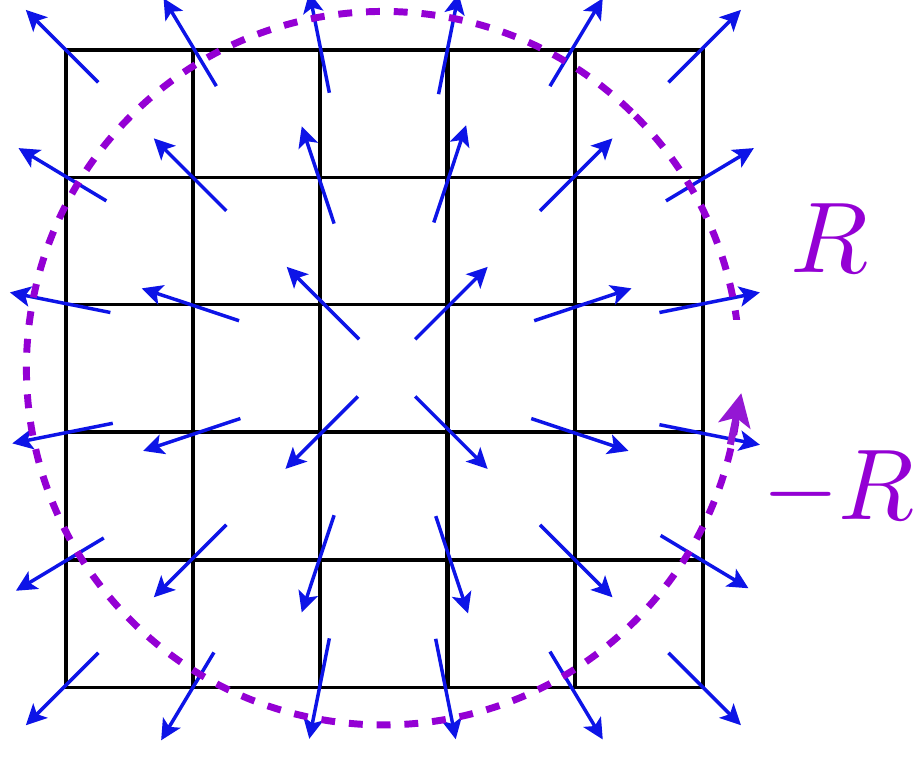}
\end{center}
\caption{A vortex defect in the antiferromagnetic SDW order, $\mathcal{N}_a$. The staggering of the underlying spins, 
associated with the $\eta_i$, is not shown. Upon parallel transport around such a vortex, the frame of reference is rotated
by $2\pi$, and correspondingly the spinor field $R$ changes sign. Thus it is not possible to consistently define
the fermion, $f_p$, in the rotated reference frame around such a vortex via Eq.~(\ref{R}).
}
\label{fig:obstacle}
\end{figure}

The argument just presented is clearly too facile: if correct, it would imply that we can always transform to a rotating reference
frame in a state with fluctuating SDW order, and find rotated fermions with reconstructed Fermi surfaces. There must be an additional
obstacle to be overcome before a consistent transformation to a rotating reference frame is possible. Indeed there is such
an obstacle, and it is illustrated in Fig.~\ref{fig:obstacle}. 
For simplicity, consider the case where the antiferromagnetic SDW order is restricted to 
lie in a single plane (we will consider the general cases in Section~\ref{sec:su2theory}). The obstacle arises when we consider
a vortex defect in the fluctuating SDW order, and attempt to find a space-dependent rotation $R$ which maps it into a uniformly
staggered Higgs field, as in Eq.~(\ref{HH0}). As is well-known, a $2 \pi$ rotation in the adjoint representation of SU(2), maps to a double-valued
spinor representation: the rotation $R$ does not remain single-valued as we transport it around the vortex, as shown in Fig.~\ref{fig:obstacle}.
Consequently, if there are any vortices in the SDW order present, we cannot find a single-valued transformation $R$ to consistently define
the $f_p$ fermions via Eq.~(\ref{R}), and an effective Hamiltonian of the form in Eq.~(\ref{Hsdwp}) is not meaningful.

So we reach some of the key conclusions of this review. In a state with fluctuating SDW order, we can consistently transform the fermions
into a rotating reference frame with uniform SDW order only if $\pm 2\pi$ vortices in the SDW order are 
expelled (for the easy-plane SDW case) \cite{SBCS16}.
In other words, using our extensive discussion of defect suppression so far, we conclude that the fluctuating SDW state needs to have topological order
with an emergent $\mathbb{Z}_2$ gauge field in this case.
And in such a fluctuating SDW state with topological order, the Fermi surface can consistently reconstruct to small pocket Fermi surfaces,
as indicated in Fig.~\ref{fig:sdwtopo}. The $\pm 2 \pi$ vortices in the SDW order become stable, gapped, vison excitations in such a phase.

It is useful to mention here the analogy to the Glashow-Weinberg-Salam SU(2)$\times$U(1) gauge theory of nuclear weak interactions. 
In that theory, the Higgs field is the origin of the masses of the fermions. In our case, the Higgs field renders the fermions at the `hot spots' gapful
via Eqs.~(\ref{HH0}) and (\ref{Hsdwp}),
and this leads to the reconstruction of the Fermi surface. The weak interaction Higgs field transforms as a SU(2) fundamental, and hence the SU(2) gauge group is fully Higgsed; in our case, the Higgs field transforms as a SU$_s$(2) adjoint, and so there is at least an unbroken $\mathbb{Z}_2$ gauge group.

\subsection{SU$_s$(2) gauge theory}
\label{sec:su2theory}

We now specify the complete SU$_s$(2) gauge theory which describes all the phases in Fig.~\ref{fig:sdwtopo}.
 
The structure of the theory of the chargons $f$, the Higgs field $H_b$, and the spinons $R$, follows
from the gauge transformations in Eqs.~(\ref{gauge}) and (\ref{HU}), and the imposition of square lattice and spin rotation
symmetries. We write the Lagrangian as
\beq
\mathcal{L}_{SU_s(2)} = \mathcal{L}_f + \mathcal{L}_Y + \mathcal{L}_R + \mathcal{L}_H \,.
\label{LQCP}
\eeq
The first term for the $f$ fermions descends the fermion hopping terms in $\mathcal{H}_{f,{\rm sdw}}$ in Eq.~(\ref{Hsdwp})
\bea
\mathcal{L}_f &=& \sum_i f_{i,p}^\dagger \left[\left(\frac{\partial}{\partial \tau}-\mu \right)\delta_{pp'} + iA_\tau^b\sigma^b_{pp'} \right]f_{i,p'} \nonumber \\
&~&~~~~~~~~~~~
+  \sum_{i,j} \tilde{t}_{ij}f^\dagger_{i,p}\bigg[e^{i\sigma^b {\bm A}^b\cdot({\bm r}_i-{\bm r}_j)}\bigg]_{pp'}f_{j,p'} \,.
\eea
We have renormalized the hopping term to $\tilde{t}_{ij}$ to account for corrections from the transformation to the
rotating reference frame \cite{SCWFGS}. But more importantly, 
we have introduced a SU$_s$(2) gauge field $A_{\mu}^b \equiv (A_\tau^b, {\bm A}^b)$
to allow for properly gauge-invariant hopping between sites.
The Yukawa coupling between the fermions and the Higgs field
\beq
\mathcal{L}_Y = - \sum_i  \, H_{ib} \, f_{i p}^\dagger \sigma^b_{pp'} f_{i p'} \label{yukawa}
\eeq
was also contained in Eq.~(\ref{Hsdwp}), and is already gauge invariant.

We will not spell out the full explicit form of the Lagrangian for $R$, $\mathcal{L}_R$.
We note that it descends mainly from the $t_{ij}$
hopping in $\mathcal{H}_U$, after transforming to the rotating reference frame, and performing a mean field factorization
on the resulting terms; see Refs.~\cite{CSS17,SCWFGS,MSSS18} for details. All the resulting terms have to be invariant
under gauge transformations (by Eq.~(\ref{gauge}), these act by right multiplication of a spacetime-dependent SU$_s$(2) matrix on $R$) 
and global spin rotations (which act by left multiplication of a spacetime-independent SU(2) matrix). We can write the SU(2) matrix $R$ in the form 
\beq
R = \left( \begin{array}{cc} z_\uparrow & -z_\downarrow^\ast \\
z_\downarrow & z_\uparrow^\ast \end{array} \right) \quad, \quad |z_\uparrow|^2 + |z_\downarrow|^2 = 1\,,
\eeq
and then the effective action for $R$ takes a form closely related to that of the $\mathbb{CP}^1$ model obtained
in the Schwinger boson approach \cite{NRSS89,SSNR91,SS09,SCWFGS,MSSS18}.

Finally, the Lagrangian for the Higgs field, $\mathcal{L}_H$, has a similar structure to those in Eqs.~(\ref{LU1}), (\ref{LXY}), (\ref{LU1o}), 
and (\ref{SXYo}), after generalizing for a SU$_s$(2) gauge invariance: the field $H_b$ transforms as an adjoint under
spacetime-dependent SU$_s$(2) gauge transformations. We have to allow the Higgs condensate to have an arbtirary spatial dependence \cite{CSS17,MSSS18},
and so cannot yet take the continuum limit here. We spell out a few terms in $\mathcal{L}_H$ on the lattice:
\beq
\mathcal{L}_H = g \sum_i H_{ib}^2 + \sum_{i<j} J_{ij} H_{ia} D_{ij,ab} H_{jb} + \ldots \label{LHsu2}
\eeq
where $D_{ij,ab}$ is the Wigner $D$-matrix of the SO(3) rotation associated with the SU$_s$(2) rotation generated by the gauge field
${\bm A}^b\cdot({\bm r}_i-{\bm r}_j)$. For our purposes here, the important information we need from
Eq.~(\ref{LHsu2}) is the spatial structure of the Higgs condensate, as this controls the nature of the topological order
in Fig.~\ref{fig:sdwtopo}: this spatial
structure is controlled by the couplings $J_{ij}$ (and higher order terms) which are ultimately connected to the exchange
interactions between the underlying electrons.
 
\begin{figure}
\begin{center}
\includegraphics[height=6cm]{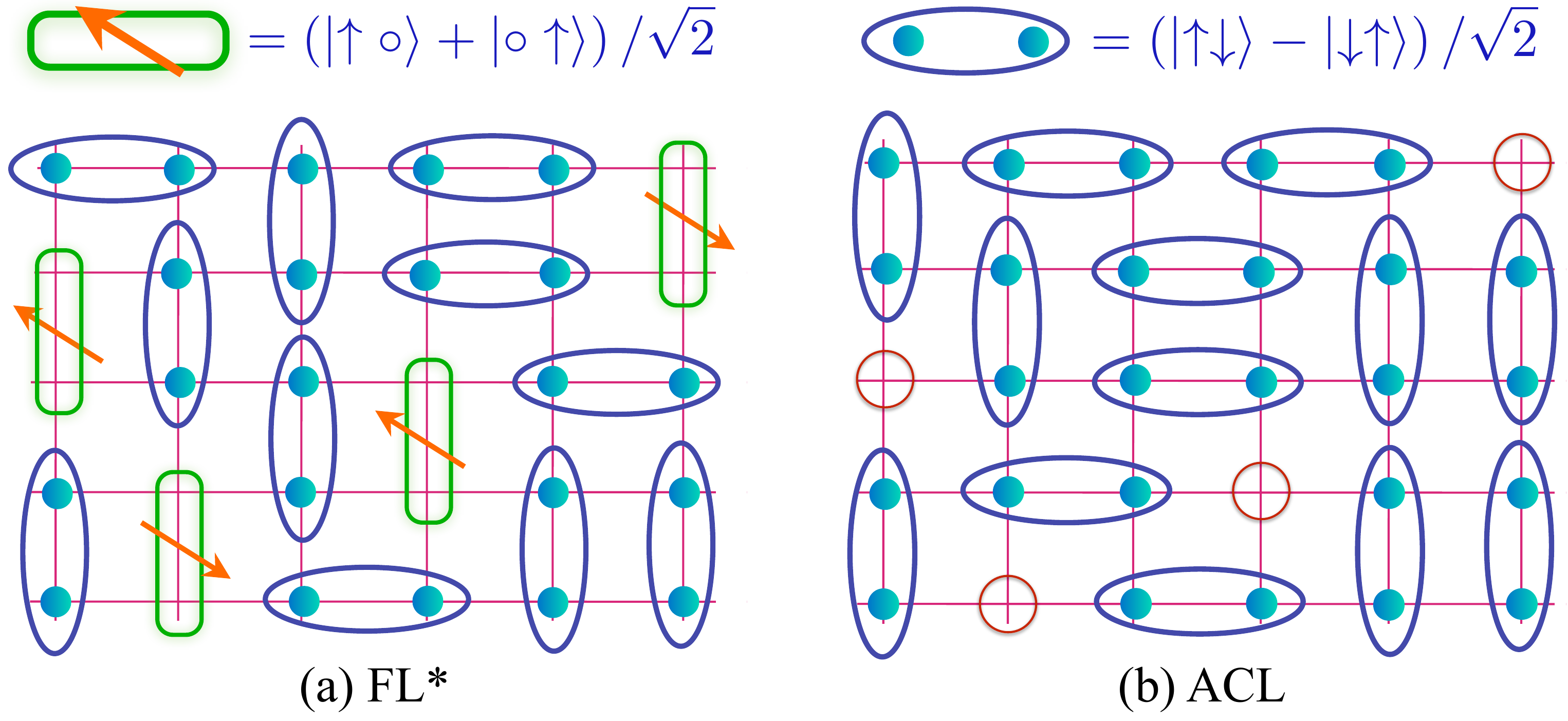}
\end{center}
\caption{(a) A component of a resonating bond wavefunction for FL* in a single-band model on the square lattice \cite{Punk15}. 
The density of the green
bonds is $p$, and these are fermions which form `reconstructed' Fermi surface of volume $p$ with electron-like quasiparticles.
(b) A component of a wavefunction for an ACL. The vacancies are the `holons', or more generally, the `chargons'; they are 
assumed to be fermions which form a Fermi surface of 
spinless quasiparticles of charge $e$.}
\label{fig:nambu}
\end{figure}

The remaining discussion here will be limited to the possible Higgs phases with topological order, realizing the state at the top
of Fig.~\ref{fig:sdwtopo}. With a Higgs condensate in $\mathcal{L}_Y$, we compute the fermion dispersion 
from $\mathcal{L}_f + \mathcal{L}_Y$, and find (as in Section~\ref{sec:sdw}) that the Fermi surface has been reconstructed, but
now has chargon quasiparticles $f_p$.
Refs.~\cite{SCSS17,CSS17,MSSS18} 
explored the distinct physical properties of a variety of Higgs condensates; some of the Higgs condensates also
break square lattice and/or time-reversal symmetries. Here we will restrict ourselves to two of the simplest condensates
which do not break {\it any\/} global symmetries: they break the gauge invariance down to U(1) and $\mathbb{Z}_2$,
and are described in the following subsections. The topological order in these states is associated with the expulsion
of distinct defects in the SDW order, and the consequent appearance of emergent deconfined 
U(1) and $\mathbb{Z}_2$ gauge fields. Both states have fractionalized gapped bosonic spinon excitations, $R$, and gapless 
fermionic chargon excitations,
$f_p$, around reconstructed Fermi surfaces, so are `algebraic charge liquids' (ACL) in the notation of Refs.~\cite{RKK07,RKK08}. 
It is very likely, given the attraction induced by the hopping term in the Hubbard model \cite{SBCS16}, that the bosonic spinons and fermionic chargons form a fermionic bound state \cite{XGWPAL96}---such a bound state 
has the same quantum
numbers as an electron. If the binding is strong enough, then the quasiparticles on the
reconstructed Fermi surfaces have electron-like quasiparticles: such a state is a fractionalized Fermi liquid or FL* \cite{TSSSMV03,TSMVSS04,QS10,Mei11,MPSS12,Punk15,SBCS16,Punk17a,Punk17b}. Intermediate states with both chargon and electron Fermi surfaces are also possible \cite{RKK08}, and there is a Luttinger-like sum rule only on the combined Fermi surfaces \cite{PSB05,CPR05,HS11}. 
In the FL* state, the chargon excitations are gapped, and so there is no chargon Fermi surface. A schematic picture (adapted from 
Ref.~\cite{SSNambu}) of the differences between the ACL and FL* states is shown in Fig.~\ref{fig:nambu}.

\subsubsection{U(1) topological order:}
\label{sec:u1higgs}
This is a state with fluctuating antiferromagnetic SDW SRO and reconstructed Fermi surfaces, 
and is obtained with a Higgs condensate which
is similar to the order parameter in Eqs.~(\ref{AFMSDW}) and (\ref{HH0})
\beq
\langle H_{ib} \rangle = \eta_i H_0 e_b \,, \label{HH0a}
\eeq
with $e_b$ a unit vector, and the strength of the condensate measured by $H_0$. 
Such a Higgs condensate leaves a U(1) subgroup of SU$_s$(2) unbroken,
corresponding to rotations in SU$_s$(2) about the $e_b$ axis. Correspondingly, a perturbative treatment of gauge fluctuations
will yield an emergent gapless U(1) photon excitation. However, non-perturbative topological effects can disrupt these 
gapless photon excitations.
In the SU$_s$(2) gauge theory, the stable defects are `instantons' (corresponding to tunneling events in the quantum system)
which are 'tHooft-Polyakov monopoles \cite{tH74,Polyakov74}. Indeed, the present Higgs field and the SU$_s$(2) gauge field have the
same structure as the Georgi-Glashow model \cite{Dunne01,ATSS18} used in these monopole computations. Equivalently, we can work in the
reduced U(1) gauge theory, and then the monopoles are Dirac monopoles in the compact U(1) gauge theory. 
And in terms of the original SDW order, the monopoles are `hedgehogs' in the 3-component vector order parameter \cite{NRSS90}.

In the half-filled Hubbard
model, the present state can be an insulator in which all $f$ excitations are gapped: in this case the monopole instantons are relevant,
and so we expect to obtain a confining 
state without topological order. It was argued that the monopole instantons acquire Berry phases, and these
Berry phases lead to VBS order in the confining phase \cite{NRSS89,NRSS90}. In the original computation, these Berry phase were obtained
from the time evolution of canonical Schwinger boson wavefunction. In the present SU$_s$(2) gauge theory, the $R$
bosons are not canonical, but the same Berry phases are obtained from the filled band of gapped $f$ chargons reacting to the monopole-induced time evolution \cite{SS09,ATSS18}. The mechanism of the 
Berry phase-induced VBS order is similar to that discussed in Sections~\ref{sec:z2odd} and~\ref{sec:xyodd}.

However, U(1) topological order, and the gapless emergent photon, 
can survive if the monopoles are expelled \cite{OMAV04}: metallic states with $f_p$ Fermi surfaces
have been shown to suppress monopoles \cite{Hermele04}. In general, monopole defects are suppressed, and U(1) topological
order is stable, as long as there are Fermi surfaces of quasiparticles carrying U(1) electric charges.

The above discussion ignores the possibility of a pairing instability to superconductivity, which is invariably present at low temperatures.
In the present case, the attractive force can be provided by the U(1) gauge field coupling to $f_p$ with opposite gauge charges \cite{Moon09}.

\subsubsection{$\mathbb{Z}_2$ topological order:}
\label{sec:z2higgs}
This is a state with fluctuating canted-antiferromagnetic SDW SRO and reconstructed Fermi surfaces, 
and is obtained with a Higgs condensate which corresponds to the canted-antiferromagnet order parameter \cite{CSS17,YangWang16}
in Eq.~(\ref{cantedSDW}):
\beq
\langle H_{ib} \rangle = \eta_i H_0 e_{1b} + H_1 e_{2b} \quad, \quad e_{1b} e_{2b} = 0\,, \label{CantedH}
\eeq
with $e_{1b}$ and $e_{2b}$ two orthonormal vectors, and the strength of the Higgs condensate measured by $H_0$ and $H_1$.
With two orthonormal vectors ($e_{1b}$ and $e_{2b}$) determining the spatial dependence of the Higgs condensate,
there is no rotation axis about which the Higgs condensate is invariant. Indeed, only a $\mathbb{Z}_2$ gauge invariance remains,
because $H_b$ transforms under the adjoint representation of SU$_s$(2), and only the $\pm$(unit matrix) gauge transformations leave it invariant.
So all gauge excitations are gapped, and there is stable $\mathbb{Z}_2$ topological order. Indeed this topological order is stable
 in both insulators \cite{NRSS91,Wen91} and metals. The unbroken gauge group implies the presence of a deconfined, but gapped,
emergent $\mathbb{Z}_2$ gauge field. 

Emergent $\mathbb{Z}_2$ gauge fields also appear for more complex Higgs field condensates than that in Eq.~(\ref{CantedH}).
These can break time-reversal, mirror plane, or other point-group symmetries of the Hamiltonian, and are discussed elsewhere \cite{SCSS17,CSS17,MSSS18}.

\begin{figure}[htb]
\begin{center}
\includegraphics[height=2.3in]{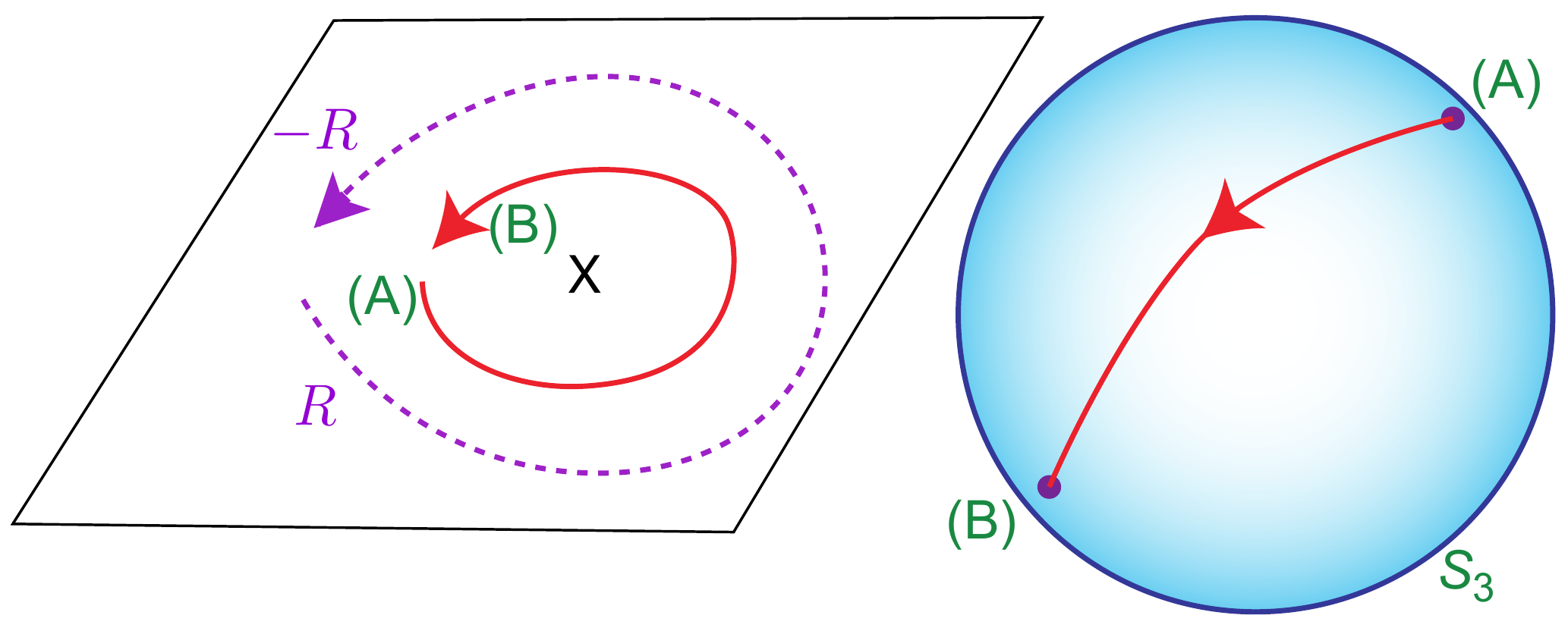}
\caption{A vison defect with a Higgs field in $S_3/\mathbb{Z}_2$.
The full line shows the
trajectory of $w_\alpha$ (defined in Eq.~(\ref{s3})) on $S_3$ around a vison defect centered at X.
All such anti-podal configurations of $w_\alpha$ are averaged over. The dashed line shows parallel transport
of a spinon, $R$, around the vison. Compare to Fig~\ref{fig:abrikosov} and Fig.~\ref{fig:obstacle}.}
\label{fig:visonsu2}
\end{center}
\end{figure}

As in previous cases, the topological order of the condensate in Eq.~(\ref{CantedH}) 
is characterized by expulsion of the $\mathbb{Z}_2$ vortex defects from the ground state,
and these defects become gapped `vison' excitations. 
The visons carry flux of the $\mathbb{Z}_2$ gauge field, and they have a 
statistical interactions with the fermionic chargons $f_p$ and bosonic spinons $R$, both of which carry unit $\mathbb{Z}_2$ electric charges.
In the context of the SU$_s$(2) gauge theory, the vison is a finite energy 
vortex solution of the SU$_s$(2) gauge theory which is analogous to the 
Abrikosov vortex solutions discussed in Section~\ref{sec:z2even}; this is shown in Fig.~\ref{fig:visonsu2}. 
To obtain a simple description of the vison vortex solution \cite{Dombre,CSS93,CSS94}, 
let us write $e_{1b}$ and $e_{2b}$ in terms of the pair of complex
numbers $w_{1,2}$ via
\begin{equation}
e_{1 b} + i e_{2b} = \epsilon_{\alpha\gamma} w_\gamma \sigma^b_{\alpha\beta} w_\beta \,.
\label{s3}
\end{equation}
Then with $|w_1|^2 + |w_2|^2 = 1$, it can be verified that the orthonormality constraints 
$e_{1b}$ and $e_{2b}$ are 
are automatically satisfied. Note that $w_\alpha$ and $-w_\alpha$ both map to the same
values of $e_{1a}$ and $e_{2b}$. So the mapping in Eq.~(\ref{s3}) is 2-to-1: the complex
number $w_\alpha$ defines the surface of a unit sphere in 4 dimensions, $S_3$, and Eq.~(\ref{s3}) establishes that
the Higgs condensate in Eq.~(\ref{CantedH}) is an element of $S_3 /\mathbb{Z}_2$. The vison defect is associated with the homotopy group
$\pi_1 (S_3/\mathbb{Z}_2) = \mathbb{Z}_2$, and is easy to identify in the $w_\alpha$ parameterization: as one encircles the defect, $w_\alpha$
moves to its anti-podal point; see Fig.~\ref{fig:visonsu2}. The core of the vison will have SU$_s$(2) gauge flux (as in the Abrikosov
vortex in Fig.~\ref{fig:abrikosov}), and this has two important consequences: ({\it i\/}) the SU$_s$(2) gauge field screens the precession
of the Higgs field far from the core of the vison, leading to a finite energy vison solution; ({\it ii\/}) the chargon and spinon pick
up a Berry phase of $\pi$ around the vison, and so become mutual semions with the vison.

\subsection{Quantum criticality without symmetry breaking}

The most interesting phase transition in Fig.~\ref{fig:sdwtopo} is the topological transition between the two SDW SRO phases of the Hubbard model. This transition does not involve any symmetry breaking order parameter, and is associated
with the onset of topological order and Fermi surface reconstruction; so it is possibly linked to the observations in 
Ref.~\cite{LTCP15}. It is useful to reason by analogy to the phenomena in Sections~\ref{sec:z2odd}
and \ref{sec:xyodd}, where static background matter induced deconfined criticality with a U(1) gauge field at the corresponding transitions in
Fig.~\ref{fig:u1a} and Fig.~\ref{fig:lxyo}. In the Hubbard model, we have dynamical fermionic matter described by $f_p$, and at half-filling
in a large gap insulator, this matter contributes the same background Berry phase as that by $\mathcal{S}_B$ in Eq.~(\ref{SXYo}).
So it is a reasonable conjecture that deconfined criticality also applies in the metallic case with dynamic, gapless, fermionic matter.
The needed theory descends directly from Eq.~(\ref{LQCP}): it has deconfined SU$_s$(2) gauge fields, a Fermi surface of $f_p$ chargons, and the coupling $g$
in Eq.~(\ref{LHsu2}) is tuned to make the Higgs field $H_b$ critical. The physical properties of such a theory were examined
in Refs.~\cite{SS09,DCSS15b,SSNambu}. In general, we need a large enough gauge group in the deconfined critical theory to accomodate
both the adjacent topological order(s) and pattern(s) of confinement. Obtaining a large Fermi surface Fermi liquid as the confining state
on one side of the 
transition appears to require a gauge group at least as large as SU$_s$(2) \cite{SS09}.

Evidence for a deconfined SU$_s$(2) gauge field at the onset of confinement has emerged in recent quantum Monte Carlo 
simulations \cite{Snir18}. This study examined a pseudogap phase at half-filling with fractionalized Dirac fermion excitations, and will
be discussed further in Section~\ref{sec:om}.

An alternate view of the transition between the two SDW SRO phases arises from the approach in Ref.~\cite{SBCS16}. In this approach, we view the 
transition from the perspective of the conventional Fermi liquid state, and only include gauge-neutral fermions, $c_\alpha$, in the low energy critical theory. We couple the the physical electrons $c_\alpha$ to the SDW order parameter,
then write the theory of the SDW order parameter 
fluctuations as a SU$_s$(2) gauge theory via the decomposition in Eq.~(\ref{SH}). On its own, such a SU$_s$(2) gauge theory,
with a Higgs field as in Eq.~(\ref{CantedH}), is similar to the Georgi-Glashow model of particle physics \cite{Kibble02}, 
and it has a transition from a confined phase to a deconfined phase with $\mathbb{Z}_2$ topological order.
This transition can be in the same universality class as the even $\mathbb{Z}_2$ gauge theory of Section~\ref{sec:ising}, 
but more subtle deconfined critical points are also possible. The coupling to the large Fermi surface of electrons $c_\alpha$ leads to marginally relevant corrections which were studied in Ref.~\cite{SSTM02,TGTS09}, but they don't alter the basic picture of a transition driven by an even $\mathbb{Z}_2$ gauge theory with no gauge-charged matter. Once we are in the deconfined ({\it i.e.\/} Higgs) phase, the fermionic excitations
can fractionalize via the converse of Eq.~(\ref{R}): the electrons $c_\alpha$ bind with the deconfined spinons to yield reconstructed Fermi surfaces of fermionic
chargons $f_p$, as was described in some detail in Ref.~\cite{SBCS16}. We can also envisage a situation in which the fractionalization of the low energy excitations in the deconfined phase occurs {\it only\/} in the bosonic sector {\it i.e.\/} $S^a$ fractionalizes into $R$ and $H_b$ as in Eq.~(\ref{SH}), while the low energy excitations near the reconstructed Fermi surfaces remain charge $e$, spin-$1/2$ $c_\alpha$ electrons, as in a FL* state. As noted above, the quantum criticality is described by the original electronic theory in Eq.~(\ref{Hsdw}) after substituting in $R$ and $H$ via Eq.~(\ref{SH}), along with $\mathcal{L}_R + \mathcal{L}_H$. Reconstructed Fermi surfaces of the $c_\alpha$ can arise in the Higgs phase of such a theory
along the lines of the computation in Ref.~\cite{MPSS12}, and this is an interesting avenue for further research.

We close this subsection by also mentioning the universality classes of the other two phase transitions in Fig.~\ref{fig:sdwtopo}.
The symmetry breaking 
transition between SDW LRO and the large Fermi surface metal is just that described in Section~\ref{sec:sdw}: this is described
by an order parameter theory of SDW fluctuations, which are damped by the Fermi surface \cite{hertz}. The symmetry breaking and topological transition between SDW LRO
and the metal with topological order reduces to a theory of the $R$ spinons alone in the O(4)$^\ast$ 
universality class \cite{CSS93,RKK08b,GS10}.

\section{Conclusions and extensions}
\label{sec:conc}

This review began, in Section~\ref{sec:ising}, with a detailed discussion of the topological order in Wegner's quantum 
$\mathbb{Z}_2$ gauge theory on a 
square lattice. Section~\ref{sec:z2even} showed that the topological order of this theory, and the phase transition to the `trivial' confining state
are conveniently described by a U(1) gauge theory with charge 2 Higgs field; the topological phase acquired a deconfined 
$\mathbb{Z}_2$ gauge field. The same topological phase with an emergent $\mathbb{Z}_2$ gauge field, along with the 
confinement transition, also appears
in a classical XY model in $D=3$ at non-zero temperature, 
or in quantum models of bosons with short-range interactions on the square lattice at integer filling at zero temperature: 
this is summarized in Fig.~\ref{fig:xytopo}, and was described in Sections~\ref{sec:xytopo} and \ref{sec:xyeven}. 
Finally, we showed that a phase diagram very similar to Fig.~\ref{fig:xytopo}, appearing in Fig.~\ref{fig:sdwtopo},
applied to the electron Hubbard model on the square lattice.

The most interesting feature of Fig.~\ref{fig:sdwtopo} is the presence of a metallic state with the topological
order of emergent gauge fields, reconstructed Fermi surfaces (with chargon ($f_p$) or electron-like quasiparticles), and no broken symmetry. We presented a simple physical argument in Section~\ref{sec:su2}
showing how such topological order can reconstruct the Fermi surface even in the absence of translational
symmetry breaking. Such a metallic state, with fluctuating SDW order leading to emergent gauge fields, is an attractive
candidate for a theory of the pseudogap state of the cuprate superconductors \cite{SS09}, and we noted its connection to a variety
of experiments \cite{ZXtopo,LTCP15,SP15,Greven16,Boebinger11} in Section~\ref{sec:intro}.
Recent theoretical work \cite{WSCSGF,SCWFGS} has compared the metallic state with an emergent gapless U(1) photon, 
described in Section~\ref{sec:u1higgs},
to cluster dynamical mean field theory (DMFT) and quantum Monte Carlo studies of the lightly hole-doped
Hubbard model appropriate for the cuprates. Good agreement was found in both the real and imaginary
parts of the electron Green's function computed from the theory in Section~\ref{sec:u1higgs}. In particular, the Higgs condensate
in Eq.~(\ref{HH0a}) was responsible for inducing a gap in the anti-nodal region of the Brillouin zone, and led to lines of approximate
zeros of the electron Green's function. 
The electron spectral function of this metallic state with emergent gauge fields can also help understand
recent photoemission observations \cite{ZXtopo} 
in the electron-doped cuprate Nd$_{2-x}$Ce$_x$CuO$_4$, which detected a reconstruction gap in the electronic
dispersion at a doping $x$ where there is no antiferromagnetic order.

\subsection{Pairing fluctuations in the pseudogap}
\label{sec:pairing}

An important question for further studies of the cuprate pseudogap phase is the role of electron pairing fluctuations. 
These have been addressed by a formally distinct SU(2) gauge theory of the pseudogap described by Lee {\em et al.\/} in Ref.~\cite{LeeWenRMP}: here we will refer to this as the SU$_c$(2) gauge theory.
However, there is a close relationship between their SU$_c$(2) gauge theory and our SU$_s$(2) gauge theory of SDW fluctuations. This relationship can be described in a unified formalism that includes both SDW and pairing
fluctuations \cite{XS10}.

To see the connection between the two approaches, it is useful to introduce a $2 \times 2$ matrix electron operator
\beq
C_i = \left(
\begin{array}{cc}
c_{i \uparrow} & - c_{i \downarrow}^\dagger \\
c_{i \downarrow} & c_{i \uparrow}^\dagger
\end{array}
\right) \,. \label{Xc}
\eeq
This matrix obeys the relation
\beq
C_i^\dagger = \sigma^y C_i^T \sigma^y \,.
\eeq
Global SU(2) spin rotations act on $C$ by left multiplication,
while right multiplication corresponds to global SU(2) Nambu pseudospin rotations. We also introduce the corresponding matrix form of the
fermionic chargons
\beq
F_i = \left(
\begin{array}{cc}
f_{i +} & - f_{i -}^\dagger \\
f_{i -} & f_{i +}^\dagger
\end{array}
\right) \,.\label{Ff}
\eeq
Then it is easy to check that the transformation to a rotating reference frame in spin space in Eq.~(\ref{R}) can be written simply as
\beq
C_i = R_i \, F_i \,.
\label{RF}
\eeq
The SU$_c$(2) gauge theory of Lee {\em et al} \cite{LeeWenRMP,ATSS18} corresponds to a transformation to a rotating reference frame in pseudospin space, and is obtained instead
by the decomposition
\beq
C_i =  F_i \, \widetilde{R}_i \,;
\label{FR}
\eeq
now the $F_i$ are interpreted as fermionic spinons, while $\widetilde{R}_i$ is a SU(2) matrix representing the bosonic chargons. 
Because the electromagnetic charge is now carried by the bosons, the approach in Eq.~(\ref{FR}) 
does not yield a zero temperature metallic state
with topological order at non-zero doping, because their non-zero density causes the $\widetilde{R}$ bosons to condense. Metallic states with topological order were obtained from the SU$_s$(2) gauge theory associated with 
Eqs.~(\ref{RF}) and (\ref{R}). 
However, it remains possible that Eq.~(\ref{FR}) could be relevant for non-zero temperature
where the $\widetilde{R}$ bosons are thermally fluctuating. 
It can be verified that the operations in Eqs.~(\ref{RF}) and (\ref{FR}) commute with each other, and the most general approach combines them in a SO(4) $\sim$ SU$_c$(2)$\times$SU$_s$(2) gauge theory of fluctuating SDW and pairing orders with \cite{XS10}
\beq
C_i = R_i \, F_i \, \widetilde{R}_i \,.
\label{RFR}
\eeq
A wide variety of Higgs fields appear
possible in such a SO(4) gauge theory, yielding interesting refinements of the fluctuating SDW theory of the pseudogap state
{\it e.g.\/} a spatially varying Higgs field in the SU$_c$(2) sector can account for pair density wave fluctuations \cite{leeprx}.

\begin{figure}[htb]
\begin{center}
\includegraphics[height=3in]{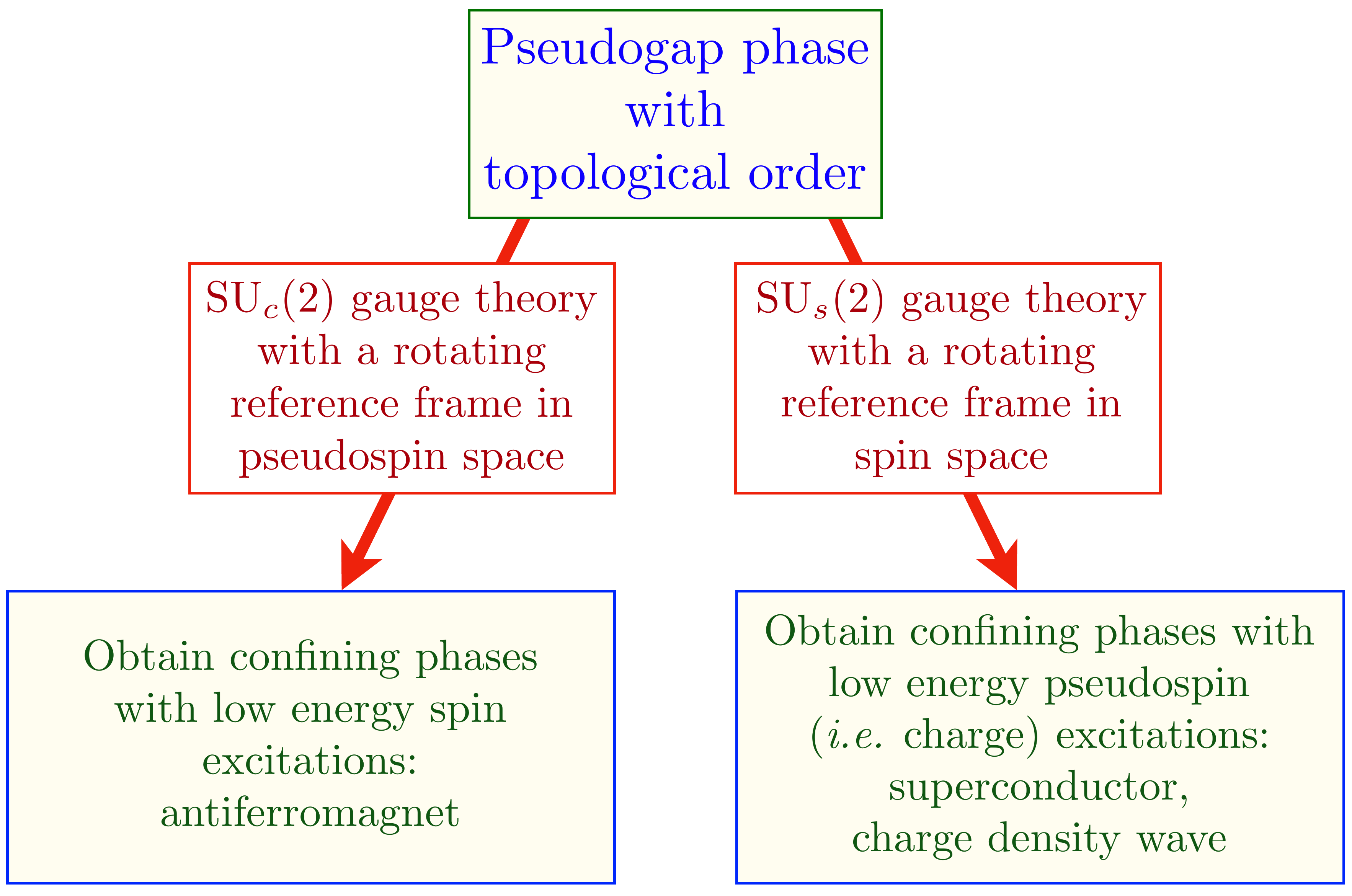}
\end{center}
\caption{Adapted from Ref.~\cite{MSSS18}. Schematic representation of routes to confinement out of the pseudogap phase. Specific realizations of quantum critical points described by SU$_s$(2)
and SU$_c$(2) gauge theories appear in the model of Gazit {\em et al.}~\cite{Snir16,Snir18}, and was studied by sign-problem-free quantum Monte Carlo simulations. The horizontal axis of the figure is proposed to be similar to increasing doping in the high temperature superconductors. 
}
\label{fig:om}
\end{figure}
Fig.~\ref{fig:om} (adapted from Ref.~\cite{MSSS18}) presents a perspective on the roles of the SU$_s$(2) and SU$_c$(2) gauge theories, in which we treat the pseudogap phase as the parent
of other phases in the cuprate phase diagram. This perspective emerged 
from studies of metallic states with topological order amenable to sign-problem-free Monte Carlo across confinement transitions \cite{Snir16,Snir18}, 
as described below in Section~\ref{sec:om}.

\subsection{Matrix Higgs fields and the orthogonal metal}
\label{sec:om}

The SU$_s$(2) gauge theory outlined in Section~\ref{sec:hubbard}, obtained by transforming to a rotating reference frame in spin space and then
Higgsing the SU$_s$(2) down to smaller groups, led to states with topological order often referred to as `algebraic charge liquids' (ACLs).
The ACLs have fermionic excitations which carry charge but not spin. Similarly, the transformations in
Section~\ref{sec:pairing}, which involve transforming to a rotating reference frame in pseudospin space, followed by Higgsing the pseudospin SU$_s$(2),
lead to states with topological order called `algebraic spin liquids' (ASLs). The ASLs have fermionic excitations which carry spin but not charge.
However, there is a third possibility: the `orthogonal metal' (OM) \cite{NMS12}, 
in which the fermions carry both spin and charge, along with a $\mathbb{Z}_2$ gauge charge
which makes them distinct from electrons. The OM can also be obtained by Higgsing either the gauged spin SU$_s$(2) or pseudospin SU$_c$(2) by a novel
{\it matrix\/} Higgs field, as was pointed out recently in Ref.~\cite{Snir18}.

We can view the Higgs field, $H_b$, of the ACL as a composite of the fermionic chargons $f_p$; the Yukawa coupling in Eq.~(\ref{yukawa})
implies that $H_b \sim f_{p}^\dagger \sigma^b_{pp'} f_p$. To obtain an OM, we have to consider a Higgs field which is a composite of the bosonic
spinons, $R$. From the definition in Eq.~(\ref{R}), we know that the SU(2) matrix field $R$ transforms as a spinor under global spin SU(2) upon left multiplication, and as a spinor under gauge SU$_s$(2) upon right multiplication. Upon considering pairs of $R$, we therefore expect fields which are
singlets or triplets under spin and gauge SU(2). All 4 possibilities are potentially realized by the fields $\mbox{Tr} \left( R R^\dagger \right)$, 
$\mbox{Tr} \left( \sigma^a R R^\dagger \right)$, $\mbox{Tr} \left( R \sigma^b R^\dagger \right)$, and $\mbox{Tr} \left( \sigma^a R \sigma^b R^\dagger \right)$. The first is a constant, the next two vanish, leaving only the matrix Higgs field
\beq
H_{ab} \sim \mbox{Tr} \left( \sigma^a R \sigma^b R^\dagger \right) \,.
\label{HRR}
\eeq
Note that the index $a$ is a triplet under the spin SU(2), while the index $b$ is a triplet under the gauge SU$_s$(2).
For the SU$_s$(2) gauge theory of the Hubbard model in Sections~\ref{sec:su2} and~\ref{sec:su2theory}, we now consider new phases where
$H_{ab}$ condenses. The phase where the only condensate is $\langle H_{ab} \rangle = H_0 \delta_{ab}$ turns out to be an OM.
Because the index $b$ is a SU$_s$(2) triplet, the condensate is invariant under the $\mathbb{Z}_2$ center of SU$_s$(2): so the phase has $\mathbb{Z}_2$ topological order. The diagonal $\delta_{ab}$ structure in the matrix space ties the gauge and spin indices, and consequently \cite{Snir18} 
the fermionic fields $f_p$ effectively
acquire a global SU(2) spin: the $f_p$ are then the fermionic excitations of the OM carrying both spin and charge, along with a $\mathbb{Z}_2$ gauge charge.

Refs.~\cite{Snir16,Snir18} also the addressed the nature of the confining transition out of the OM where the Higgs condensate $\langle H_{ab} \rangle$ vanishes. The case with the SU$_s$(2) gauge theory corresponding to a rotating reference frame in spin space leads to confining phases with superconducting and/or charge density wave order; see Fig.~\ref{fig:om}. They also considered an alternative model in which the OM is defined by transforming to a rotating reference frame in pseudospin space, using the $\widetilde{R}$ matrix in Eq.~(\ref{FR}), and a corresponding Higgs field (replacing Eq.~(\ref{HRR}))
\beq
 \widetilde{H}_{ab} \sim \mbox{Tr} \left( \sigma^a \widetilde{R} \sigma^b \widetilde{R}^\dagger \right) \,.
\label{tHRR}
\eeq
The $\widetilde{H}_{ab}$ condensate leads to the same OM state as the $H_{ab}$ condensate, but the confining state beyond the Higgs critical point is different: it has antiferromagnetic order, as shown in Fig.~\ref{fig:om}. 
So if we view the pseudogap phase as an OM, then two distinct theories are needed to reach the antiferromagnet or the superconductor/charge density wave, both illustrated in Fig.~\ref{fig:om}. For the antiferromagnetic state with spin order, we should gauge the pseudospin using the $\widetilde{R}$ matrix. Conversely, for the superconducting/charge density wave (or pair density wave) 
states with pseudospin order, we should gauge the spin using the $R$ matrix. We have followed the latter $R$ approach in the present paper, 
because we are interested in the evolution from the pseudogap to the larger doping superconducting, charge/pair density wave \cite{Edkins18}, 
and Fermi liquid states.

\ack
I thank Erez Berg, Shubhayu Chatterjee, Yoni Schattner, and Mathias Scheurer 
for recent collaborations on the material reviewed in Section~\ref{sec:hubbard}, and Fakher Assaad, Snir Gazit, and Ashvin Vishwanath for the collaboration reviewed in Section~\ref{sec:conc}.
I also acknowledge many stimulating discussions with the participants of the  
34th Jerusalem Winter School in Theoretical Physics at the Israel Institute for Advanced Studies.
This research was supported by NSF Grant DMR-1664842. Research at Perimeter Institute is supported by the Government of Canada through Industry Canada and by the Province of Ontario through the Ministry of Research and Innovation.
The author also acknowledges support from Cenovus Energy at Perimeter Institute, and from the Hanna Visiting Professor program at Stanford University.

\section*{References}
\bibliography{z2}

\providecommand{\href}[2]{#2}\begingroup\raggedright\begin{thebibliography}{100}

\bibitem{Luttinger}
J.~M. Luttinger and J.~C. Ward, \emph{{Ground-State Energy of a Many-Fermion
  System. II}}, \href{https://doi.org/10.1103/PhysRev.118.1417}{\emph{Phys.
  Rev.} {\bfseries 118} (Jun, 1960) 1417--1427}.

\bibitem{MO00}
M.~{Oshikawa}, \emph{{Topological Approach to Luttinger's Theorem and the Fermi
  Surface of a Kondo Lattice}},
  \href{https://doi.org/10.1103/PhysRevLett.84.3370}{\emph{Phys. Rev. Lett.}
  {\bfseries 84} (Apr., 2000) 3370},
  [\href{https://arxiv.org/abs/cond-mat/0002392}{{\ttfamily
  cond-mat/0002392}}].

\bibitem{TSMVSS04}
T.~{Senthil}, M.~{Vojta} and S.~{Sachdev}, \emph{{Weak magnetism and non-Fermi
  liquids near heavy-fermion critical points}},
  \href{https://doi.org/10.1103/PhysRevB.69.035111}{\emph{Phys. Rev. B}
  {\bfseries 69} (Jan., 2004) 035111},
  [\href{https://arxiv.org/abs/cond-mat/0305193}{{\ttfamily
  cond-mat/0305193}}].

\bibitem{APAV04}
A.~{Paramekanti} and A.~{Vishwanath}, \emph{{Extending Luttinger's theorem to
  $\mathbb{Z}_{2}$ fractionalized phases of matter}},
  \href{https://doi.org/10.1103/PhysRevB.70.245118}{\emph{Phys. Rev. B}
  {\bfseries 70} (Dec., 2004) 245118},
  [\href{https://arxiv.org/abs/cond-mat/0406619}{{\ttfamily
  cond-mat/0406619}}].

\bibitem{SBCS16}
S.~{Sachdev}, E.~{Berg}, S.~{Chatterjee} and Y.~{Schattner}, \emph{{Spin
  density wave order, topological order, and Fermi surface reconstruction}},
  \href{https://doi.org/10.1103/PhysRevB.94.115147}{\emph{Phys. Rev. B}
  {\bfseries 94} (Sept., 2016) 115147},
  [\href{https://arxiv.org/abs/1606.07813}{{\ttfamily 1606.07813}}].

\bibitem{ZXtopo}
J.-F. {He}, C.~R. {Rotundu}, M.~S. {Scheurer}, Y.~{He}, M.~{Hashimoto}, K.~{Xu}
  et~al., \emph{{Fermi surface reconstruction in electron-doped cuprates
  without antiferromagnetic long-range order}}, {\emph{preprint} (2018) }.

\bibitem{LTCP15}
S.~{Badoux}, W.~{Tabis}, F.~{Lalibert{\'e}}, G.~{Grissonnanche}, B.~{Vignolle},
  D.~{Vignolles} et~al., \emph{{Change of carrier density at the pseudogap
  critical point of a cuprate superconductor}},
  \href{https://doi.org/10.1038/nature16983}{\emph{Nature} {\bfseries 531}
  (Mar., 2016) 210--214}, [\href{https://arxiv.org/abs/1511.08162}{{\ttfamily
  1511.08162}}].

\bibitem{EMSY16}
A.~{Eberlein}, W.~{Metzner}, S.~{Sachdev} and H.~{Yamase}, \emph{{Fermi Surface
  Reconstruction and Drop in the Hall Number due to Spiral Antiferromagnetism
  in High-T$_{c}$ Cuprates}},
  \href{https://doi.org/10.1103/PhysRevLett.117.187001}{\emph{Phys. Rev. Lett.}
  {\bfseries 117} (Oct., 2016) 187001},
  [\href{https://arxiv.org/abs/1607.06087}{{\ttfamily 1607.06087}}].

\bibitem{CSE17}
S.~{Chatterjee}, S.~{Sachdev} and A.~{Eberlein}, \emph{{Thermal and electrical
  transport in metals and superconductors across antiferromagnetic and
  topological quantum transitions}},
  \href{https://doi.org/10.1103/PhysRevB.96.075103}{\emph{Phys. Rev. B}
  {\bfseries 96} (Aug., 2017) 075103},
  [\href{https://arxiv.org/abs/1704.02329}{{\ttfamily 1704.02329}}].

\bibitem{SP15}
S.~E. {Sebastian} and C.~{Proust}, \emph{{Quantum Oscillations in Hole-Doped
  Cuprates}},
  \href{https://doi.org/10.1146/annurev-conmatphys-030212-184305}{\emph{Annual
  Review of Condensed Matter Physics} {\bfseries 6} (Mar., 2015) 411--430},
  [\href{https://arxiv.org/abs/1507.01315}{{\ttfamily 1507.01315}}].

\bibitem{Greven16}
M.~K. {Chan}, N.~{Harrison}, R.~D. {McDonald}, B.~J. {Ramshaw}, K.~A. {Modic},
  N.~{Bari{\v s}i{\'c}} et~al., \emph{{Single reconstructed Fermi surface
  pocket in an underdoped single-layer cuprate superconductor}},
  \href{https://doi.org/10.1038/ncomms12244}{\emph{Nature Communications}
  {\bfseries 7} (July, 2016) 12244},
  [\href{https://arxiv.org/abs/1606.02772}{{\ttfamily 1606.02772}}].

\bibitem{Boebinger11}
S.~C. {Riggs}, O.~{Vafek}, J.~B. {Kemper}, J.~B. {Betts}, A.~{Migliori}, F.~F.
  {Balakirev} et~al., \emph{{Heat capacity through the magnetic-field-induced
  resistive transition in an underdoped high-temperature superconductor}},
  \href{https://doi.org/10.1038/nphys1921}{\emph{Nature Physics} {\bfseries 7}
  (Apr., 2011) 332--335}, [\href{https://arxiv.org/abs/1008.1568}{{\ttfamily
  1008.1568}}].

\bibitem{ACS14}
A.~{Allais}, D.~{Chowdhury} and S.~{Sachdev}, \emph{{Connecting high-field
  quantum oscillations to zero-field electron spectral functions in the
  underdoped cuprates}}, \href{https://doi.org/10.1038/ncomms6771}{\emph{Nature
  Communications} {\bfseries 5} (Dec., 2014) 5771},
  [\href{https://arxiv.org/abs/1406.0503}{{\ttfamily 1406.0503}}].

\bibitem{HasanRMP}
M.~Z. {Hasan} and C.~L. {Kane}, \emph{{Colloquium: Topological insulators}},
  \href{https://doi.org/10.1103/RevModPhys.82.3045}{\emph{Rev. Mod. Phys.}
  {\bfseries 82} (Oct., 2010) 3045--3067},
  [\href{https://arxiv.org/abs/1002.3895}{{\ttfamily 1002.3895}}].

\bibitem{ZhangRMP}
X.-L. {Qi} and S.-C. {Zhang}, \emph{{Rev. Mod. Phys.}},
  \href{https://doi.org/10.1103/RevModPhys.83.1057}{\emph{Rev. Mod. Phys.}
  {\bfseries 83} (Oct., 2011) 1057--1110},
  [\href{https://arxiv.org/abs/1008.2026}{{\ttfamily 1008.2026}}].

\bibitem{HasanARCMP}
M.~Z. {Hasan} and J.~E. {Moore}, \emph{{Three-Dimensional Topological
  Insulators}},
  \href{https://doi.org/10.1146/annurev-conmatphys-062910-140432}{\emph{Annual
  Review of Condensed Matter Physics} {\bfseries 2} (Mar., 2011) 55--78},
  [\href{https://arxiv.org/abs/1011.5462}{{\ttfamily 1011.5462}}].

\bibitem{wegner71}
F.~J. Wegner, \emph{{Duality in Generalized Ising Models and Phase Transitions
  without Local Order Parameters}},
  \href{https://doi.org/10.1063/1.1665530}{\emph{J. Math. Phys.} {\bfseries 12}
  (1971) 2259--2272}.

\bibitem{kogut79}
J.~B. Kogut, \emph{{An introduction to lattice gauge theory and spin systems}},
  \href{https://doi.org/10.1103/RevModPhys.51.659}{\emph{Rev. Mod. Phys.}
  {\bfseries 51} (Oct, 1979) 659--713}.

\bibitem{FradkinShenker}
E.~Fradkin and S.~H. Shenker, \emph{{Phase diagrams of lattice gauge theories
  with Higgs fields}},
  \href{https://doi.org/10.1103/PhysRevD.19.3682}{\emph{Phys. Rev. D}
  {\bfseries 19} (Jun, 1979) 3682--3697}.

\bibitem{Berezinskii1}
V.~L. {Berezinski{\v i}}, \emph{{Destruction of Long-range Order in
  One-dimensional and Two-dimensional Systems having a Continuous Symmetry
  Group I. Classical Systems}}, {\emph{Soviet Journal of Experimental and
  Theoretical Physics} {\bfseries 32} (1971) 493}.

\bibitem{Berezinskii2}
V.~L. {Berezinski{\v i}}, \emph{{Destruction of Long-range Order in
  One-dimensional and Two-dimensional Systems Possessing a Continuous Symmetry
  Group. II. Quantum Systems}}, {\emph{Soviet Journal of Experimental and
  Theoretical Physics} {\bfseries 34} (1972) 610}.

\bibitem{KT73}
J.~M. Kosterlitz and D.~J. Thouless, \emph{{Ordering, metastability and phase
  transitions in two-dimensional systems}},
  \href{https://doi.org/10.1088/0022-3719/6/7/010}{\emph{Journal of Physics C:
  Solid State Physics} {\bfseries 6} (1973) 1181}.

\bibitem{KT74}
J.~M. Kosterlitz, \emph{{The critical properties of the two-dimensional XY
  model}}, \href{https://doi.org/10.1088/0022-3719/7/6/005}{\emph{Journal of
  Physics C: Solid State Physics} {\bfseries 7} (1974) 1046}.

\bibitem{RJSS91}
R.~A. Jalabert and S.~Sachdev, \emph{{Spontaneous alignment of frustrated bonds
  in an anisotropic, three-dimensional Ising model}},
  \href{https://doi.org/10.1103/PhysRevB.44.686}{\emph{Phys. Rev. B} {\bfseries
  44} (Jul, 1991) 686--690}.

\bibitem{SSMV99}
S.~{Sachdev} and M.~{Vojta}, \emph{{Translational symmetry breaking in
  two-dimensional antiferromagnets and superconductors}}, {\emph{J. Phys. Soc.
  Jpn {\bf 69}, Supp. B, 1} (Oct., 1999) },
  [\href{https://arxiv.org/abs/cond-mat/9910231}{{\ttfamily
  cond-mat/9910231}}].

\bibitem{SM02}
T.~{Senthil} and O.~I. {Motrunich}, \emph{{Microscopic models for
  fractionalized phases in strongly correlated systems}},
  \href{https://doi.org/10.1103/PhysRevB.66.205104}{\emph{Phys. Rev. B}
  {\bfseries 66} (Nov., 2002) 205104},
  [\href{https://arxiv.org/abs/cond-mat/0201320}{{\ttfamily
  cond-mat/0201320}}].

\bibitem{SM02PRL}
O.~I. {Motrunich} and T.~{Senthil}, \emph{{Exotic Order in Simple Models of
  Bosonic Systems}},
  \href{https://doi.org/10.1103/PhysRevLett.89.277004}{\emph{Phys. Rev. Lett.}
  {\bfseries 89} (Dec., 2002) 277004},
  [\href{https://arxiv.org/abs/cond-mat/0205170}{{\ttfamily
  cond-mat/0205170}}].

\bibitem{DRSK88}
D.~Rokhsar and S.~A. Kivelson, \emph{{Superconductivity and the Quantum
  Hard-Core Dimer Gas}},
  \href{https://doi.org/10.1103/PhysRevLett.61.2376}{\emph{Phys. Rev. Lett.}
  {\bfseries 61} (Nov, 1988) 2376--2379}.

\bibitem{EFSK90}
E.~Fradkin and S.~A. Kivelson, \emph{Short range resonating valence bond
  theories and superconductivity},
  \href{https://doi.org/10.1142/S0217984990000295}{\emph{Mod. Phys. Lett. B}
  {\bfseries 04} (1990) 225--232}.

\bibitem{NRSS90}
N.~Read and S.~Sachdev, \emph{{Spin-Peierls, valence-bond solid, and N\'eel
  ground states of low-dimensional quantum antiferromagnets}},
  \href{https://doi.org/10.1103/PhysRevB.42.4568}{\emph{Phys. Rev. B}
  {\bfseries 42} (Sep, 1990) 4568--4589}.

\bibitem{EH13}
A.~M. {Essin} and M.~{Hermele}, \emph{{Classifying fractionalization: Symmetry
  classification of gapped Z$_{2}$ spin liquids in two dimensions}},
  \href{https://doi.org/10.1103/PhysRevB.87.104406}{\emph{Phys. Rev. B}
  {\bfseries 87} (Mar., 2013) 104406},
  [\href{https://arxiv.org/abs/1212.0593}{{\ttfamily 1212.0593}}].

\bibitem{SPTSET}
X.~{Chen}, Z.-C. {Gu}, Z.-X. {Liu} and X.-G. {Wen}, \emph{{Symmetry protected
  topological orders and the group cohomology of their symmetry group}},
  \href{https://doi.org/10.1103/PhysRevB.87.155114}{\emph{Phys. Rev. B}
  {\bfseries 87} (Apr., 2013) 155114},
  [\href{https://arxiv.org/abs/1106.4772}{{\ttfamily 1106.4772}}].

\bibitem{HPS11}
Y.~{Huh}, M.~{Punk} and S.~{Sachdev}, \emph{{Vison states and confinement
  transitions of Z$_{2}$ spin liquids on the kagome lattice}},
  \href{https://doi.org/10.1103/PhysRevB.84.094419}{\emph{Phys. Rev. B}
  {\bfseries 84} (Sept., 2011) 094419},
  [\href{https://arxiv.org/abs/1106.3330}{{\ttfamily 1106.3330}}].

\bibitem{NRSS89}
N.~Read and S.~Sachdev, \emph{{Valence-bond and spin-Peierls ground states of
  low-dimensional quantum antiferromagnets}},
  \href{https://doi.org/10.1103/PhysRevLett.62.1694}{\emph{Phys. Rev. Lett.}
  {\bfseries 62} (Apr, 1989) 1694--1697}.

\bibitem{senthil1}
T.~{Senthil}, A.~{Vishwanath}, L.~{Balents}, S.~{Sachdev} and M.~P.~A.
  {Fisher}, \emph{{Deconfined Quantum Critical Points}},
  \href{https://doi.org/10.1126/science.1091806}{\emph{Science} {\bfseries 303}
  (Mar., 2004) 1490--1494},
  [\href{https://arxiv.org/abs/cond-mat/0311326}{{\ttfamily
  cond-mat/0311326}}].

\bibitem{senthil2}
T.~{Senthil}, L.~{Balents}, S.~{Sachdev}, A.~{Vishwanath} and M.~P.~A.
  {Fisher}, \emph{{Quantum criticality beyond the Landau-Ginzburg-Wilson
  paradigm}}, \href{https://doi.org/10.1103/PhysRevB.70.144407}{\emph{Phys.
  Rev. B} {\bfseries 70} (Oct., 2004) 144407},
  [\href{https://arxiv.org/abs/cond-mat/0312617}{{\ttfamily
  cond-mat/0312617}}].

\bibitem{SS09}
S.~{Sachdev}, M.~A. {Metlitski}, Y.~{Qi} and C.~{Xu}, \emph{{Fluctuating spin
  density waves in metals}},
  \href{https://doi.org/10.1103/PhysRevB.80.155129}{\emph{Phys. Rev. B}
  {\bfseries 80} (Oct., 2009) 155129},
  [\href{https://arxiv.org/abs/0907.3732}{{\ttfamily 0907.3732}}].

\bibitem{SSRoyal}
S.~{Sachdev}, \emph{{Emergent gauge fields and the high-temperature
  superconductors}},
  \href{https://doi.org/10.1098/rsta.2015.0248}{\emph{Philosophical
  Transactions of the Royal Society of London Series A} {\bfseries 374} (Aug.,
  2016) 20150248}, [\href{https://arxiv.org/abs/1512.00465}{{\ttfamily
  1512.00465}}].

\bibitem{SSNambu}
S.~{Sachdev} and D.~{Chowdhury}, \emph{{The novel metallic states of the
  cuprates: Fermi liquids with topological order and strange metals}},
  \href{https://doi.org/10.1093/ptep/ptw110}{\emph{Progress of Theoretical and
  Experimental Physics} {\bfseries 2016} (Dec., 2016) 12C102},
  [\href{https://arxiv.org/abs/1605.03579}{{\ttfamily 1605.03579}}].

\bibitem{SCSS17}
S.~{Chatterjee} and S.~{Sachdev}, \emph{{Insulators and metals with topological
  order and discrete symmetry breaking}},
  \href{https://doi.org/10.1103/PhysRevB.95.205133}{\emph{Phys. Rev. B}
  {\bfseries 95} (May, 2017) 205133},
  [\href{https://arxiv.org/abs/1703.00014}{{\ttfamily 1703.00014}}].

\bibitem{CSS17}
S.~Chatterjee, S.~Sachdev and M.~Scheurer, \emph{{Intertwining topological
  order and broken symmetry in a theory of fluctuating spin density waves}},
  \href{https://doi.org/10.1103/PhysRevLett.119.227002}{\emph{Phys. Rev. Lett.}
  {\bfseries 119} (2017) 227002},
  [\href{https://arxiv.org/abs/1705.06289}{{\ttfamily 1705.06289}}].

\bibitem{WSCSGF}
W.~{Wu}, M.~S. {Scheurer}, S.~{Chatterjee}, S.~{Sachdev}, A.~{Georges} and
  M.~{Ferrero}, \emph{{Pseudogap and Fermi-Surface Topology in the
  Two-Dimensional Hubbard Model}},
  \href{https://doi.org/10.1103/PhysRevX.8.021048}{\emph{Phys. Rev. X}
  {\bfseries 8} (Apr., 2018) 021048},
  [\href{https://arxiv.org/abs/1707.06602}{{\ttfamily 1707.06602}}].

\bibitem{SCWFGS}
M.~S. {Scheurer}, S.~{Chatterjee}, W.~{Wu}, M.~{Ferrero}, A.~{Georges} and
  S.~{Sachdev}, \emph{{Topological order in the pseudogap metal}},
  \href{https://doi.org/10.1073/pnas.1720580115}{\emph{Proc. Nat. Acad. Sci.}
  {\bfseries 115} (May, 2018) E3665},
  [\href{https://arxiv.org/abs/1711.09925}{{\ttfamily 1711.09925}}].

\bibitem{MSSS18}
M.~S. {Scheurer} and S.~{Sachdev}, \emph{{Orbital currents in insulating and
  doped antiferromagnets}}, {\emph{ArXiv e-prints} (Aug., 2018) },
  [\href{https://arxiv.org/abs/1808.04826}{{\ttfamily 1808.04826}}].

\bibitem{XS10}
C.~{Xu} and S.~{Sachdev}, \emph{{Majorana Liquids: The Complete
  Fractionalization of the Electron}},
  \href{https://doi.org/10.1103/PhysRevLett.105.057201}{\emph{Phys. Rev. Lett.}
  {\bfseries 105} (July, 2010) 057201},
  [\href{https://arxiv.org/abs/1004.5431}{{\ttfamily 1004.5431}}].

\bibitem{wilsonfisher72}
K.~G. Wilson and M.~E. Fisher, \emph{{Critical Exponents in 3.99 Dimensions}},
  \href{https://doi.org/10.1103/PhysRevLett.28.240}{\emph{Phys. Rev. Lett.}
  {\bfseries 28} (Jan, 1972) 240--243}.

\bibitem{2016PhRvL.117u0401S}
M.~{Schuler}, S.~{Whitsitt}, L.-P. {Henry}, S.~{Sachdev} and A.~M.
  {L{\"a}uchli}, \emph{{Universal Signatures of Quantum Critical Points from
  Finite-Size Torus Spectra: A Window into the Operator Content of
  Higher-Dimensional Conformal Field Theories}},
  \href{https://doi.org/10.1103/PhysRevLett.117.210401}{\emph{Phys. Rev. Lett.}
  {\bfseries 117} (Nov., 2016) 210401},
  [\href{https://arxiv.org/abs/1603.03042}{{\ttfamily 1603.03042}}].

\bibitem{2016PhRvB..94h5134W}
S.~{Whitsitt} and S.~{Sachdev}, \emph{{Transition from the $\mathbb{Z}_{2}$
  spin liquid to antiferromagnetic order: Spectrum on the torus}},
  \href{https://doi.org/10.1103/PhysRevB.94.085134}{\emph{Phys. Rev. B}
  {\bfseries 94} (Aug., 2016) 085134},
  [\href{https://arxiv.org/abs/1603.05652}{{\ttfamily 1603.05652}}].

\bibitem{NRSS91}
N.~Read and S.~Sachdev, \emph{{Large $N$ expansion for frustrated quantum
  antiferromagnets}},
  \href{https://doi.org/10.1103/PhysRevLett.66.1773}{\emph{Phys. Rev. Lett.}
  {\bfseries 66} (Apr, 1991) 1773--1776}.

\bibitem{Wen91}
X.~G. Wen, \emph{{Mean-field theory of spin-liquid states with finite energy
  gap and topological orders}},
  \href{https://doi.org/10.1103/PhysRevB.44.2664}{\emph{Phys. Rev. B}
  {\bfseries 44} (Aug, 1991) 2664--2672}.

\bibitem{Bais92}
F.~A. Bais, P.~van Driel and M.~de~Wild~Propitius, \emph{Quantum symmetries in
  discrete gauge theories},
  \href{https://doi.org/10.1016/0370-2693(92)90773-W}{\emph{Phys. Lett. B}
  {\bfseries 280} (1992) 63 -- 70}.

\bibitem{MMS01}
J.~M. Maldacena, G.~W. Moore and N.~Seiberg, \emph{{D-brane charges in
  five-brane backgrounds}},
  \href{https://doi.org/10.1088/1126-6708/2001/10/005}{\emph{JHEP} {\bfseries
  10} (2001) 005}, [\href{https://arxiv.org/abs/hep-th/0108152}{{\ttfamily
  hep-th/0108152}}].

\bibitem{Kitaev03}
A.~Y. {Kitaev}, \emph{{Fault-tolerant quantum computation by anyons}},
  \href{https://doi.org/10.1016/S0003-4916(02)00018-0}{\emph{Annals of Physics}
  {\bfseries 303} (Jan., 2003) 2--30},
  [\href{https://arxiv.org/abs/quant-ph/9707021}{{\ttfamily
  quant-ph/9707021}}].

\bibitem{Nayak04}
M.~{Freedman}, C.~{Nayak}, K.~{Shtengel}, K.~{Walker} and Z.~{Wang}, \emph{{A
  class of P, T-invariant topological phases of interacting electrons}},
  \href{https://doi.org/10.1016/j.aop.2004.01.006}{\emph{Annals of Physics}
  {\bfseries 310} (Apr., 2004) 428--492},
  [\href{https://arxiv.org/abs/cond-mat/0307511}{{\ttfamily
  cond-mat/0307511}}].

\bibitem{Hansson04}
T.~H. {Hansson}, V.~{Oganesyan} and S.~L. {Sondhi}, \emph{{Superconductors are
  topologically ordered}},
  \href{https://doi.org/10.1016/j.aop.2004.05.006}{\emph{Annals of Physics}
  {\bfseries 313} (Oct., 2004) 497--538},
  [\href{https://arxiv.org/abs/cond-mat/0404327}{{\ttfamily
  cond-mat/0404327}}].

\bibitem{Kivelson89}
S.~Kivelson, \emph{Statistics of holons in the quantum hard-core dimer gas},
  \href{https://doi.org/10.1103/PhysRevB.39.259}{\emph{Phys. Rev. B} {\bfseries
  39} (Jan, 1989) 259--264}.

\bibitem{RC89}
N.~Read and B.~Chakraborty, \emph{Statistics of the excitations of the
  resonating-valence-bond state},
  \href{https://doi.org/10.1103/PhysRevB.40.7133}{\emph{Phys. Rev. B}
  {\bfseries 40} (Oct, 1989) 7133--7140}.

\bibitem{SenthilFisher}
T.~{Senthil} and M.~P.~A. {Fisher}, \emph{{$\mathbb{Z}_{2}$ gauge theory of
  electron fractionalization in strongly correlated systems}},
  \href{https://doi.org/10.1103/PhysRevB.62.7850}{\emph{Phys. Rev. B}
  {\bfseries 62} (Sept., 2000) 7850--7881},
  [\href{https://arxiv.org/abs/cond-mat/9910224}{{\ttfamily
  cond-mat/9910224}}].

\bibitem{Fisher67}
M.~E. Fisher and R.~J. Burford, \emph{{Theory of Critical-Point Scattering and
  Correlations. I. The Ising Model}},
  \href{https://doi.org/10.1103/PhysRev.156.583}{\emph{Phys. Rev.} {\bfseries
  156} (Apr, 1967) 583--622}.

\bibitem{SSNR91}
S.~Sachdev and N.~Read, \emph{{Large $N$ expansion for frustrated and doped
  quantum antiferromagnets}},
  \href{https://doi.org/10.1142/S0217979291000158}{\emph{Int. J. Mod. Phys. B}
  {\bfseries 5} (1991) 219--249},
  [\href{https://arxiv.org/abs/cond-mat/0402109}{{\ttfamily
  cond-mat/0402109}}].

\bibitem{LRT93}
P.~E. Lammert, D.~S. Rokhsar and J.~Toner, \emph{Topology and nematic
  ordering}, \href{https://doi.org/10.1103/PhysRevLett.70.1650}{\emph{Phys.
  Rev. Lett.} {\bfseries 70} (Mar, 1993) 1650--1653}.

\bibitem{CSS93}
A.~V. {Chubukov}, T.~{Senthil} and S.~{Sachdev}, \emph{{Universal magnetic
  properties of frustrated quantum antiferromagnets in two dimensions}},
  \href{https://doi.org/10.1103/PhysRevLett.72.2089}{\emph{Phys. Rev. Lett.}
  {\bfseries 72} (Mar., 1994) 2089--2092},
  [\href{https://arxiv.org/abs/cond-mat/9311045}{{\ttfamily
  cond-mat/9311045}}].

\bibitem{LRT95a}
P.~E. {Lammert}, D.~S. {Rokhsar} and J.~{Toner}, \emph{{Topology and nematic
  ordering. I. A gauge theory}},
  \href{https://doi.org/10.1103/PhysRevE.52.1778}{\emph{Phys. Rev. E}
  {\bfseries 52} (Aug., 1995) 1778--1800},
  [\href{https://arxiv.org/abs/cond-mat/9501101}{{\ttfamily
  cond-mat/9501101}}].

\bibitem{LRT95b}
J.~{Toner}, P.~E. {Lammert} and D.~S. {Rokhsar}, \emph{{Topology and nematic
  ordering. II. Observable critical behavior}},
  \href{https://doi.org/10.1103/PhysRevE.52.1801}{\emph{Phys. Rev. E}
  {\bfseries 52} (Aug., 1995) 1801--1810},
  [\href{https://arxiv.org/abs/cond-mat/9501100}{{\ttfamily
  cond-mat/9501100}}].

\bibitem{SP01}
S.~{Sachdev} and K.~{Park}, \emph{{Ground States of Quantum Antiferromagnets in
  Two Dimensions}}, \href{https://doi.org/10.1006/aphy.2002.6232}{\emph{Annals
  of Physics} {\bfseries 298} (May, 2002) 58--122},
  [\href{https://arxiv.org/abs/cond-mat/0108214}{{\ttfamily
  cond-mat/0108214}}].

\bibitem{SSS02}
R.~D. {Sedgewick}, D.~J. {Scalapino} and R.~L. {Sugar}, \emph{{Fractionalized
  phase in an XY-Z$_{2}$ gauge model}},
  \href{https://doi.org/10.1103/PhysRevB.65.054508}{\emph{Phys. Rev. B}
  {\bfseries 65} (Feb., 2002) 054508},
  [\href{https://arxiv.org/abs/cond-mat/0012028}{{\ttfamily
  cond-mat/0012028}}].

\bibitem{PS02}
K.~{Park} and S.~{Sachdev}, \emph{{Bond and N{\'e}el order and
  fractionalization in ground states of easy-plane antiferromagnets in two
  dimensions}}, \href{https://doi.org/10.1103/PhysRevB.65.220405}{\emph{Phys.
  Rev. B} {\bfseries 65} (June, 2002) 220405},
  [\href{https://arxiv.org/abs/cond-mat/0112003}{{\ttfamily
  cond-mat/0112003}}].

\bibitem{MSF01}
R.~{Moessner}, S.~L. {Sondhi} and E.~{Fradkin}, \emph{{Short-ranged resonating
  valence bond physics, quantum dimer models, and Ising gauge theories}},
  \href{https://doi.org/10.1103/PhysRevB.65.024504}{\emph{Phys. Rev. B}
  {\bfseries 65} (Jan., 2002) 024504},
  [\href{https://arxiv.org/abs/cond-mat/0103396}{{\ttfamily
  cond-mat/0103396}}].

\bibitem{Polyakov77}
A.~M. Polyakov, \emph{{Quark confinement and topology of gauge theories}},
  \href{https://doi.org/https://doi.org/10.1016/0550-3213(77)90086-4}{\emph{Nuclear
  Physics B} {\bfseries 120} (1977) 429 -- 458}.

\bibitem{Peskin78}
M.~E. Peskin, \emph{{Mandelstam-'t Hooft duality in abelian lattice models}},
  \href{https://doi.org/https://doi.org/10.1016/0003-4916(78)90252-X}{\emph{Annals
  of Physics} {\bfseries 113} (1978) 122 -- 152}.

\bibitem{DH81}
C.~Dasgupta and B.~I. Halperin, \emph{{Phase Transition in a Lattice Model of
  Superconductivity}},
  \href{https://doi.org/10.1103/PhysRevLett.47.1556}{\emph{Phys. Rev. Lett.}
  {\bfseries 47} (Nov, 1981) 1556--1560}.

\bibitem{TKPS08}
I.~S. {Tupitsyn}, A.~{Kitaev}, N.~V. {Prokof'Ev} and P.~C.~E. {Stamp},
  \emph{{Topological multicritical point in the phase diagram of the toric code
  model and three-dimensional lattice gauge Higgs model}},
  \href{https://doi.org/10.1103/PhysRevB.82.085114}{\emph{Phys. Rev. B}
  {\bfseries 82} (Aug., 2010) 085114},
  [\href{https://arxiv.org/abs/0804.3175}{{\ttfamily 0804.3175}}].

\bibitem{LFS01}
C.~{Lannert}, M.~P. {Fisher} and T.~{Senthil}, \emph{{Quantum confinement
  transition in a d-wave superconductor}},
  \href{https://doi.org/10.1103/PhysRevB.63.134510}{\emph{Phys. Rev. B}
  {\bfseries 63} (Apr., 2001) 134510},
  [\href{https://arxiv.org/abs/cond-mat/0007002}{{\ttfamily
  cond-mat/0007002}}].

\bibitem{SS04review}
S.~{Sachdev}, \emph{{Quantum Phases and Phase Transitions of Mott Insulators}},
   in \emph{Quantum Magnetism} (U.~{Schollw{\"o}ck}, J.~{Richter}, D.~J.~J.
  {Farnell} and R.~F. {Bishop}, eds.), vol.~645 of \emph{Lecture Notes in
  Physics, Berlin Springer Verlag}, p.~381, 2004,
  \href{https://arxiv.org/abs/cond-mat/0401041}{{\ttfamily cond-mat/0401041}},
  \href{https://doi.org/10.1007/BFb0119599}{DOI}.

\bibitem{dwave}
A.~D. {King}, J.~{Carrasquilla}, J.~{Raymond}, I.~{Ozfidan}, E.~{Andriyash},
  A.~{Berkley} et~al., \emph{{Observation of topological phenomena in a
  programmable lattice of 1,800 qubits}},
  \href{https://doi.org/10.1038/s41586-018-0410-x}{\emph{Nature} {\bfseries
  560} (Aug., 2018) 456--460},
  [\href{https://arxiv.org/abs/1803.02047}{{\ttfamily 1803.02047}}].

\bibitem{BBBSS05}
L.~{Balents}, L.~{Bartosch}, A.~{Burkov}, S.~{Sachdev} and K.~{Sengupta},
  \emph{{Putting competing orders in their place near the Mott transition}},
  \href{https://doi.org/10.1103/PhysRevB.71.144508}{\emph{Phys. Rev. B}
  {\bfseries 71} (Apr., 2005) 144508},
  [\href{https://arxiv.org/abs/cond-mat/0408329}{{\ttfamily
  cond-mat/0408329}}].

\bibitem{PCAS16}
A.~A. Patel, D.~Chowdhury, A.~Allais and S.~Sachdev, \emph{{Confinement
  transition to density wave order in metallic doped spin liquids}},
  \href{https://doi.org/10.1103/PhysRevB.93.165139}{\emph{Phys. Rev. B}
  {\bfseries 93} (2016) 165139},
  [\href{https://arxiv.org/abs/1602.05954}{{\ttfamily 1602.05954}}].

\bibitem{Haldane88}
F.~D.~M. Haldane, \emph{{O(3) Nonlinear $\sigma$ Model and the Topological
  Distinction between Integer- and Half-Integer-Spin Antiferromagnets in Two
  Dimensions}}, \href{https://doi.org/10.1103/PhysRevLett.61.1029}{\emph{Phys.
  Rev. Lett.} {\bfseries 61} (Aug, 1988) 1029--1032}.

\bibitem{MSC99}
R.~{Moessner}, S.~L. {Sondhi} and P.~{Chandra}, \emph{{Two-Dimensional Periodic
  Frustrated Ising Models in a Transverse Field}},
  \href{https://doi.org/10.1103/PhysRevLett.84.4457}{\emph{Phys. Rev. Lett.}
  {\bfseries 84} (May, 2000) 4457--4460},
  [\href{https://arxiv.org/abs/cond-mat/9910499}{{\ttfamily
  cond-mat/9910499}}].

\bibitem{RMSLS01}
R.~{Moessner} and S.~L. {Sondhi}, \emph{{Resonating Valence Bond Phase in the
  Triangular Lattice Quantum Dimer Model}},
  \href{https://doi.org/10.1103/PhysRevLett.86.1881}{\emph{Phys. Rev. Lett.}
  {\bfseries 86} (Feb., 2001) 1881},
  [\href{https://arxiv.org/abs/cond-mat/0007378}{{\ttfamily
  cond-mat/0007378}}].

\bibitem{MSC01}
R.~{Moessner}, S.~L. {Sondhi} and P.~{Chandra}, \emph{{Phase diagram of the
  hexagonal lattice quantum dimer model}},
  \href{https://doi.org/10.1103/PhysRevB.64.144416}{\emph{Phys. Rev. B}
  {\bfseries 64} (Oct., 2001) 144416},
  [\href{https://arxiv.org/abs/cond-mat/0106288}{{\ttfamily
  cond-mat/0106288}}].

\bibitem{VBS04}
A.~{Vishwanath}, L.~{Balents} and T.~{Senthil}, \emph{{Quantum criticality and
  deconfinement in phase transitions between valence bond solids}},
  \href{https://doi.org/10.1103/PhysRevB.69.224416}{\emph{Phys. Rev. B}
  {\bfseries 69} (June, 2004) 224416},
  [\href{https://arxiv.org/abs/cond-mat/0311085}{{\ttfamily
  cond-mat/0311085}}].

\bibitem{FHMOS04}
E.~{Fradkin}, D.~A. {Huse}, R.~{Moessner}, V.~{Oganesyan} and S.~L. {Sondhi},
  \emph{{Bipartite Rokhsar Kivelson points and Cantor deconfinement}},
  \href{https://doi.org/10.1103/PhysRevB.69.224415}{\emph{Phys. Rev. B}
  {\bfseries 69} (June, 2004) 224415},
  [\href{https://arxiv.org/abs/cond-mat/0311353}{{\ttfamily
  cond-mat/0311353}}].

\bibitem{OMAV04}
O.~I. {Motrunich} and A.~{Vishwanath}, \emph{{Emergent photons and transitions
  in the O(3) sigma model with hedgehog suppression}},
  \href{https://doi.org/10.1103/PhysRevB.70.075104}{\emph{Phys. Rev. B}
  {\bfseries 70} (Aug., 2004) 075104},
  [\href{https://arxiv.org/abs/cond-mat/0311222}{{\ttfamily
  cond-mat/0311222}}].

\bibitem{RKK07}
R.~K. {Kaul}, A.~{Kolezhuk}, M.~{Levin}, S.~{Sachdev} and T.~{Senthil},
  \emph{{Hole dynamics in an antiferromagnet across a deconfined quantum
  critical point}},
  \href{https://doi.org/10.1103/PhysRevB.75.235122}{\emph{Phys. Rev. B}
  {\bfseries 75} (June, 2007) 235122},
  [\href{https://arxiv.org/abs/cond-mat/0702119}{{\ttfamily
  cond-mat/0702119}}].

\bibitem{RKK08}
R.~K. {Kaul}, Y.~B. {Kim}, S.~{Sachdev} and T.~{Senthil}, \emph{{Algebraic
  charge liquids}}, \href{https://doi.org/10.1038/nphys790}{\emph{Nature
  Physics} {\bfseries 4} (Jan., 2008) 28--31},
  [\href{https://arxiv.org/abs/0706.2187}{{\ttfamily 0706.2187}}].

\bibitem{XGWPAL96}
X.-G. {Wen} and P.~A. {Lee}, \emph{{Theory of Underdoped Cuprates}},
  \href{https://doi.org/10.1103/PhysRevLett.76.503}{\emph{Phys. Rev. Lett.}
  {\bfseries 76} (Jan., 1996) 503--506},
  [\href{https://arxiv.org/abs/cond-mat/9506065}{{\ttfamily
  cond-mat/9506065}}].

\bibitem{TSSSMV03}
T.~{Senthil}, S.~{Sachdev} and M.~{Vojta}, \emph{{Fractionalized Fermi
  Liquids}}, \href{https://doi.org/10.1103/PhysRevLett.90.216403}{\emph{Phys.
  Rev. Lett.} {\bfseries 90} (May, 2003) 216403},
  [\href{https://arxiv.org/abs/cond-mat/0209144}{{\ttfamily
  cond-mat/0209144}}].

\bibitem{QS10}
Y.~{Qi} and S.~{Sachdev}, \emph{{Effective theory of Fermi pockets in
  fluctuating antiferromagnets}},
  \href{https://doi.org/10.1103/PhysRevB.81.115129}{\emph{Phys. Rev. B}
  {\bfseries 81} (Mar., 2010) 115129},
  [\href{https://arxiv.org/abs/0912.0943}{{\ttfamily 0912.0943}}].

\bibitem{Mei11}
J.-W. {Mei}, S.~{Kawasaki}, G.-Q. {Zheng}, Z.-Y. {Weng} and X.-G. {Wen},
  \emph{{Luttinger-volume violating Fermi liquid in the pseudogap phase of the
  cuprate superconductors}},
  \href{https://doi.org/10.1103/PhysRevB.85.134519}{\emph{Phys. Rev. B}
  {\bfseries 85} (Apr., 2012) 134519},
  [\href{https://arxiv.org/abs/1109.0406}{{\ttfamily 1109.0406}}].

\bibitem{MPSS12}
M.~{Punk} and S.~{Sachdev}, \emph{{Fermi surface reconstruction in hole-doped
  $t$-$J$ models without long-range antiferromagnetic order}},
  \href{https://doi.org/10.1103/PhysRevB.85.195123}{\emph{Phys. Rev. B}
  {\bfseries 85} (May, 2012) 195123},
  [\href{https://arxiv.org/abs/1202.4023}{{\ttfamily 1202.4023}}].

\bibitem{Punk15}
M.~Punk, A.~Allais and S.~Sachdev, \emph{{A quantum dimer model for the
  pseudogap metal}}, \href{https://doi.org/10.1073/pnas.1512206112}{\emph{Proc.
  Nat. Acad. Sci.} {\bfseries 112} (2015) 9552},
  [\href{https://arxiv.org/abs/1501.00978}{{\ttfamily 1501.00978}}].

\bibitem{Punk17a}
S.~{Huber}, J.~{Feldmeier} and M.~{Punk}, \emph{{Electron spectral functions in
  a quantum dimer model for topological metals}},
  \href{https://doi.org/10.1103/PhysRevB.97.075144}{\emph{Phys. Rev. B}
  {\bfseries 97} (Feb., 2018) 075144},
  [\href{https://arxiv.org/abs/1710.00012}{{\ttfamily 1710.00012}}].

\bibitem{Punk17b}
J.~{Feldmeier}, S.~{Huber} and M.~{Punk}, \emph{{Exact solution of a
  two-species quantum dimer model for pseudogap metals}},
  \href{https://doi.org/10.1103/PhysRevLett.120.187001}{\emph{Phys. Rev. Lett.}
  {\bfseries 120} (May, 2018) 187001},
  [\href{https://arxiv.org/abs/1712.01854}{{\ttfamily 1712.01854}}].

\bibitem{PSB05}
S.~{Powell}, S.~{Sachdev} and H.~P. {B{\"u}chler}, \emph{{Depletion of the
  Bose-Einstein condensate in Bose-Fermi mixtures}},
  \href{https://doi.org/10.1103/PhysRevB.72.024534}{\emph{Phys. Rev. B}
  {\bfseries 72} (July, 2005) 024534},
  [\href{https://arxiv.org/abs/cond-mat/0502299}{{\ttfamily
  cond-mat/0502299}}].

\bibitem{CPR05}
P.~{Coleman}, I.~{Paul} and J.~{Rech}, \emph{{Sum rules and Ward identities in
  the Kondo lattice}},
  \href{https://doi.org/10.1103/PhysRevB.72.094430}{\emph{Phys. Rev. B}
  {\bfseries 72} (Sept., 2005) 094430},
  [\href{https://arxiv.org/abs/cond-mat/0503001}{{\ttfamily
  cond-mat/0503001}}].

\bibitem{HS11}
L.~Huijse and S.~Sachdev, \emph{{Fermi surfaces and gauge-gravity duality}},
  \href{https://doi.org/10.1103/PhysRevD.84.026001}{\emph{Phys. Rev. D}
  {\bfseries 84} (2011) 026001},
  [\href{https://arxiv.org/abs/1104.5022}{{\ttfamily 1104.5022}}].

\bibitem{tH74}
G.~'t~Hooft, \emph{Magnetic monopoles in unified gauge theories},
  \href{https://doi.org/10.1016/0550-3213(74)90486-6}{\emph{Nucl. Phys. B}
  {\bfseries 79} (1974) 276 -- 284}.

\bibitem{Polyakov74}
A.~M. Polyakov, \emph{Particle spectrum in quantum field theory}, {\emph{JETP
  Lett.} {\bfseries 20} (1974) 194}.

\bibitem{Dunne01}
G.~V. Dunne, I.~I. Kogan, A.~Kovner and B.~Tekin, \emph{{Deconfining phase
  transition in (2+1)-dimensions: The Georgi-Glashow model}},
  \href{https://doi.org/10.1088/1126-6708/2001/01/032}{\emph{JHEP} {\bfseries
  01} (2001) 032}, [\href{https://arxiv.org/abs/hep-th/0010201}{{\ttfamily
  hep-th/0010201}}].

\bibitem{ATSS18}
A.~{Thomson} and S.~{Sachdev}, \emph{{Fermionic Spinon Theory of Square Lattice
  Spin Liquids near the N{\'e}el State}},
  \href{https://doi.org/10.1103/PhysRevX.8.011012}{\emph{Phys. Rev. X}
  {\bfseries 8} (Jan., 2018) 011012},
  [\href{https://arxiv.org/abs/1708.04626}{{\ttfamily 1708.04626}}].

\bibitem{Hermele04}
M.~{Hermele}, T.~{Senthil}, M.~P.~A. {Fisher}, P.~A. {Lee}, N.~{Nagaosa} and
  X.-G. {Wen}, \emph{{Stability of U(1) spin liquids in two dimensions}},
  \href{https://doi.org/10.1103/PhysRevB.70.214437}{\emph{Phys. Rev. B}
  {\bfseries 70} (Dec., 2004) 214437},
  [\href{https://arxiv.org/abs/cond-mat/0404751}{{\ttfamily
  cond-mat/0404751}}].

\bibitem{Moon09}
E.~G. {Moon} and S.~{Sachdev}, \emph{{Competition between spin density wave
  order and superconductivity in the underdoped cuprates}},
  \href{https://doi.org/10.1103/PhysRevB.80.035117}{\emph{Phys. Rev. B}
  {\bfseries 80} (Jul, 2009) 035117},
  [\href{https://arxiv.org/abs/0905.2608}{{\ttfamily 0905.2608}}].

\bibitem{YangWang16}
X.~{Yang} and F.~{Wang}, \emph{{Schwinger boson spin-liquid states on square
  lattice}}, \href{https://doi.org/10.1103/PhysRevB.94.035160}{\emph{Phys. Rev.
  B} {\bfseries 94} (July, 2016) 035160},
  [\href{https://arxiv.org/abs/1507.07621}{{\ttfamily 1507.07621}}].

\bibitem{Dombre}
T.~Dombre and N.~Read, \emph{{Nonlinear $\sigma$ models for triangular quantum
  antiferromagnets}},
  \href{https://doi.org/10.1103/PhysRevB.39.6797}{\emph{Phys. Rev. B}
  {\bfseries 39} (Apr, 1989) 6797--6801}.

\bibitem{CSS94}
A.~V. {Chubukov}, S.~{Sachdev} and T.~{Senthil}, \emph{{Quantum phase
  transitions in frustrated quantum antiferromagnets}},
  \href{https://doi.org/10.1016/0550-3213(94)90023-X}{\emph{Nucl. Phys. B}
  {\bfseries 426} (Sept., 1994) 601--643},
  [\href{https://arxiv.org/abs/cond-mat/9402006}{{\ttfamily
  cond-mat/9402006}}].

\bibitem{DCSS15b}
D.~{Chowdhury} and S.~{Sachdev}, \emph{{Higgs criticality in a two-dimensional
  metal}}, \href{https://doi.org/10.1103/PhysRevB.91.115123}{\emph{Phys. Rev.
  B} {\bfseries 91} (Mar., 2015) 115123},
  [\href{https://arxiv.org/abs/1412.1086}{{\ttfamily 1412.1086}}].

\bibitem{Snir18}
S.~{Gazit}, F.~{Assaad}, S.~{Sachdev}, A.~{Vishwanath} and C.~{Wang},
  \emph{{Confinement transition of the orthogonal semi-metal: emergent QCD$_3$
  and SO(5) symmetry}},
  \href{https://doi.org/10.1073/pnas.1806338115}{\emph{Proc. Nat. Acad. Sci.}
  (2018) E6987}, [\href{https://arxiv.org/abs/1804.01095}{{\ttfamily
  1804.01095}}].

\bibitem{Kibble02}
A.~C. Davis, A.~Hart, T.~W.~B. Kibble and A.~Rajantie, \emph{{Monopole mass in
  the three-dimensional Georgi-Glashow model}},
  \href{https://doi.org/10.1103/PhysRevD.65.125008}{\emph{Phys. Rev. D}
  {\bfseries 65} (Jun, 2002) 125008}.

\bibitem{SSTM02}
S.~{Sachdev} and T.~{Morinari}, \emph{{Strongly coupled quantum criticality
  with a Fermi surface in two dimensions: Fractionalization of spin and charge
  collective modes}},
  \href{https://doi.org/10.1103/PhysRevB.66.235117}{\emph{Phys. Rev. B}
  {\bfseries 66} (Dec., 2002) 235117},
  [\href{https://arxiv.org/abs/cond-mat/0207167}{{\ttfamily
  cond-mat/0207167}}].

\bibitem{TGTS09}
T.~{Grover} and T.~{Senthil}, \emph{{Quantum phase transition from an
  antiferromagnet to a spin liquid in a metal}},
  \href{https://doi.org/10.1103/PhysRevB.81.205102}{\emph{Phys. Rev. B}
  {\bfseries 81} (May, 2010) 205102},
  [\href{https://arxiv.org/abs/0910.1277}{{\ttfamily 0910.1277}}].

\bibitem{hertz}
J.~A. Hertz, \emph{{Quantum critical phenomena}},
  \href{https://doi.org/10.1103/PhysRevB.14.1165}{\emph{Phys. Rev. B}
  {\bfseries 14} (Aug, 1976) 1165--1184}.

\bibitem{RKK08b}
R.~K. {Kaul}, M.~A. {Metlitski}, S.~{Sachdev} and C.~{Xu}, \emph{{Destruction
  of N{\'e}el order in the cuprates by electron doping}},
  \href{https://doi.org/10.1103/PhysRevB.78.045110}{\emph{Phys. Rev. B}
  {\bfseries 78} (July, 2008) 045110},
  [\href{https://arxiv.org/abs/0804.1794}{{\ttfamily 0804.1794}}].

\bibitem{GS10}
T.~{Grover} and T.~{Senthil}, \emph{{Quantum phase transition from an
  antiferromagnet to a spin liquid in a metal}},
  \href{https://doi.org/10.1103/PhysRevB.81.205102}{\emph{Phys. Rev. B}
  {\bfseries 81} (May, 2010) 205102},
  [\href{https://arxiv.org/abs/0910.1277}{{\ttfamily 0910.1277}}].

\bibitem{LeeWenRMP}
P.~A. Lee, N.~Nagaosa and X.-G. Wen, \emph{{Doping a Mott insulator: Physics of
  high-temperature superconductivity}},
  \href{https://doi.org/10.1103/RevModPhys.78.17}{\emph{Rev. Mod. Phys.}
  {\bfseries 78} (Jan, 2006) 17--85},
  [\href{https://arxiv.org/abs/cond-mat/0410445}{{\ttfamily
  cond-mat/0410445}}].

\bibitem{leeprx}
P.~A. {Lee}, \emph{{Amperean Pairing and the Pseudogap Phase of Cuprate
  Superconductors}},
  \href{https://doi.org/10.1103/PhysRevX.4.031017}{\emph{Phys. Rev. X}
  {\bfseries 4} (July, 2014) 031017},
  [\href{https://arxiv.org/abs/1401.0519}{{\ttfamily 1401.0519}}].

\bibitem{Snir16}
S.~{Gazit}, M.~{Randeria} and A.~{Vishwanath}, \emph{{Charged fermions coupled
  to $\mathbb{Z}_2$ gauge fields: Superfluidity, confinement and emergent Dirac
  fermions}}, \href{https://doi.org/10.1038/nphys4028}{\emph{Nature Physics}
  {\bfseries 13} (2017) 484},
  [\href{https://arxiv.org/abs/1607.03892}{{\ttfamily 1607.03892}}].

\bibitem{NMS12}
R.~{Nandkishore}, M.~A. {Metlitski} and T.~{Senthil}, \emph{{Orthogonal metals:
  The simplest non-Fermi liquids}},
  \href{https://doi.org/10.1103/PhysRevB.86.045128}{\emph{Phys. Rev. B}
  {\bfseries 86} (July, 2012) 045128},
  [\href{https://arxiv.org/abs/1201.5998}{{\ttfamily 1201.5998}}].

\bibitem{Edkins18}
S.~D. {Edkins}, A.~{Kostin}, K.~{Fujita}, A.~P. {Mackenzie}, H.~{Eisaki}, S.-I.
  {Uchida} et~al., \emph{{Magnetic-field Induced Pair Density Wave State in the
  Cuprate Vortex Halo}}, {\emph{ArXiv e-prints} (Feb., 2018) },
  [\href{https://arxiv.org/abs/1802.04673}{{\ttfamily 1802.04673}}].

\end{thebibliography}\endgroup

\end{document}